\documentclass[12pt]{iopart}

\expandafter\let\csname equation*\endcsname\relax
\expandafter\let\csname endequation*\endcsname\relax
\usepackage{amsmath} \usepackage{amssymb} \usepackage{hyperref}
\usepackage{graphicx} \usepackage{mathrsfs} %\usepackage{fullpage}
\usepackage[usenames,dvipsnames]{color} \usepackage{ulem}
\usepackage{caption} \usepackage{subcaption}

\newcommand{\scri}{\mathscr{I}}

\usepackage[top=1in, bottom=1in, left=1.5in, right=1in]{geometry}
\numberwithin{equation}{section}

\begin{document}

\title{Spectral Cauchy Characteristic Extraction of strain, news and
  gravitational radiation flux} \author{Casey J. Handmer${}^{1}$,
  B\'{e}la Szil\'{a}gyi${}^{1,2}$, Jeffrey Winicour${}^3$}
\address{${}^{1}$TAPIR, Walter Burke Institute for Theoretical
  Physics, MC 350-17, California Institute of Technology, 1200 E
  California Blvd, Pasadena CA 91125, USA \\ ${}^{2}$Jet Propulsion
  Laboratory, California Institute of Technology, 4800 Oak Grove
  Dr. Pasadena CA, 91109, USA \\${}^{3}$Department of Physics and
  Astronomy University of Pittsburgh, Pittsburgh, PA 15260, USA}
\ead{chandmer@caltech.edu}

\begin{abstract}

We present a new approach for the Cauchy-characteristic extraction of
gravitational radiation strain, news function, and the flux of the
energy-momentum, supermomentum and angular momentum associated with
the Bondi-Metzner-Sachs asymptotic symmetries. In
Cauchy-characteristic extraction, a characteristic evolution code
takes numerical data on an inner worldtube supplied by a Cauchy
evolution code, and propagates it outwards to obtain the space-time
metric in a neighborhood of null infinity. The metric is first
determined in a scrambled form in terms of coordinates determined by
the Cauchy formalism. In prior treatments, the waveform is first
extracted from this metric and then transformed into an asymptotic
inertial coordinate system. This procedure provides the physically
proper description of the waveform and the radiated energy but it does
not generalize to determine the flux of angular momentum or
supermomentum.  Here we formulate and implement a new approach which
transforms the full metric into an asymptotic inertial frame and
provides a uniform treatment of all the radiation fluxes associated
with the asymptotic symmetries.  Computations are performed and
calibrated using the Spectral Einstein Code (\texttt{SpEC}).

\end{abstract}


\maketitle

\section{Introduction}

The strong gravitational radiation produced in the inspiral and merger
of binary black holes has been a dominant motivation for the
construction of gravitational wave observatories. This effort has
recently been brought to fruition by the observation of a binary
inspiral and merger by the LIGO gravitational wave
detectors~\cite{gw2016}.  The details of the gravitational waveform
supplied by numerical simulation is a key theoretical tool to fully
complement the sensitivity of the LIGO and Virgo
observatories~\cite{Waldman2011,Accadia:2009zz,Grote:2010zz,Somiya:2012},
by enhancing the detection and the scientific interpretation of the
gravitational signal. Besides the gravitational waveform, the flux of
energy-momentum carried off by the waves has important astrophysical
effects on the binary system. In particular, the recoil ``kick'' on
the binary due the radiative loss of momentum can possibly eject the
final black hole from a galactic center. The strength of such kicks
has been computed by various
means~\cite{Tichy:2007hk,Lousto:2011kp,Gonzalez2007,Favata2004,Baker2008,Healy2008}.
The most unambiguous and accurate approach is in terms of the Bondi news
function~\cite{Bondi1962}, which supplies the gravitational energy and
momentum flux at future null infinity $\scri^+$.

This can be carried out via Cauchy-characteristic extraction (CCE), in
which the Cauchy evolution is used to supply the boundary data on a
timelike inner worldtube, which then allows a characteristic
evolution extending to $\scri^+$ where the radiation is computed
using the geometric methods developed by Bondi {\it et al.}~\cite{Bondi1962},
Sachs~\cite{Sachs1962} and Penrose~\cite{Penrose1963}. For a review,
see~\cite{Winicour2009}. A version of this initial-boundary value
problem based upon a timelike worldtube~\cite{TamburinoWinicour1966}
has been implemented as a characteristic evolution code, the PITT null
code~\cite{Isaacson:1983,Bishop:1997ik,BabiucEtAl2008,Babiuc:2010ze},
and more recently as the \texttt{SpEC} characteristic
code~\cite{Handmer:2014,Handmer:2015}, both of which incorporate a
Penrose compactification of the exterior space-time extending to $\scri^+$.
In this way, characteristic evolution coupled to Cauchy evolution has been
implemented to give an accurate numerical computation of the Bondi
news function, which determines both the waveform and the radiated
energy-momentum.

\begin{figure}[h!]
  \centering
  \includegraphics[width=0.5\textwidth]{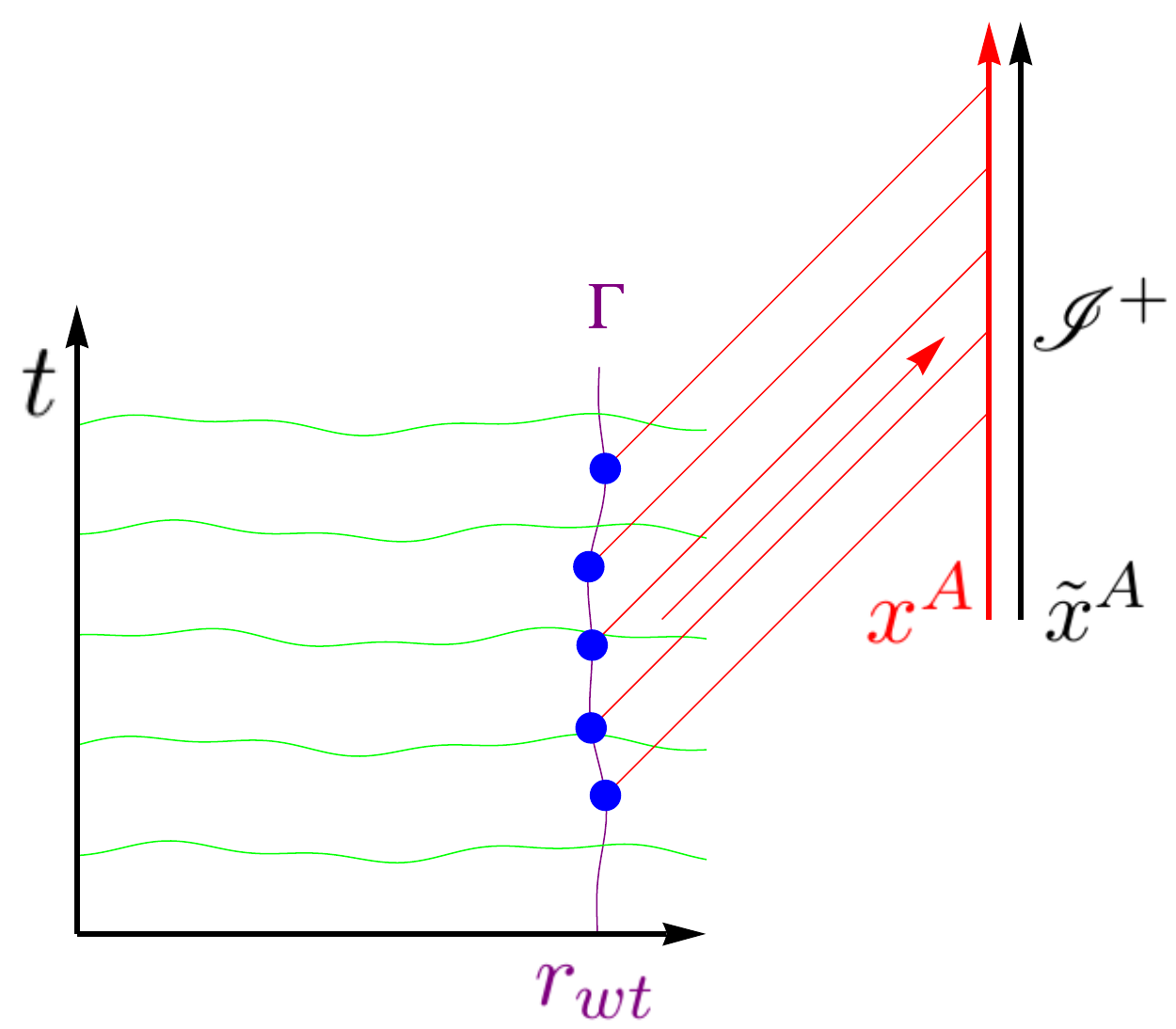}
  \caption{\small{Cauchy characteristic extraction. A Cauchy evolution
      of the Einstein field equation proceeds on a space-like
      foliation (green). A finite topologically spherical worldtube
      \(\Gamma\) at areal radius \(r_{wt}\) forms the inner boundary
      to a characteristic evolution on a null foliation (red). Based
      on a spherical coordinate system \(x^A\) constructed from Cauchy
      coordinates on the worldtube, gravitational information is
      propagated to compactified future null infinity \(\scri^+\). At
      \(\scri^+\), an inertial coordinate system \(\tilde x^A\) is
      co-evolved, in which the desired physical waveform can be
      expressed.}}
  \label{fig:AGNewsInertialDiagram}
\end{figure}

One technical complication introduced by CCE is that the coordinates
induced on $\scri^+$ are related to the Cauchy coordinates on the
inner worldtube. Consequently, these computational coordinates do not
correspond to inertial observers at $\scri^+$, i.e.  to the
coordinates intrinsic to a distant freely falling and non-rotating
observatory. Thus, the gravitational waveform first obtained in the
``computational coordinates'' of CCE is in a scrambled form. Because
the news function is an invariant irrespective of coordinate system,
the procedure up to now has been to compute it first in the
computational coordinates.\footnote{More accurately, the news tensor
$N_{ab}$ is a gauge invariant
  field on $\scri^+$~\cite{Geroch1977}. The news function
  $N=\frac{1}{4}N_{ab} q^a q^b$ depends upon a choice of complex
  polarization dyad $q^a$, which has gauge freedom. But this freedom
  can be trivially unwrapped by the construction $N_{ab}=N \bar q_a
  \bar q_b + \bar N q_a q_b$.} 
It is then unscrambled by constructing
the transformation between code coordinates and inertial coordinates
on $\scri^+$, as portrayed in Fig.~\ref{fig:AGNewsInertialDiagram}.
Likewise, a physically relevant calculation of the radiation waveform
must also be referred to such inertial coordinates on $\scri^+$.

In addition to energy-momentum loss, the gravitational radiation of
angular momentum has important consequences for the evolution of
a relativistic binary system. For a historic account of attempts at
a universally accepted definition of
angular momentum for radiating systems in general relativity
see~\cite{Szabados2004}. At spatial infinity,
reasonable asymptotic conditions establish the Poincar{\' e} group as
the asymptotic symmetry group. This allows a Poincar{\' e} covariant
definition of angular momentum in which the translation freedom mixes
angular momentum with linear momentum in the standard
manner~\cite{Ashtekar-Hansen}.  However, the Bondi-Metzner-Sachs (BMS)
asymptotic symmetry group~\cite{sachsbms} at $\scri^+$ has an infinite
supertranslation subgroup. Fortunately, the translations form an
invariant subgroup of the BMS group, which leads to an unambiguous
definition of energy-momentum.  However, although the Lorentz group is
a subgroup of the BMS group, the supertranslations lead to a mixing of
the associated supermomentum with angular momentum, the physical
consequences of which have not been fully explored. For a stationary epoch in
the neighborhood of $\scri^+$, this supertranslation freedom can be
removed and the BMS group reduced to the Poincar{\' e} group, in which
case angular momentum can be well-defined.  However, for a system
which makes a stationary to stationary transition, the two Poincar{\'e}
groups obtained at early and late times can be shifted by a
supertranslation.  Such supertranslation shifts could lead to a
distinctly general relativistic mechanism for a system to lose angular
momentum. See~\cite{NewmanPenrose1966,jwangular,Geroch1981} for
discussions.  This in fact occurs if the intervening gravitational
radiation produces a non-zero gravitational memory effect.  A non-zero
gravitational radiation memory is equivalent to such a
supertranslation shift~\cite{jwemem}.  In this paper, we develop a
unified algorithm for the computation of the gravitational fluxes of
energy-momentum, angular momentum and supermomentum to infinity.

There are two distinct approaches for obtaining flux-conserved
quantities which form a representation of the BMS asymptotic symmetry
group. One approach consists of the BMS linkage integrals
$L_\xi(\Sigma)$~\cite{TamburinoWinicour1966,win1968,Geroch1981}, which
for each spherical cross-section $\Sigma$ of $\scri^+$ generalize the
Komar integrals~\cite{Komar:1958wp} for exact symmetries to the case
of asymptotic Killing vectors $\xi^a$.  Associated with the linkage
integrals are locally defined fluxes ${}^L F_\xi$ whose integral
determines the change $L_\xi(\Sigma_2)$-$L_\xi(\Sigma_1)$ between two
cross-sections.  The second approach, originated by Ashetkar and
Streubel, is based upon the Hamiltonian phase space of gravitational
radiation modes at $\scri^+$~\cite{ashtekar-streubel}. The Hamiltonian
densities generating a BMS transformation also geometrically define
local fluxes ${}^H F_\xi$. In the case of the supertranslations
$\alpha^a$, Ashetkar and Streubel showed that these Hamiltonian fluxes
could be integrated to obtain flux conserved charges
$Q_\alpha(\Sigma)$. In the case of the BMS time and space
translations, $\xi^a=\tau^a$, the corresponding linkage and
Hamiltonian energy-momentum integrals and their fluxes are identical,
i.e. $L_\tau(\Sigma) = Q_\tau(\Sigma)$ and ${}^L F_\tau={}^H
F_\tau$. However, their supermomentum fluxes differ locally.
Subsequently, Wald and Zoupas~\cite{waldzoupas} generalized the
Hamiltonian approach and obtained flux conserved quantities for all
the Ashetkar-Streubel Hamiltonian fluxes, including angular
momentum. They showed that these flux conserved quantities, including
angular momentum, were identical to previous expressions proposed by
Dray and Streubel~\cite{dray1984angular}.

The relation between the linkage and Hamiltonian BMS fluxes has been
examined in~\cite{ashtekar-win}. Although their construction and their local
values are completely different, it has been shown that the integrals
of all the linkage and Hamiltonian fluxes between cross-sections of
$\scri^+$ agree, including the angular momentum and  supermomentum fluxes, 
in the most physically relevant case when $\Sigma_2$ and $\Sigma_1$ are shear-free
cross-sections in the limits of infinite future and past retarded
time, respectively. Although, for the rotations and boosts a
factor-of-two anomaly in the linkages must be taken into account, i.e. $\int_{\Sigma}
{}^L F_\xi dS=2 \int_{\Sigma} {}^H F_\xi dS$ .

There have been other approaches to defining energy-momentum and
angular momentum at $\scri^+$ which do not appeal to the BMS
symmetries. Some~\cite{dray1984angular,helfer} have been based upon an
asymptotic version of Penrose's construction of quasi--local
energy-momentum and angular momentum using twistor
theory~\cite{penrose1982quasi}. Another has been based upon the modification
of the BMS group to a quasigroup of asymptotic
symmetries~\cite{nesterov1997quasigroups}.
Other important work on the computation
of the physical properties of radiation at $\scri^+$ has been presented
in~\cite{deshbish,Bishop:2013, boyle16,helfer10}.

Here, we concentrate on the linkage approach because it
is readily adapted to previous computational infrastructure treating
$\scri^+$. However, the computational methods presented here
should also be useful in computing the Hamiltonian charges
and fluxes, as formulated by Wald and Zoupas, which have all
the desired physical properties. Although the linkage approach is geometrically well defined,
its main physical shortcoming is that the associated supermomentum fluxes do not
locally vanish in Minkowksi space, although their integrals between
cross-sections of $\scri^+$ determined by two Minkowski space light
cones do vanish. As a result, the local physical significance of these
supermomentum fluxes is unclear, although the local time rate of
change of all the linkage fluxes ${}^L \dot F_\xi$ do vanish in
Minkowski space.  

Furthermore, as shown in Sec.~\ref{sec:flux}, in a
general radiating spacetime ${}^L \dot F_\xi=0$ for all BMS generators
in any region where the Newman-Penrose radiative component of the Weyl
tensor $\psi_4^0$ vanishes. In particular, ${}^L \dot F_\xi
\rightarrow 0$ for all BMS symmetries in the infinite retarded time past on
$\scri^+$.  It is precisely in this limit that the supertranslation
freedom can also be eliminated and a preferred Poincar{\' e} subgroup
can be identified~\cite{NewmanPenrose1966}. As a result, the
energy-momentum and angular momentum fluxes ${}^L
F_\xi$ can be uniquely defined by a retarded time integration
of ${}^L \dot F_\xi$, using the initial
value ${}^L F_\xi=0$ at $u=-\infty$. Similarly, the energy momentum and angular
momentum can also be uniquely defined for such systems by a
retarded time integration of ${}^L F_\xi$, using their initial
values at $u=-\infty$, or at spatial infinity. However, there remains the possibility of
carrying out a similar construction at $\scri^+$ in the infinite retarded time
future $u=+\infty$. This leads to the same unresolved issue discussed above for a
stationary to stationary transition, i.e. to a net supertranslation
shift between the future and past Poincar{\' e} groups. The identical
supertranslation ambiguity exists in the Hamiltonian description of angular momentum.

As already discussed, in previous applications of CCE to compute
energy-momentum loss, the Bondi news function was first computed from
the radiation field in the computational coordinates by a gauge
invariant method. The news and radiation waveform were then
transformed to inertial coordinates on $\scri^+$. In that approach, it
was only necessary to find the 3-dimensional transformation to inertial
coordinates intrinsic to $\scri^+$ itself. This procedure is possible because the
news function can be defined geometrically, without reference to the
BMS symmetries.  However, this procedure is not feasible in computing
the angular momentum and supermomentum fluxes. Here, we present a
unified approach to computing all the BMS fluxes of energy-momentum,
angular momentum and supermomentum to infinity by carrying out the
transformation to inertial coordinates in a full 4-dimensional neighborhood of
$\scri^+$. We are able to accomplish this by constructing a surprisingly simple
transformation between the computational and inertial coordinates. The
metric is then transformed to the inertial coordinates, in which the
BMS symmetries are readily identified and the corresponding fluxes
computed. We formulate a simple
computational algorithm for carrying out this transformation.

In this procedure, there remains the freedom of the BMS group in the
choice of inertial observers.  In special relativistic theories, the
corresponding freedom reduces to the translations and Lorentz
transformations of the Poincar{\' e} group. The BMS supertranslations
introduce a gauge freedom in the radiation strain.

\section{Waveforms at \(\scri^+\)}
\label{sec:waveforms}

The characteristic formalism is based upon a family of outgoing null
hypersurfaces emanating from an inner worldtube and extending to
infinity, where they foliate $\scri^+$ into spherical slices.  Let $u$
label these null hypersurfaces, $x^A$ $(A=2,3)$ be angular coordinates
which label the null rays and $r$ be a surface area coordinate. Using
the notion of~\cite{TamburinoWinicour1966}, in the resulting
$(u,r,x^A)$ coordinates, the metric takes the Bondi-Sachs form
\begin{eqnarray}
   ds^2 & = & -\left(e^{2\beta}\frac{V}{r} -r^2h_{AB}U^AU^B\right)du^2
   -2e^{2\beta}dudr -2r^2 h_{AB}U^Bdudx^A \nonumber \\ & + &
   r^2h_{AB}dx^Adx^B,
	\label{eq:bmet}
\end{eqnarray}
where $h^{AB}h_{BC}=\delta^A_C$ and $\det(h_{AB})=\det(q_{AB})$, with
$q_{AB}$ a unit sphere metric.

As described in more detail
in~\cite{TamburinoWinicour1966,Winicour1983}, in this formalism
Einstein's equations decompose into a system which propagate boundary
data for the metric variables $(\beta, V, U^A, h_{AB})$ on an inner worldtube
to a solution at $\scri^+$. In the Pitt null code and in the
\texttt{SpEC} characteristic code, this solution is computed in a
Penrose compactification of $\scri^+$, e.g. in terms of the
coordinates \(x^\mu=(u,\ell,x^A)\), where \(\ell=1/r\), \(\ell=0\) at
\(\scri^+\).  Then the conformal metric $\hat g_{\mu\nu}=\ell^{2}
g_{\mu\nu}$ is smooth at \(\scri^+\) and takes the form
\begin{equation}
   \hat g_{\mu\nu}dx^\mu dx^\nu= -\left(\ell^3 e^{2\beta} V
   -h_{AB}U^AU^B\right)du^2 +2e^{2\beta}dud\ell -2 h_{AB}U^Bdudx^A +
   h_{AB}dx^Adx^B\;.
   \label{eq:lmet}
\end{equation}

General conditions on the asymptotic behavior of the metric variables
follow from the vacuum Einstein
equations~\cite{TamburinoWinicour1966},
\begin{equation}
    \beta=H+ O(\ell^2) \;,
    \label{eq:betah}
\end{equation}
\begin{equation}
    U^A= L^A+2\ell e^{2H} H^{AB}D_B H+O(\ell^2) \;,
\end{equation} 

\begin{equation}
    \ell^2 V= D_A L^A +\ell (e^{2H}{\cal R}/2 +D_A D^A
    e^{2H})+O(\ell^2)\;,
\end{equation}
and
\begin{equation}
   h_{AB}= H_{AB}+\ell c_{AB}+O(\ell^2)\;,
\end{equation}
where ${\cal R}$ and $D_A$ are the 2-dimensional curvature scalar and
covariant derivative associated with $H_{AB}$ and the determinant
condition implies
\begin{equation}
         H^{AB} c_{AB}=0\;.
         \label{eq:detcond}
\end{equation}
The expansion coefficients $H$, $H_{AB}$, $c_{AB}$ and $L^A$ (all
functions of $u$ and $x^A$) completely determine the radiation field.

One can further specialize the Bondi coordinates to be {\it inertial}
at \(\scri^+\), i.e. have asymptotic Minkowski form, in which case $H=L^A=0$,
$H_{AB}|_{\scri^+}=q_{AB}$ (the unit sphere metric) so that the
radiation field is completely determined by $c_{AB}$, which describes
the asymptotic shear of the outgoing null cones or, equivalently, the
radiation strain. In these inertial coordinates, the retarded time
derivative $\partial_u c_{AB}$ determines the Bondi news function
$N(u,x^A)$. However, the characteristic extraction of the waveform is
carried out in computational coordinates derived from the Cauchy
coordinates on the inner worldtube, so this inertial simplification
cannot be assumed.

In previous work the Bondi news function $N$ was first computed in the
computational $\hat g_{\mu\nu}$ frame. It was then transformed to
inertial coordinates $(\tilde u, \tilde x^A)$ on $\scri^+$ to
determine the physical dependence of the waveform on retarded time and
angle. Here we construct the transformation to a compactified version
of inertial coordinates $(\tilde u,\tilde \ell, \tilde x^A)$ in a
full neighborhood of $\scri^+$.

\section{Construction of inertial coordinates}
\label{sec:inertial}

First, we recall some basic elements of Penrose compactification.  In
a spacetime with metric $g_{\mu\nu}$, the vacuum Einstein equations
$G_{\mu\nu}=0$ expressed in terms of a conformally related metric $
\hat g_{\mu\nu}=\Omega^2 g_{\mu\nu}$, where $\Omega=0\) on \(\scri^+$,
take the form
 \begin{equation}
   \Omega^2 \hat G_{\mu\nu} + 2\Omega \hat \nabla_\mu \hat \nabla _\nu
   \Omega - \hat g_{\mu\nu} \bigg (2\Omega \hat \nabla^\rho \hat
   \nabla_\rho \Omega -3( \hat \nabla^\rho \Omega \hat \nabla_\rho
   \Omega) \bigg ) =0\;.
\end{equation}
It immediately follows that
\begin{equation}
   (\hat \nabla^\rho \Omega ) \hat \nabla_\rho \Omega|_{\scri^+} =
  0\;,
\end{equation}  
 so that \(\scri^+\) is a null hypersurface and that
\begin{equation}
     [ \hat \nabla_\mu \hat \nabla _\nu \Omega -\frac{1}{4}\hat
       g_{\mu\nu} \hat \nabla^\rho \hat \nabla_\rho \Omega
     ]|_{\scri^+} =0\;.
\end{equation} 

With respect to this general frame, we now choose $\Omega=\ell$ and
computational coordinates $(u,\ell,x^A)$, as in
Sec.~\ref{sec:waveforms}, and proceed to construct an inertial
conformal frame as follows. We introduce a new conformal factor
$\tilde \Omega =\omega \Omega =\omega \ell$, with \(\tilde g_{\mu\nu}
=\omega^2 \hat g_{\mu\nu}\) such that
\begin{equation}
    \tilde \nabla^\rho \tilde \nabla_\rho \tilde \Omega |_{\scri^+} =0
    \;,
    \label{eq:divfr}
\end{equation}
by requiring
\begin{equation}
  [2 \hat n^\sigma \partial_\sigma \omega +\omega \hat \nabla_\sigma
    \hat n^\sigma]|_{\scri^+} =0\;, \quad \hat n^\sigma =\hat
  g^{\rho\sigma}\nabla_\rho \ell \;.
      \label{eq:omev}
\end{equation}
It then follows that
\begin{equation}
      \tilde \nabla_\mu \tilde \nabla _\nu \tilde \Omega |_{\scri^+}
      =0\;, \quad \tilde \nabla^\rho \tilde \nabla_\rho \tilde \Omega
      |_{\scri^+} = 0\;,
        \label{eq:nabnabom}
\end{equation}        
i.e. in the $\tilde g_{\mu\nu}$ conformal frame $\scri^+$ is null,
shear-free and divergence-free.  It also follows that
\begin{equation}
      \tilde n^\sigma \tilde \nabla _\sigma \tilde n^\nu |_{\scri^+}
      =0\;, \quad \tilde n^\sigma =\tilde
      g^{\rho\sigma}\tilde\nabla_\rho \tilde\Omega\;,
\end{equation} 
 i.e. in the \(\tilde g_{\mu\nu}\) frame \(\tilde n^\sigma\) is an
 affinely parametrized null generator of~\(\scri^+\).

We now construct inertial conformal coordinates \((\tilde u, \tilde x^
A)\) on \(\scri^+\), by first assigning angular coordinates \(\tilde
x^A\) to each point of some initial spacelike spherical slice \(\tilde
u=u_0\) of \(\scri^+\).  We then propagate these coordinates along the
null geodesics generating \(\scri^+\) according to
\begin{equation}
      \tilde n^\rho \partial_\rho \tilde x^A |_{\scri^+} =
      \omega^{-1} \hat n^\rho \partial_\rho \tilde x^A |_{\scri^+} =
      0\;.
\end{equation}
In addition, we require
\begin{equation}
      \tilde n^\rho \partial_\rho \tilde u |_{\scri^+} = \omega^{-1}
      \hat n^\rho \partial_\rho \tilde u |_{\scri^+} = 1\;,
\end{equation}
so that \(\tilde u\) is an affine parameter along the generators in
the \(\tilde g_{\mu\nu}\) conformal frame.

This determines the transformation from the computational coordinates
$x^\mu=(u, \ell, x^A)$ to inertial coordinates $(\tilde u (u,x^B),
\tilde x^A(u,x^B))$ on $\scri^+$, which allows the news function and
the extracted waveform to be re-expressed in the physically relevant
coordinates of a detector.  However, this is not sufficient to
identify the BMS symmetries and their associated fluxes. The remaining
complication is that after transforming to inertial coordinates the
metric on the spherical cross-sections of $\scri^+$,
\begin{equation}
     H_{\tilde A \tilde B} = \hat g_{\tilde A \tilde B} |_{\scri^+}=
     \frac{\partial x^\mu}{\partial \tilde x^A} \frac{\partial
       x^\nu}{\partial \tilde x^B}\hat g_{\mu \nu} |_{\scri^+} \;,
\end{equation} 
does not reduce to a unit sphere metric. As a result, the
identification of the translations as a subgroup of the
supertranslation group is complicated; essentially one must solve an
elliptic equation to identify the curved 2-space version of the
$\ell=0$ and $\ell=1$ spherical harmonics on the spherical cross-sections.

For this purpose, it is simplest to proceed by determining the
conformal factor $\omega$ relating $H_{\tilde A \tilde B} $ to a
unit-sphere metric $Q_{\tilde A \tilde B}$,
\begin{equation}
    Q_{\tilde A \tilde B} = \omega^2 H_{\tilde A \tilde B}\;.
\end{equation} 
We determine \(\omega\) by solving the elliptic equation governing the
conformal transformation of the curvature scalar to the unit sphere
curvature,
\begin{equation}
     {\cal R}=2(\omega^2+D_A D^A \log \omega) \;,
\label{eq:conf}
\end{equation}
where ${\cal R}$ and $D_A$ are the curvature scalar and covariant
derivative associated with the 2-metric $H_{AB}$. Since this is a
scalar equation it can be solved in the computational coordinates.

The elliptic equation (\ref{eq:conf}) need only be solved at the
initial time. Then, the shear-free property of the null geodesics
generating $\scri^+$ implies that $\omega$ may be propagated along the
generators by means of (\ref{eq:nabnabom}), which takes the explicit
form
\begin{equation}
     2\hat n^\alpha \partial_{\alpha} \log \omega =-e^{-2H}D_AL^A \;.
    \label{eq:omegev}
\end{equation}
After initialization of \(\omega\) so that the initial slice of
\(\scri^+\) has unit sphere geometry, it then follows that all
cross-sections of \(\scri^+\) have unit sphere geometry. In terms of
standard spherical coordinates \(\tilde x^A=(\tilde \theta,\tilde \phi)\), the
induced metric on the cross-sections of \(\scri^+\) has components
\begin{equation}
      \tilde g_{\tilde A \tilde B}(\tilde u, \tilde x^A) |_{\scri^+} =
      Q_{\tilde A \tilde B} \; , \quad \tilde g^{\tilde A \tilde
        B}(\tilde u, \tilde x^A) |_{\scri^+} = Q^{\tilde A \tilde B}\;,
\end{equation}  
where, in these coordinates,
\begin{equation}
        Q_{\tilde A \tilde B}d \tilde x^A d \tilde x^B =d\tilde \theta^2 +
        \sin^2 \tilde \theta d\tilde \phi^2\;.
\end{equation}

Given such inertial coordinates on $\scri^+$, we then extend them to
coordinates $\tilde x^\mu = (\tilde u,\tilde \ell, \tilde x^A)$ in a
neighborhood of $\scri^+$, where
\begin{equation}
          \tilde \ell =\omega(u,x^A) \ell =\tilde \Omega \;.
          \label{eq:tildeell}
\end{equation}
The Jacobian of the transformation from computational to
inertial coordinates has the simple property
\begin{equation}
   \partial_{\ell} \tilde u = \partial_{\ell} \tilde x^A =0\;, \quad
   \partial_\ell \tilde \ell =\omega\;.
\end{equation}
As a result it immediately follows that at $\scri^+$ the metric reduces
to the simple form
\begin{equation}
      \tilde g_{\tilde A \tilde B}(\tilde u, \tilde x^A) |_{\scri^+} =
      Q_{\tilde A \tilde B} \; , \quad \tilde g^{\tilde A \tilde
        B}(\tilde u, \tilde x^A) |_{\scri^+} = Q^{\tilde A \tilde
        B}\;,
      \label{eq:inducedq}
\end{equation}  
\begin{equation}
      \tilde g_{\tilde u \tilde u}(\tilde u, \tilde x^A) |_{\scri^+} =
      \tilde g_{\tilde u \tilde A} (\tilde u, \tilde x^A) |_{\scri^+}
      =\tilde g^{\tilde \ell \tilde \ell}(\tilde u, \tilde x^A)
      |_{\scri^+} =\tilde g^{\tilde \ell \tilde A}(\tilde u, \tilde
      x^A) |_{\scri^+} = 0\; .
      \label{eq:inducedm}
\end{equation}  
This transformation to inertial coordinates also determines the metric
in a neighborhood of $\scri^+$, which simplifies the identification of
the BMS group and the computation of the radiation strain and BMS
fluxes.

\section{The BMS group}
\label{sec:BMS}

In the inertial coordinates and conformal frame constructed in
Sec.~\ref{sec:inertial}, the asymptotic Killing vectors composing the
BMS group can be described by
by~\cite{Sachs1962,TamburinoWinicour1966}
\begin{equation}
         \xi^{\tilde \rho}\partial_{\tilde \rho} |_{\scri^+} = \bigg (
         \alpha(\tilde x^A) + \frac{1}{2} \tilde u f^{\tilde
           A}_{:\tilde A} \bigg )\partial_{\tilde u} + f^{\tilde
           A}\partial_{\tilde A}\;,
        \label{eq:bms}
\end{equation}
where a ``colon'' denotes the covariant derivative with respect to the
unit sphere metric $Q_{\tilde A \tilde B}$, $\alpha(\tilde x^A)$
represents the supertranslation freedom and \(f^{\tilde A}(\tilde
x^B)\) is a conformal Killing vector on the unit sphere,
\begin{equation}
         f^{(\tilde A: \tilde B)}=\frac {1}{2}Q^{\tilde A \tilde B}
         f^{\tilde C}_{:\tilde C}\;.
\end{equation}
This conformal group is isomorphic to the Lorentz group.

The additional property that \(\xi^\mu\) is tangent to $\scri^+$,
\begin{equation}
  \xi^{\tilde \ell} |_{\scri^+}=\xi^\rho \partial_\rho \tilde
  \Omega|_{\scri^+} =0\;,
\end{equation}
implies that \(\xi^\mu\) satisfies the asymptotic Killing equation in
the physical space geometry,
\begin{equation}
 \tilde  \Omega^{-2} \nabla^{(\mu} \xi^{\nu )} |_{\scri^+}=
  [\tilde  \nabla^{(\tilde \mu} \xi^{\tilde\nu )} 
      - \tilde\Omega^{-1} \tilde g^{\tilde \mu\tilde \nu} \xi^{\tilde \rho}
     \tilde \nabla_{\tilde \rho} \tilde \Omega]|_{\scri^+} =0\;.
            \label{eq:asymkill}
\end{equation}

The supertranslations form an invariant subgroup of the BMS group for
which \(f^{\tilde A}=0\). The translations are an invariant
4-parameter subgroup of the supertranslations for which
\(\alpha(\tilde x^A)\) is composed of \(\ell=0\) and \(\ell=1\)
spherical harmonics. The rotation subgroup intrinsic to a
cross-section \(\Sigma^+\) of \(\scri^+\) consists of the BMS
transformations which map \(\Sigma^+\) onto itself. Without
introducing any artificially preferred structure on \(\scri^+\), there
is no invariant way to extract a rotation group, Lorentz group or
Poincar{\' e} group from the BMS group.

\begin{figure}[h!]
  \centering
  \includegraphics[width=0.5\textwidth]{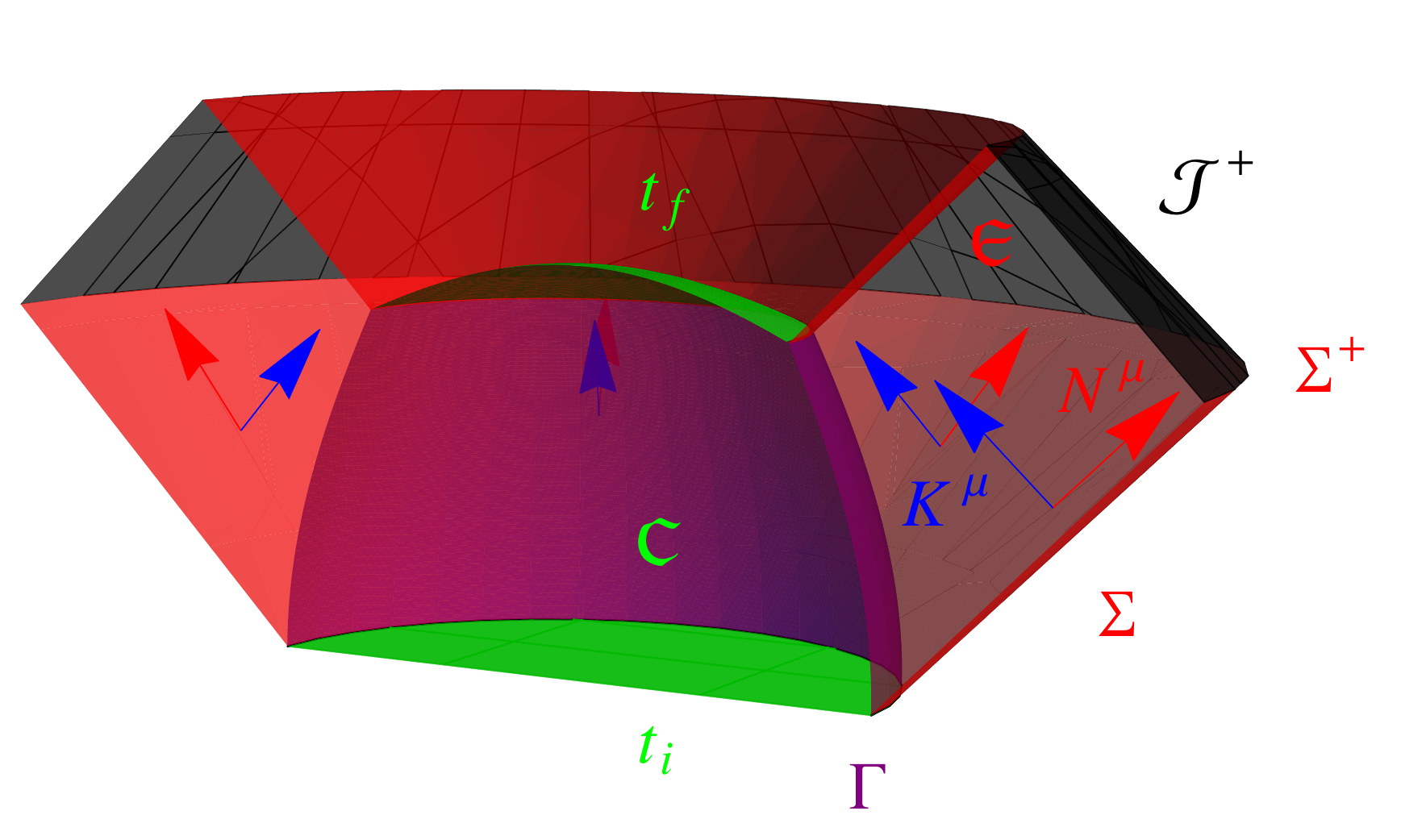}
  \caption{\small{Diagram of the computational domain under Penrose compactification,
  with angle \(\tilde \theta\) suppressed. The Cauchy evolution $\mathcal{C}$ runs from time
  $t_i$ to $t_f$ within the worldtube $\Gamma$. The characteristic evolution $\mathcal{E}$
  is performed on null hypersurfaces extending from $\Gamma$ to future null infinity $\scri^+$.
  Cross-sections $\Sigma$ of a null hypersurface (topologically a 2-sphere) approach the limit
  $\Sigma^+$ at $\scri^+$, where the linkage integral is defined. Ingoing ($K^\mu$) and
  outgoing ($N^\mu$) null vectors are normal to $\Sigma$. Gauge independent news, strain, $\psi_4^0$
   and flux are computed in inertial coordinates at $\scri^+$.}}
  \label{fig:KomarDiagram}
\end{figure}

Given a cross-section $\Sigma^+$ of $\scri^+$ and a generator
$\xi^\mu$ of the BMS group, the linkage integral $L_\xi (\Sigma^+)$
which generalizes the Komar integral for an exact symmetry is given in
terms of the physical space geometry by
\begin{equation}
     L_\xi (\Sigma^+) =\lim_{\Sigma \rightarrow \Sigma^+}
     \oint_{\Sigma} ( \nabla^{[\mu}\xi^{\mu]} -K^{[\mu}N^{\nu]}
     \nabla_\rho \xi^\rho) dS_{\mu\nu} \;,
\end{equation}
where $K^{\mu}$ and $N^\mu$ are, respectively, ingoing and outgoing
null vectors normal to $\Sigma$, normalized by $K^\mu N_\mu=-1$.  The
limit is taken along the outgoing null hypersurface ${\cal N}$
emanating from $\Sigma$ to $\Sigma^+$, as shown in Fig.~\ref{fig:KomarDiagram}.
The value of this limit depends
upon how the BMS generator is extended off $\scri^+$. The asymptotic
symmetry condition (\ref{eq:asymkill}) allows the freedom in this
extension of the form
\begin{equation}
        \xi^\mu \rightarrow \xi^\mu + \tilde \Omega^2 v^\mu \;.
             \label{eq:xifreed}
\end{equation}
In the original formulation of the
linkages~\cite{TamburinoWinicour1966}, $v^\mu$ was determined by a
null hypersurface propagation law on ${\cal N}$,
\begin{equation}
    ( \nabla^{(\mu}\xi^{\nu)}- \frac{1}{2} g^{\mu\nu} \nabla_\rho
  \xi^\rho)K_\mu =0 \;.
\end{equation}
In \cite{Geroch1981}, other choices of propagation law were
considered. In the next section. we adopt a simple extension
in which the generators $\xi^\mu$ only depend linearly on
the inertial conformal factor $\tilde \ell$.

\section{The BMS fluxes}
\label{sec:flux}

Our focus is on the radiation flux $F_\xi$ across $\scri^+$ which
governs the change of linkage between two cross-sections,
\begin{equation}
        L_\xi (\Sigma_2) - L_\xi (\Sigma_1) =
        \int_{\Sigma^+_1}^{\Sigma^+_2} F_\xi dV \;.
\end{equation}
In a conformal frame in which $\scri^+$ is
divergence-free, it was shown in \cite{Geroch1981} that there is a
local geometric expression for $F_\xi$ which is independent of the
freedom (\ref{eq:xifreed}) governing the extension of $\xi^\mu$ off
$\scri^+$.  The resulting flux is uniquely determined by (i) the
choice of $\xi^\mu$ on $\scri^+$ and (ii) the local conformal geometry
near $\scri^+$; and (iii) it is independent of any choice of
cross-section or other extraneous constructs.

In the $\tilde g_{\tilde \mu \tilde \nu}$ conformal inertial frame
described in Sec.~\ref{sec:inertial}, this flux is constructed as
follows. The transformation from the computational coordinates
$x^\mu(u, \ell, x^A)$ to inertial coordinates $(\tilde u (u,x^B),
\tilde x^A(u,x^B))$ on $\scri^+$ is first extended to coordinates
$\tilde x^\mu = (\tilde u ,\tilde \ell, \tilde x^A)$ in a neighborhood
of $\scri^+$ where, as before,
\begin{equation}
          \tilde \ell =\omega(u,x^A) \ell =\tilde \Omega \;.
\end{equation}
This transformation determines the metric in the extension of the
inertial frame to a neighborhood of $\scri^+$,
\begin{equation}
     \tilde g^{\tilde \mu \tilde \nu}(\tilde u, \tilde \ell, \tilde
     x^A)=\omega^{-2} \frac {\partial \tilde x^\mu}{\partial x^\alpha}
     \frac {\partial \tilde x^\nu}{\partial x^\beta}\hat
     g^{\alpha\beta}(u, \ell, x^A) \;.
 \end{equation}
In addition to (\ref{eq:inducedq}) and (\ref{eq:inducedm}), this
implies
\begin{eqnarray}
      \tilde g^{\tilde u \tilde \ell}(\tilde u,0, \tilde x^A)&=&1 \;,
      \nonumber \\ \tilde g^{\tilde u \tilde u}(\tilde u,0, \tilde
      x^A) &=& \omega^{-2} \frac{\partial \tilde u}{\partial x^A}
      \frac{\partial \tilde u}{\partial x^B}\hat g^{AB} \;,
        \label{eq:gup} \\
      \tilde g^{\tilde u \tilde A}(\tilde u,0, \tilde x^A) &=&
      \omega^{-2} \frac{\partial \tilde u}{\partial x^B}
      \frac{\partial \tilde x^A}{\partial x^C}\hat g^{BC} \nonumber
      \;.
\end{eqnarray}
The corresponding covariant components are
\begin{eqnarray}
      \tilde g_{\tilde u \tilde \ell}(\tilde u,0, \tilde x^A)&=&1 \;,
      \nonumber \\ \tilde g_{\tilde \ell \tilde \ell}(\tilde u,0,
      \tilde x^A) &=& -\tilde g^{\tilde u \tilde u} +Q_{\tilde A
        \tilde B} \tilde g^{\tilde u\tilde A} \tilde g^{\tilde u\tilde
        B} \;,\\ \tilde g_{\tilde \ell \tilde A}(\tilde u,0, \tilde
      x^A) &=& - Q_{\tilde A \tilde B} \tilde g^{\tilde u\tilde B}
      \nonumber \;.
\end{eqnarray}
Note that the $\tilde x^\mu$ coordinates are not null coordinates.
While the $\tilde u=const$ cross-sections of $\scri^+$ are space-like,
(\ref{eq:gup}) implies in general that $\tilde g^{\tilde u \tilde u}>0$ so that
the $\tilde u=const$ hypersurfaces in the neighborhood of $\scri^+$
are asymptotically time-like.  This somewhat surprising feature stems
from the requirement that the transformation from computational
coordinates to inertial coordinates is linear in $\ell$.  The
transformation to the $\tilde x^\mu$ inertial coordinates determines
the essential geometric quantities associated with the inertial frame
metric \(\tilde g_{\tilde \mu\tilde \nu}\), e.g. the associated
covariant derivative $\tilde \nabla_{\tilde \mu}$ and curvature scalar
$\tilde R$.

Restrictions on the $\tilde \ell$-derivatives of the inertial metric
at $\scri^+$ arise from (\ref{eq:nabnabom}),
\begin{eqnarray}
      \partial_{\tilde \ell} \tilde g_{\tilde u \tilde
        u}|_{\scri^+}&=&0 \;, \nonumber \\ \partial_{\tilde \ell}
      \tilde g_{\tilde u \tilde \ell}|_{\scri^+} &=& \frac{1}{2}
      \partial_{\tilde u} \tilde g_{\tilde \ell \tilde
        \ell}|_{\scri^+} \;, \\ \partial_{\tilde \ell} \tilde
      g_{\tilde u \tilde A}|_{\scri^+} &=& \partial_{\tilde u} \tilde
      g_{\tilde A \tilde \ell} |_{\scri^+} \;.  \nonumber
\end{eqnarray}

In addition,
\begin{equation}
         \tilde g_{\tilde u \tilde \ell} = \frac {\partial u}
                {\partial\tilde u}\omega e^{2\beta} \;,
\end{equation}
so since (\ref{eq:betah}) implies $\partial_{\tilde \ell}
\beta|_{\scri^+}=0$, we have
\begin{equation}
              \partial_{\tilde \ell} \tilde g_{\tilde u \tilde
                \ell}|_{\scri^+} = \partial_{\tilde u} \tilde
              g_{\tilde \ell \tilde \ell}|_{\scri^+} =0 \; .
\end{equation}
The determinant condition (\ref{eq:detcond}), i.e.
$h^{AB}\partial_\ell h_{AB}|_{\scri^+} =0$ implies
\begin{equation}
  \bigg\{ Q^{\tilde A \tilde B} \partial_{\tilde \ell} \tilde
  g_{\tilde A \tilde B} +2 \tilde g^{\tilde u \tilde
    A}\partial_{\tilde \ell} \tilde g_{\tilde u \tilde A} \bigg
  \}|_{\scri^+} =0 \;,
\end{equation}
so that
\begin{equation}
  \bigg\{ Q^{\tilde A \tilde B} \partial_{\tilde \ell} \tilde
  g_{\tilde A \tilde B} +2 \tilde g^{\tilde u \tilde
    A}\partial_{\tilde u} \tilde g_{\tilde \ell \tilde A} \bigg
  \}|_{\scri^+} = \bigg\{ Q^{\tilde A \tilde B} \partial_{\tilde \ell}
  \tilde g_{\tilde A \tilde B} -2 Q_{\tilde A \tilde B} \tilde
  g^{\tilde u \tilde A}\partial_{\tilde u} \tilde g^{\tilde u \tilde
    B} \bigg \}|_{\scri^+} =0\;.
\end{equation}

The corresponding contravariant components satisfy
\begin{align}
  \partial_{\tilde \ell} \tilde g^{\tilde \ell \tilde
    \ell}|_{\scri^+}=&\;0 \;, \nonumber \\ \partial_{\tilde \ell}
  \tilde g^{\tilde \ell \tilde A}|_{\scri^+} =&\; \partial_{\tilde u}
  \tilde g^{\tilde u \tilde A} |_{\scri^+} \;, \\ \partial_{\tilde
    \ell} \tilde g^{\tilde u \tilde \ell}|_{\scri^+} = \frac{1}{2}
  \partial_{\tilde u} \tilde g^{\tilde u \tilde u}|_{\scri^+} =&\;
  Q_{\tilde A \tilde B}g^{\tilde u \tilde A} \partial_{\tilde u}
  \tilde g^{\tilde u \tilde B} |_{\scri^+}\;, \nonumber \\ Q_{\tilde A
    \tilde B} \partial_{\tilde \ell} \tilde g^{\tilde A \tilde
    B}|_{\scri^+} =& \;0\;. \nonumber
  \label{eq:contrad}
\end{align}
     
In general, since $\xi^{\tilde \mu}$ is tangent to $\scri^+$, we can
define a smooth scalar field $K=\tilde \Omega^{-1} \xi^{\tilde
  \mu}\partial_{\tilde \mu} \tilde \Omega$ and the asymptotic Killing
equation (\ref{eq:asymkill}) implies we can define a smooth field
$X^{\tilde \mu \tilde \nu}$ according to
\begin{equation}
   \tilde \nabla^{(\tilde \mu}\xi^{\tilde \nu)} = K \tilde g^{\tilde
     \mu \tilde \nu} +\tilde \Omega X^{\tilde \mu \tilde \nu} \;.
   \label{eq:tildeaskill}
\end{equation}
We write $X=X^{\tilde \mu}_{ \tilde \mu}$.
In the $\tilde g_{\tilde \mu \tilde \nu}$ inertial frame,
\begin{equation}
    K=\tilde \ell^{-1}\xi^{\tilde \ell} \;,
\label{eq:K}
\end{equation}     
so that
\begin{equation}
         K |_{\scri^+}=\frac {\partial \xi^{\tilde \ell}}{\partial
           \tilde \ell}|_{\scri^+} \;.
\end{equation}

Evaluation of (\ref{eq:tildeaskill}) at $\scri^+$ gives, in addition
to (\ref{eq:bms}),
\begin{eqnarray}
    \frac{\partial \xi^{\tilde \ell}}{\partial \tilde \ell}|_{\scri^+}
    &=& \frac{\partial \xi^{\tilde u}}{\partial \tilde u }|_{\scri^+}
    =\frac{1}{2} \xi^{\tilde A}_{:\tilde A}|_{\scri^+} \nonumber \;,
    \\ \frac{\partial \xi^{\tilde u}}{\partial \tilde \ell}
    |_{\scri^+}&=& \{ -\tilde g^{\tilde u \tilde A} \frac {\partial
      \xi^{\tilde u}}{\partial x^{\tilde A}} +\frac{1}{2} \xi^{\tilde
      u} \frac{\partial \tilde g^{\tilde u \tilde u}}{\partial \tilde
      u } +\frac{1}{2} \xi^{\tilde A} \frac{\partial \tilde g^{\tilde
        u \tilde u}}{\partial x^{\tilde A}} \}|_{\scri^+}
                \label{eq:ellderivs}  \;, \\
      \frac{\partial \xi^{\tilde A}}{\partial \tilde \ell}|_{\scri^+}
      &=& \{ \tilde g^{\tilde A \tilde u} \frac {\partial \xi^{\tilde
          u}}{\partial \tilde u} -\tilde g^{\tilde u \tilde B}
           {\xi^{\tilde A}}_{:\tilde B} + \xi^{\tilde B} { g^{\tilde u
               \tilde A}}_{:\tilde B } + \xi^{\tilde u} \frac{\partial
             \tilde g^{\tilde u \tilde A}}{\partial \tilde u }
           -Q^{\tilde A \tilde B} \frac {\partial \xi^{\tilde
               u}}{\partial x^{\tilde B}} \}|_{\scri^+} \;. \nonumber
\end{eqnarray}

As shown in~\cite{Geroch1981}, it also follows that
\begin{equation}
       X^{\tilde \mu \tilde \nu}\partial_{\tilde \nu} \tilde \Omega
       |_{\scri^+} =0 \;,
\end{equation}
which results from a straightforward calculation using
(\ref{eq:inducedm}), (\ref{eq:K}) and (\ref{eq:ellderivs}).  Thus we
can further define a smooth field $X^{\tilde \mu}$ according to
\begin{equation}
       X^{\tilde \mu}= \tilde \Omega^{-1} X^{\tilde \mu \tilde
         \nu}\partial_{\tilde \nu} \tilde \Omega \;.
\end{equation}

Let $Q^{\tilde A}(\tilde x^B)$ be a complex polarization dyad
satisfying
\begin{equation}
           Q^{(\tilde A}\bar Q^{\tilde B )} =Q^{\tilde A \tilde B} \;.
\end{equation}
We have
\begin{equation}
  Q^{\tilde A} Q^{\tilde B} \partial_{\tilde \ell} \tilde g_{\tilde A
    \tilde B} |_{\scri^+} = - Q_{\tilde A} Q_{\tilde B} \bigg
  \{\partial_{\tilde \ell} \tilde g^{\tilde A \tilde B} -2\tilde
  g^{\tilde u \tilde A }\partial_{\tilde u} \tilde g^{\tilde u \tilde
    B} \bigg \}|_{\scri^+}\;.
\end{equation}
The outgoing null vector normal to the $(\tilde u=const, \tilde \ell=const)$
2-surfaces is
\begin{equation}
  N_{\tilde \alpha} = \partial_{\tilde \alpha}\tilde u -\frac{1}{2}
  \tilde g^{\tilde u \tilde u} \partial_{\tilde \alpha} \tilde \ell
  \;.
\end{equation}
The shear or asymptotic strain $h$ on a constant inertial time
cross-section of $\scri^+$ is then given by
\begin{equation}
  h = \frac{1}{2} Q^{\tilde A} Q^{\tilde B} \tilde \nabla_{\tilde A}
  N_{\tilde B} |_{\scri^+} \, ,
\end{equation}
so that
\begin{align}
  h=&\frac{1}{4} Q^{\tilde A} Q^{\tilde B} \bigg \{\partial_{\tilde
    \ell} \tilde g_{\tilde A \tilde B} -2 \tilde g_{\tilde \ell \tilde
    A:\tilde B}\bigg \}|_{\scri^+}\;,
     \\ =&\frac{1}{4} Q_{\tilde A}
  Q_{\tilde B}\bigg \{ 2\tilde g^{\tilde A \tilde u} \partial_{\tilde
    u} \tilde g^{\tilde B \tilde u} + 2 \tilde g^{\tilde u \tilde A
    :\tilde B} -\partial_{\tilde \ell} \tilde g^{\tilde A \tilde B}
  \bigg \}|_{\scri^+}\;.
\label{eq:strain}
\end{align}
Here,
\begin{equation}
  \partial_{\tilde \ell} \tilde g^{\tilde A \tilde B} = \omega^{-2}
  x^{\tilde A},_C x^{\tilde B},_D \ell,_{\tilde \ell} \partial_\ell
  \hat{g}^{CD} = \omega^{-3} x^{\tilde A},_C x^{\tilde B},_D
  \partial_\ell \hat{g}^{CD}\;.
\end{equation}

The strain $h$ has gauge freedom corresponding to the supertranslation
freedom in the choice of slicing of \(\scri^+\). A gauge independent
description of the radiation waveform is given by the news function
\begin{equation}
      N= \frac{1}{2} Q_{\tilde A} Q_{\tilde B} X^{\tilde A \tilde
        B}|_{\scri^+} =\partial_{\tilde u} h \;,
\label{eq:newsh}
\end{equation}
where $X^{\tilde A \tilde B}$ corresponds to the translation

\begin{equation}
\xi^{\tilde \alpha}|_{\scri^+} =\tilde \nabla^{\tilde \alpha} \tilde
\ell |_{\scri^+}=(1,0,0,0)\;,
\label{eq:T0}
\end{equation}
with
\begin{equation}
   \partial_{\tilde \ell} \xi^{\tilde \alpha}|_{\scri^+}
   =\partial_{\tilde u} (\frac{1}{2} \tilde g^{\tilde u \tilde
     u},0,\tilde g^{\tilde u \tilde A})|_{\scri^+} \;.
\label{eq:T0dl}
\end{equation}
A short calculation gives
\begin{equation}
      N= \frac{1}{4}Q_{\tilde A} Q_{\tilde B} \partial_{\tilde u} ( 2
      \tilde g^{\tilde u \tilde A:\tilde B} + 2\tilde g^{\tilde u
        \tilde A} \partial_{\tilde u} \tilde g^{\tilde u \tilde B} -
      \partial_{\tilde \ell} \tilde g^{\tilde A \tilde B} )\;,
\label{eq:T0News}
\end{equation}
in accordance with (\ref{eq:strain}) and~(\ref{eq:newsh}).

The absolute square of the news function determines the flux of energy
and momentum. The flux corresponding to a general asymptotic symmetry
is given by~\cite{Geroch1981}
\begin{equation}
     F_\xi = -\tilde \nabla_{\tilde \mu}\tilde \nabla_{\tilde \nu}
     X^{\tilde \mu \tilde \nu} +3\tilde \nabla_{\tilde \mu} X^{\tilde
       \mu} +\frac{3}{4} \tilde \nabla_{\tilde \mu}\tilde
     \nabla^{\tilde \mu} X +\frac{1}{24}\tilde R X \;.
         \label{eq:flux}
\end{equation}
Note that $F_\xi $ is a scalar so that it could be evaluated in any
coordinate system. However, its physical properties are only manifest
in the inertial $\tilde x^\mu$ coordinates.

Since $F_\xi $ is independent of the freedom (\ref{eq:xifreed}), it
suffices to extend the BMS generators to a neighborhood of $\scri^+$
with the linear $\tilde \ell$-dependence
\begin{equation}
          \xi^{\tilde \mu}(\tilde u, \tilde \ell, \tilde x^A)=
          \xi^{\tilde \mu}(\tilde u, 0, \tilde x^A) +\tilde \ell
          \frac {\partial \xi^{\tilde \mu}}{\partial \tilde \ell}
          (\tilde u, 0, \tilde x^A) \;,
\end{equation}
where the coefficients are determined by (\ref{eq:bms}) and
(\ref{eq:ellderivs}). This determines the asymptotic Killing vectors
in the neighborhood of ${\scri^+}$ for the purpose of computing the flux in
terms of (\ref{eq:flux}).

It is also possible to compute the time derivative of the flux in
terms of its relation to the Weyl tensor~\cite{Geroch1981},
\begin{equation}
   \dot F_\xi:= \tilde n^\mu \partial_\mu F_\xi = -\Omega^{-1}\tilde
   C_{\alpha\beta\gamma\delta} \tilde n^\beta \tilde n^\delta
   X^{\alpha\gamma}|_{\scri^+} \;,
\end{equation}
where asymptotic flatness implies $C_{\alpha\beta\gamma\delta}
=O(\Omega)$.  In inertial coordinates, in which $ X^{\tilde \alpha
  \tilde \ell}|_{\scri^+} =0$, this reduces to
\begin{equation}
     \dot F_\xi =\partial_{\tilde u} F_\xi = -\partial_{\tilde \ell}
     \tilde C_{\tilde A \tilde u \tilde B \tilde u} X^{\tilde A \tilde
       B}|_{\scri^+} \;.
\end{equation}
By virtue of the trace-free property of the Weyl tensor this may be
rewritten
\begin{equation}
    \partial_{\tilde u} F_\xi = -\frac{1}{4}\{ \partial_{\tilde \ell}
    \tilde C_{\tilde A \tilde u \tilde B \tilde u}Q^{\tilde
      A}Q^{\tilde B} X^{\tilde C \tilde D}\bar Q_{\tilde C}\bar
    Q_{\tilde D} +\partial_{\tilde \ell} \tilde C_{\tilde A \tilde u
      \tilde B \tilde u}\bar Q^{\tilde A}\bar Q^{\tilde B} X^{\tilde C
      \tilde D} Q_{\tilde C} Q_{\tilde D} \} |_{\scri^+} \;,
\end{equation}
or
\begin{equation}
\partial_{\tilde u} F_\xi = \Psi \bar{\mathcal{X}} + \bar{\Psi}
\mathcal{X}\;,
\label{eq:fluxdot}
\end{equation}
where \(\mathcal{X} = \frac{1}{2}X^{\tilde A \tilde B} Q_{\tilde A}
Q_{\tilde B} | _{\scri^+}\) and \(\Psi = -\frac{1}{2}\partial_{\tilde
  \ell} \tilde C_{\tilde A \tilde u \tilde B \tilde u} Q^{\tilde A}
Q^{\tilde B}|_{\scri^+}\). In Newman-Penrose notation $\bar \Psi$ is
the asymptotic \(\psi_4^0\) component of the Weyl tensor.

A straightforward computation based upon (\ref{eq:tildeaskill}) gives
 \begin{eqnarray}
         {\mathcal X}&=& \frac{1}{2} X^{\tilde A \tilde B} Q_{\tilde
           A} Q_{\tilde B}|_{\scri^+} \;,\nonumber
         \\ &=&\frac{1}{4}Q_{\tilde A} Q_{\tilde B} \{
         2(\partial_{\tilde \ell} \tilde g^{\tilde A \tilde C})
         {\xi^{\tilde B}}_{:\tilde C} -(\partial_{\tilde \ell} \tilde
         g^{\tilde A \tilde B})_{:\tilde C} {\xi^{\tilde C}} -
         \frac{3}{2}( \partial_{\tilde \ell} \tilde g^{\tilde A \tilde
           B} ){\xi^{\tilde C}}_{:\tilde C} +2 (\partial_{\tilde \ell}
         \xi^{\tilde B})^{:\tilde A} \nonumber \\ &+&
         2(\partial_{\tilde \ell} \tilde g^{\tilde A \tilde
           \ell})\partial_{\tilde \ell} \xi^{\tilde B} +2 \tilde
         g^{\tilde A \tilde u}\partial_{\tilde u}\partial_{\tilde
           \ell} \xi^{\tilde B} -( \partial_{\tilde u}
         \partial_{\tilde \ell} \tilde g^{\tilde A \tilde B}
         )\xi^{\tilde u} \}|_{\scri^+} \;.  \nonumber \\
 \end{eqnarray}
Here the $\partial_{\tilde \ell} \xi^{\tilde A}$ derivative of the Killing field
is supplied from (\ref{eq:ellderivs}).  Using the
properties of $\xi^\alpha |_{\scri^+} $ and $\partial_{\tilde \ell}
\tilde g^{\tilde A \tilde \ell}|_{\scri^+}$, this reduces to
\begin{eqnarray}
         {\mathcal X} &=&\frac{1}{4}Q_{\tilde A} Q_{\tilde B} \{ (
         \partial_{\tilde \ell} \tilde g^{\tilde A \tilde B}
         )\xi^{\tilde C:\tilde D}Q_{[C}\bar Q_{D]} -(\partial_{\tilde
           \ell} \tilde g^{\tilde A \tilde B})_{:\tilde C}
                 {\xi^{\tilde C}} \nonumber \\ &+&2 (\partial_{\tilde
                   \ell} \xi^{\tilde B})^{:\tilde A} +
                 \partial_{\tilde u}(2 \tilde g^{\tilde A \tilde
                   u}\partial_{\tilde \ell} \xi^{\tilde B} -
                 \xi^{\tilde u} \partial_{\tilde \ell} \tilde
                 g^{\tilde A \tilde B} )   \}|_{\scri^+} \;. \nonumber \\
 \end{eqnarray}
These expressions simplify for specific BMS symmetries.

For a supertranslation $\xi^{\tilde u} =\alpha(\tilde x^A)$,
$f^{\tilde A}=0$ and we have
\begin{equation}
  \frac{\partial \xi^{\tilde A}}{\partial \tilde \ell}|_{\scri^+} = [
    \alpha \frac{\partial \tilde g^{\tilde u \tilde A}}{\partial
      \tilde u } - \alpha^{: \tilde A} ]|_{\scri^+} \;,
\end{equation}
with the result that
\begin{equation}
     {\mathcal X}= \alpha \partial _{\tilde u} h - \frac{1}{2}
     Q_{\tilde A} Q_{\tilde B} \alpha^{: \tilde A \tilde B} \;.
\end{equation}

For the translations, for which \(\alpha(\tilde x^A)\) is an \(\ell
=0\) or \(\ell = 1\) spherical harmonic, \(Q^{\tilde A} Q^{\tilde B}
\alpha_{:\tilde A \tilde B} =0\). Since \(\partial_{\tilde u} h = N\),
we have
\begin{equation}
\mathcal{X} = \alpha N\;.
\end{equation}

For the rotations,
$\alpha=0$ and $f^{\tilde A}=\epsilon^{\tilde A \tilde B} \Phi_{:\tilde B}$,
where $\Phi$ is an $\ell=1$ harmonic satisfying
$\Phi^{:\tilde A}{}_{:\tilde A}=-2\Phi$ with $\epsilon^{\tilde A \tilde
  B}= i Q^{[\tilde A} Q^{\tilde B]}$, $\epsilon^{\tilde A \tilde
  C}\epsilon_{\tilde B \tilde C}=\delta^{\tilde A}_{\tilde B}$.  As a
result, $f^{\tilde A : \tilde B}=-\Phi \epsilon^{\tilde A \tilde B}$,
and
\begin{equation}
       \partial_{\tilde \ell} \xi^{\tilde A} |_{\scri^+}= \{-
       g^{\tilde u \tilde B}{ f^{\tilde A}}_{:\tilde B}  +f^{\tilde
         B}{ g^{\tilde u \tilde A}}_{:\tilde B} \}|_{\scri^+}\;.
\end{equation}
We define a strain tensor according to
\begin{equation}
           h^{\tilde A \tilde B}=\frac{1}{4}\bigg \{ -
           \partial_{\tilde \ell} \tilde g^{\tilde A \tilde B}
           +2\tilde g^{\tilde A \tilde u} \partial_{\tilde u} \tilde
           g^{\tilde u \tilde B } +2 \tilde g^{\tilde u (\tilde A
             :\tilde B) } \bigg \}|_{\scri^+} \;,
\end{equation}
so that $ Q_{\tilde A} Q_{\tilde B}h^{AB} =h$.  Then a straightforward
calculation gives
\begin{eqnarray}
     {\mathcal X}&= &-2i \Phi h +Q_{\tilde A} Q_{\tilde B} { h^{\tilde
         A \tilde B}}_{:\tilde C} f^{\tilde C} \nonumber \\ &+&
     Q_{\tilde A} Q_{\tilde B} \{ f^{\tilde C} { g^{\tilde B \tilde
         u}}_{:\tilde C}{} ^{:\tilde A} - f^{\tilde C} g^{\tilde B \tilde
       u}{} ^{:\tilde A}{}_{:\tilde C} + g^{\tilde A \tilde u} f^{\tilde B}
     \}|_{\scri^+} \;.
\end{eqnarray}
But the second line vanishes due to the identity
\begin{equation}
         V_{B:CA} - V_{B:AC} = Q_{AB}V_C - Q_{BC}V_A\;,
\end{equation}
for the commutator of covariant derivatives with respect to the unit
sphere metric $Q_{AB}$.  Thus, for a rotation,
\begin{equation}
          {\mathcal X}= -2i \Phi h + Q_{\tilde A} Q_{\tilde B} {
            h^{\tilde A \tilde B}}_{:\tilde C} f^{\tilde C} \;.
                                \label{eq:rot}
\end{equation}

For the boosts, $\alpha=0$, $f^{\tilde A}= \Gamma^{:\tilde A}$, where
$\Gamma^{:\tilde A \tilde B}=-\Gamma Q^{\tilde A \tilde B}$,
$\Gamma^{:\tilde A}{}_{: \tilde A}=-2\Gamma$, $\xi^{\tilde u} = -u
\Gamma$ and
\begin{equation}
       \partial_{\tilde \ell} \xi^{\tilde A} |_{\scri^+}= \{- \tilde u
       \Gamma \partial_{\tilde u} g^{\tilde u \tilde A}
       +\Gamma^{:\tilde B}{ g^{\tilde u \tilde A}}_{:\tilde B} +\tilde
       u \Gamma^{:\tilde A} \}|_{\scri^+} \;.
\end{equation}
Then by the analogous calculation leading to (\ref{eq:rot})
\begin{equation}
     {\mathcal X}= -\Gamma \partial_{\tilde u} (\tilde u h) +
     Q_{\tilde A} Q_{\tilde B} { h^{\tilde A \tilde B}}_{:\tilde C}
     \Gamma^{: \tilde C} \;.
\end{equation}

\section{Results}

Our results are based upon the same generic precessing binary black
hole run taken from Taylor {\it et al.}~\cite{Taylor:2013zia}, which
was also used in \cite{Handmer:2014} to calibrate the \texttt{SpEC}
characteristic code and in \cite{Handmer:2015} to compare waveform
extraction using the \texttt{SpEC} and Pitt null codes using the
gauge invariant version of the of the news function computed in
computational coordinates. The physical parameters are mass ratio
\(q=3\), black hole dimensionless spins \(\chi_1 =
(0.7,0,0.7)/\sqrt{2}\) and \(\chi_2 = (-0.3,0,0.3)/\sqrt{2}\), number
of orbits 26, total time \(T=7509M\), initial eccentricity
\(10^{-3}\), initial frequency \(\omega_{ini} = 0.032/M\) and
extraction radius \(R=100M\), where $M$ is the total mass of the black
holes.

Extraction was carried out at the three different resolutions elaborated
in Table~\ref{tab:ALLTHETHINGS} to assess convergence.

\begin{table}[h]
\centering
  \begin{tabular}{| r | c c c |}
    \hline Run & Low & Med & High \\ \hline \(N_r\) & 10 & 12 & 14
    \\ \(L\) & 12 & 14 & 17 \\ \(\Delta t/M\) & 1.0 & \(0.666\dots\) &
    0.5 \\ \hline
  \end{tabular}
  \caption{\small{Resolution parameters used for code convergence
      comparisons, with time steps \(\Delta t\). \(N_r\) represents
      the radial grid sizes. The \texttt{SpEC} code has \(2 L^2\)
      total angular grid points.}}
  \label{tab:ALLTHETHINGS}
\end{table}

\subsection{Verification of strain, news, and radiative Weyl component  $\Psi$}

Here we show that the news function computed using the inertial
coordinate algorithm (inertial news) agrees with the gauge independent
news (gauge-free news) computed in~\cite{Handmer:2015} using the
computational coordinates. We also show that strain and and $\Psi$
(corresponding to the Newman-Penrose component $\bar \psi^0_4$) are
consistent with the computation of the news function.

Comparison of the relative error \(E_{rel}\) between dataset \(A\) and
dataset \(B\) is computed according to
\begin{equation}
E_{rel} = \log_{10}\left(\frac{|A-B|}{|B|}\right) \; ,
\end{equation}
where in the convergence tests \(B\) is the highest resolution
dataset.  For the strain, news function and radiative Weyl component
$\Psi$, the real parts of the \((\ell,m)=(2,2)\) spherical harmonic
modes are compared. 

Spatial convergence is at an exponential rate expected of a spectral
code, while time convergence is $4^{th}$ order. The convergence test
shows which error source dominates the simulation. Given an increase
in resolution by some constant factor, spectral convergence results in
reducing the error by a constant factor on a log scale.  In contrast,
high-order polynomial convergence such as $E \propto \Delta t^{4}$
will yield logarithmic convergence on a log scale under the same
increase in resolution.  For these simulations, $4^{th}$ order time
convergent error dominates over the spatial spectral error.

\begin{figure}[h!]
  \centering
  \includegraphics[width=0.9\textwidth]{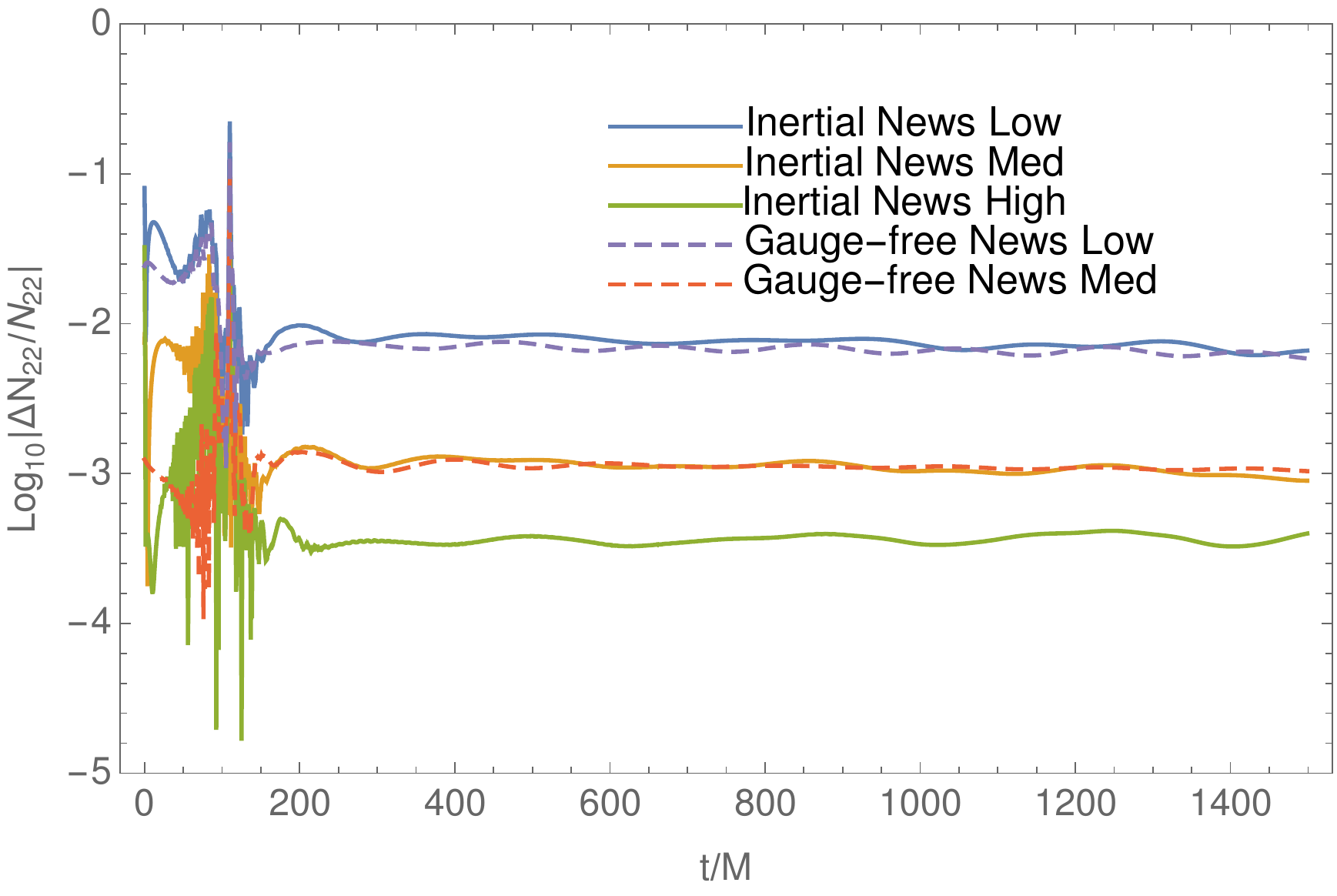}
  \caption{\small{$4^{th}$ order convergence of the news function computed two different ways.
  Solid lines are the error in the new, inertial coordinates computation.
  Dashed lines are error in the gauge-free news computed in computational coordinates.
  Here, error is computed with reference to the highest resolution grid,
  showing both self convergence {\it and} agreement between the two methods.}}
  \label{fig:NewsCvge}
\end{figure}

In Fig.~\ref{fig:NewsCvge}, we computed the news function in two
different ways at the three different resolutions outlined
in Table~\ref{tab:ALLTHETHINGS}, for a total of six datasets. The inertial
news (computed directly in the inertial coordinate system) is compared
with the high resolution results for the gauge-free news (computed
first in the computational coordinates, as in~\cite{Handmer:2015}). Both
versions of the news agree and display $4^{th}$ order convergence.

\begin{figure}[h!]
  \centering
  \includegraphics[width=0.9\textwidth]{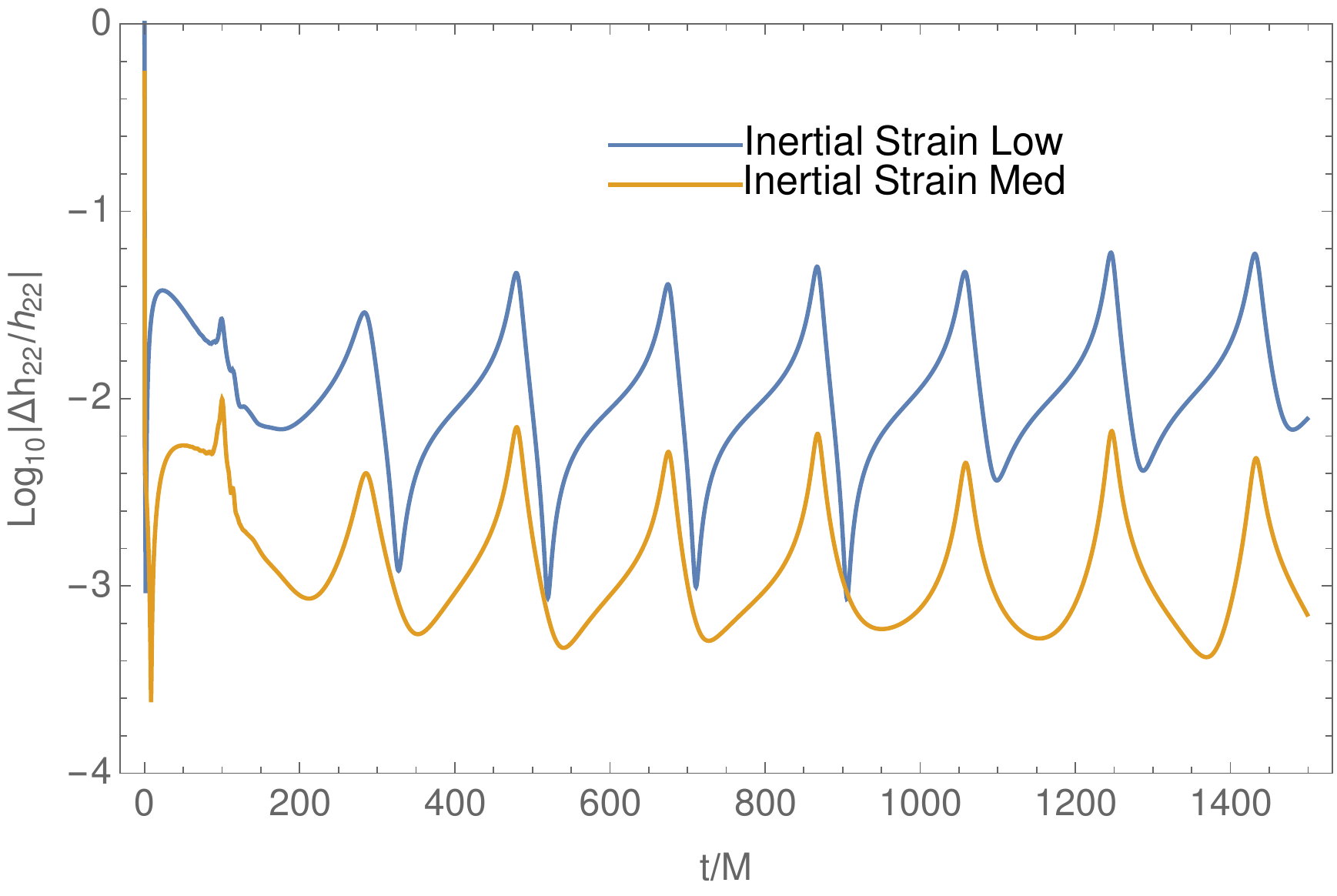}
  \caption{\small{Here, we compare error in the low and medium
      resolution calculations of the strain computed in the inertial
      coordinates (inertial strain) by comparing them to the medium
      and high resolution runs, respectively. Comparison between the
      errors in the medium and low resolution runs is consistent with
      the $4^{th}$ order convergence arising from the time
      integrator.}}
  \label{fig:StrainCvge}
\end{figure}

Similarly, we computed the error in the strain by comparing adjacent
resolutions. The strain remains convergent over the entire run, as
shown in Fig.~\ref{fig:StrainCvge}. With only a 20\% increase in
resolution, the error in the medium resolution run decreases by 85\%,
consistent with the underlying spatial spectral method and the
$4^{th}$ order time integrator.

\begin{figure}[h!]
  \centering
  \includegraphics[width=0.9\textwidth]{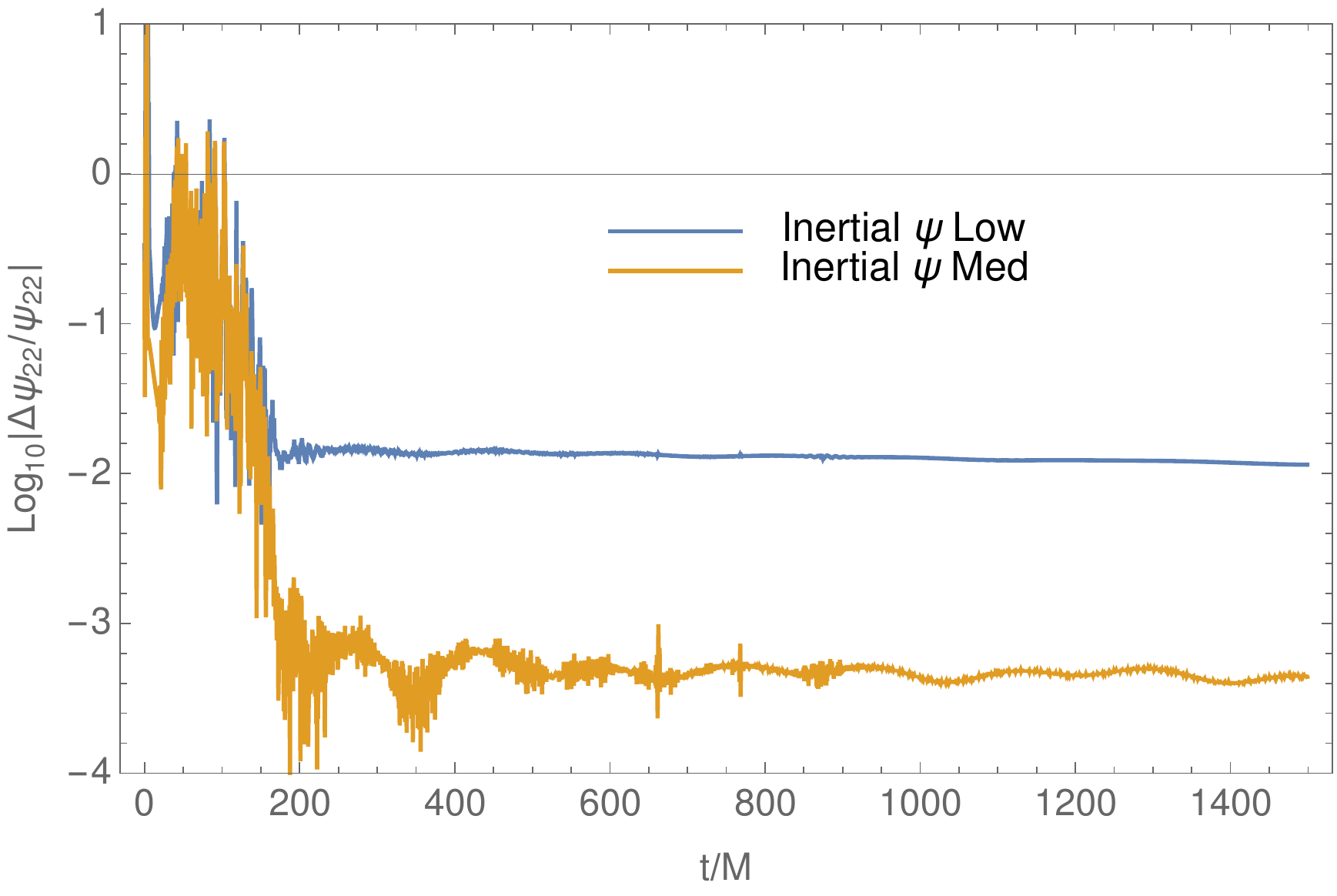}
  \caption{\small{The errors in the inertial coordinate computation of
      $\Psi$ for low and medium resolution runs are compared with the
      medium and high resolution runs, respectively. The results
      confirm $4^{th}$ order convergence following the initial phase
      of junk radiation.}}
  \label{fig:Psi4Cvge}
\end{figure}

The higher time derivatives in \(\Psi\) make it more sensitive to the
numerical noise produced by the junk radiation in the initial phase of
the run.  After this initial period, the error in \(\Psi\) displays
the same rate of convergence as the news and strain, as shown in
Fig.~\ref{fig:Psi4Cvge}.

\begin{figure}[h!]
  \centering
  \includegraphics[width=\textwidth]{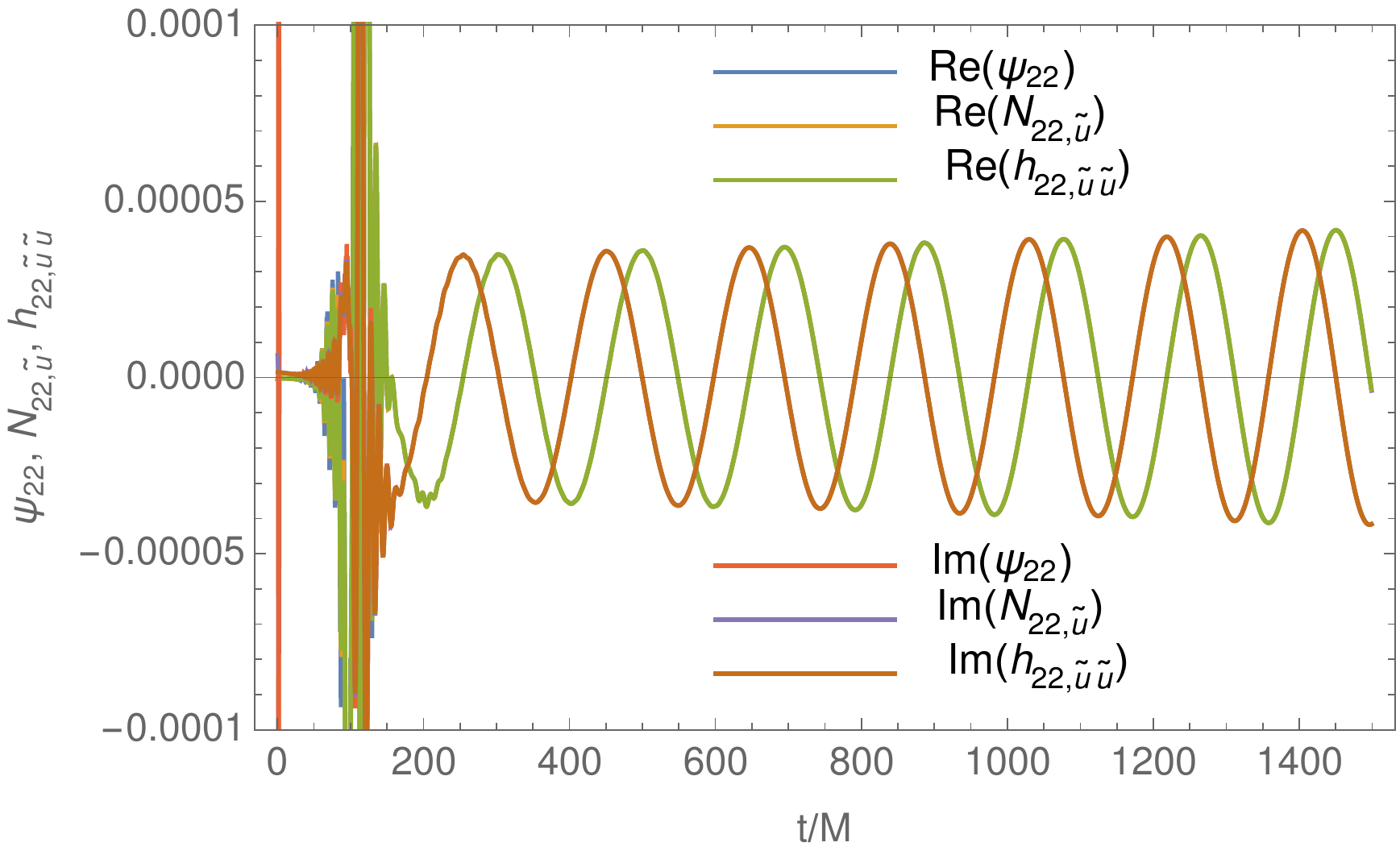}
  \caption{\small{Comparison of $\Psi = N,_{\tilde u} = h,_{\tilde u
        \tilde u}$ (real and imaginary parts).  Following the initial
      phase of junk radiation, the data overlap completely, hiding $\Psi$ and $N,_{\tilde u}$ behind $h,_{\tilde u \tilde u}$.}}
  \label{fig:Psi4NewsStrain}
\end{figure}

Finally, we verify the inertial frame relationships \(\Psi =
N,_{\tilde u} = h,_{\tilde u \tilde u}\).  Computed in the highest
resolution inertial domain, these quantities agree, as shown in
Fig.~\ref{fig:Psi4NewsStrain}.

\subsection{Flux}

In the inertial frame at \(\scri^+\), the asymptotic Killing vectors
which generate the BMS group have the form (\ref{eq:bms}), i.e.
\begin{equation}
         \xi^{\tilde \rho}\partial_{\tilde \rho} |_{\scri^+} = \bigg (
         \alpha(\tilde x^A) + \frac{1}{2} \tilde u f^{\tilde
           A}_{:\tilde A} \bigg )\partial_{\tilde u} + f^{\tilde
           A}\partial_{\tilde A}\;,
\end{equation}
where $ f^{\tilde A} (\tilde x^B)$ is a conformal killing vector for
the unit sphere metric.  Here we consider the fluxes corresponding to
the BMS generators for the time translation $T_u$, the three spatial
translations ($T_x$, $T_y$, $T_z$), the three rotations ($R_x$, $R_y$, $R_z$), and the
three boosts ($B_x$, $B_y$, $B_z$), with respect to the corresponding axes of
the asymptotic inertial frame, as well as a sample supertranslation
(ST), totaling 11 asymptotic symmetries.

We compute the energy flux by calculating the absolute square $|N|^2$
of the news function. For the remaining flux calculations, we first
use (\ref{eq:fluxdot}) to compute the flux rate of change and then
carry out a retarded inertial time integral. As a result, the numerical noise
in $\Psi$ during the initial phase of junk radiation (see
Fig.~\ref{fig:Psi4Cvge}) introduces some non-convergent
error. Consequently, although the plots of the time dependence of the
fluxes show good agreement for the three resolutions, convergence of
the error is not as clean as for the energy flux computed directly
from the news function.  Therefore, for the purpose of convergence
studies, we concentrate on the error in the retarded time derivative
of the flux, although both rates of convergence are shown for
completeness.

In all cases, we plot the waveform and convergence of the strongest
mode. Part (a) of each plot is a spherical representation of the functional form of
the corresponding BMS generator. Spatial components are shown with vector arrows,
while the time component is demarcated with a color gradient. 

\subsubsection{Energy and momentum flux}

\begin{figure}
  \centering
  \begin{subfigure}[b]{0.3\textwidth}
    \caption{} \includegraphics[width=\textwidth]{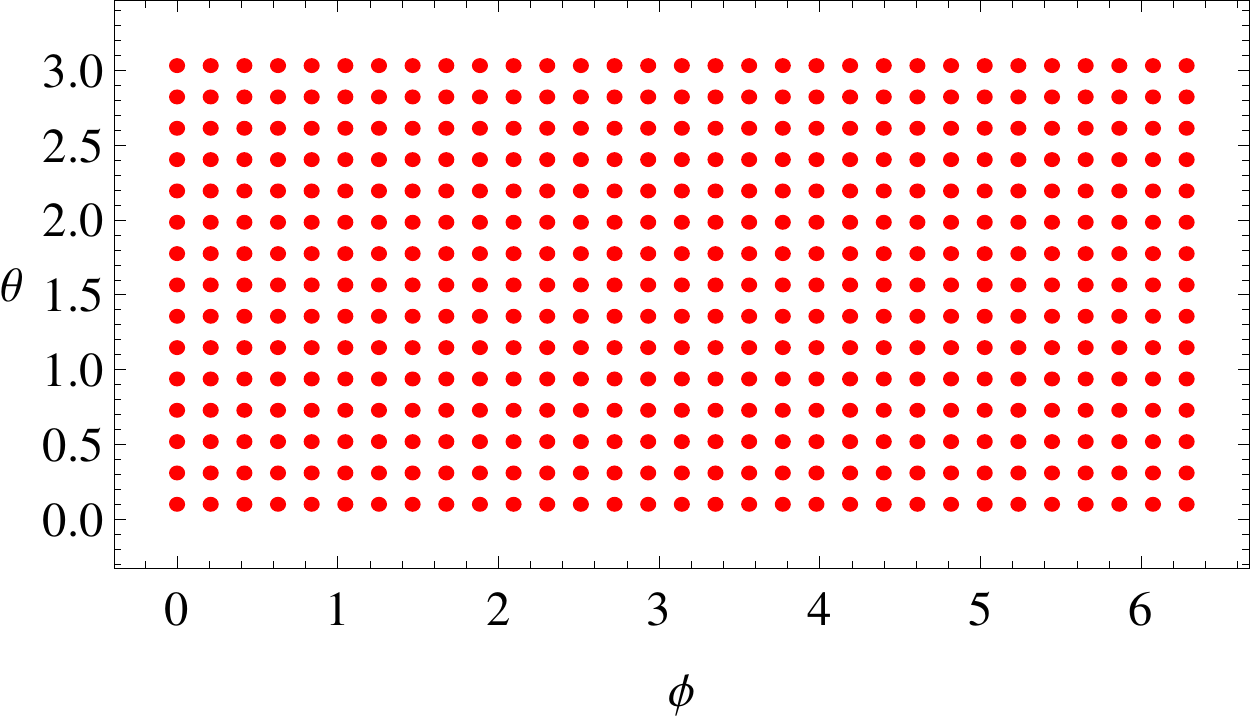}
    \label{fig:T0:KV}
  \end{subfigure}
  \begin{subfigure}[b]{0.3\textwidth}
    \caption{}
    \includegraphics[width=\textwidth]{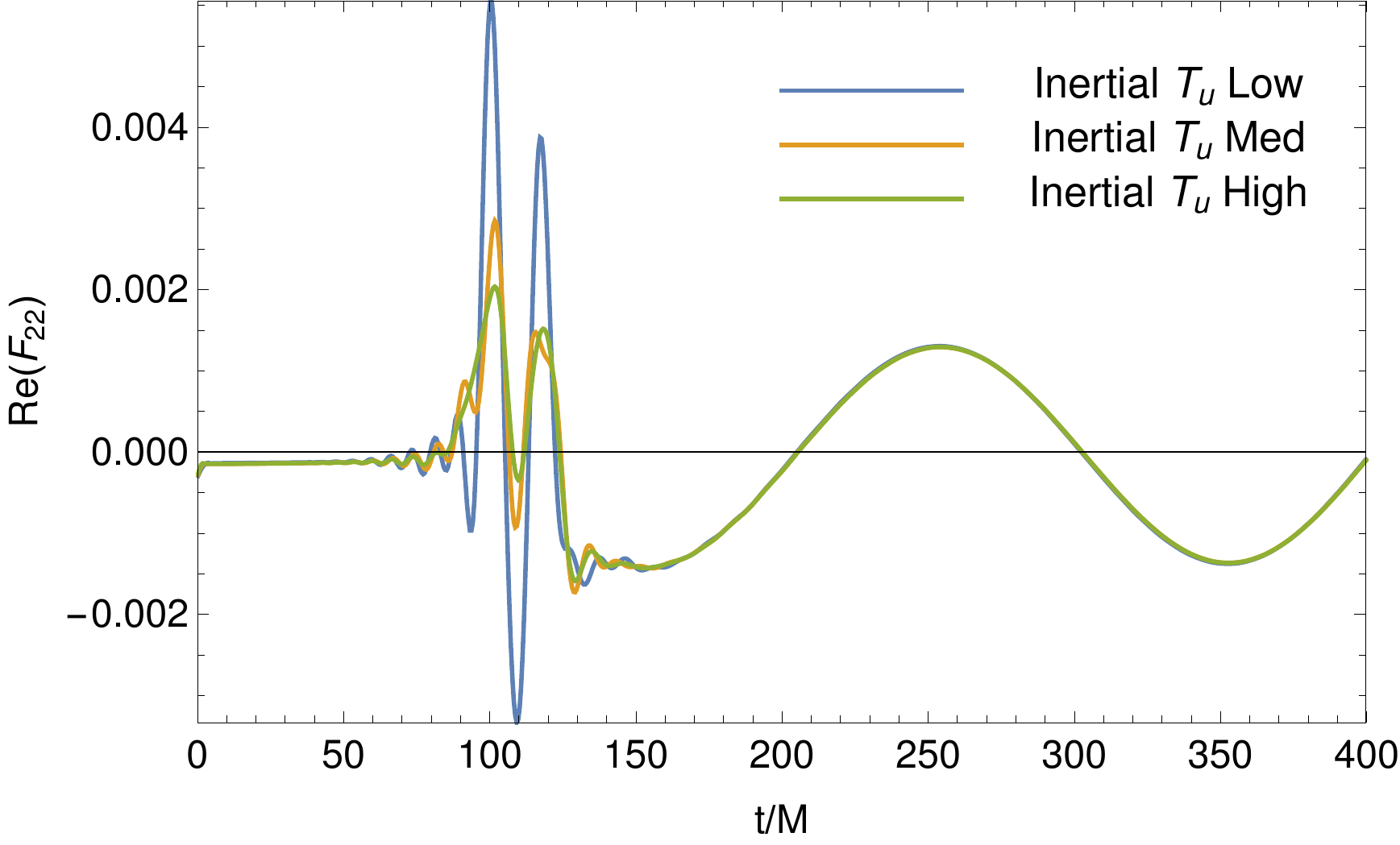}
    \label{fig:T0:Flux}
  \end{subfigure}

  \begin{subfigure}[b]{0.3\textwidth}
    \caption{}
    \includegraphics[width=\textwidth]{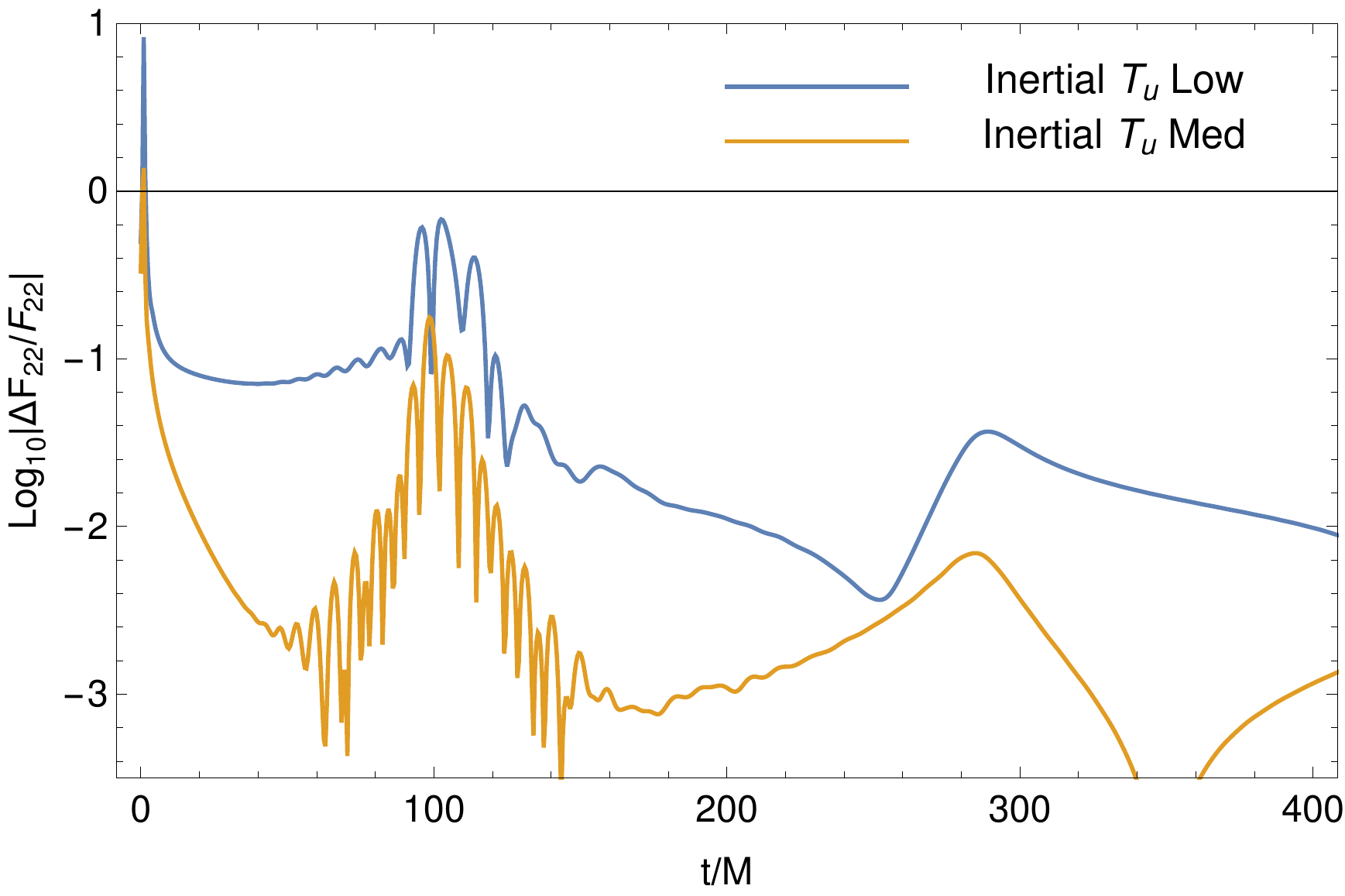}
    \label{fig:T0:FluxConv}
  \end{subfigure}
  \begin{subfigure}[b]{0.3\textwidth}
    \caption{}
    \includegraphics[width=\textwidth]{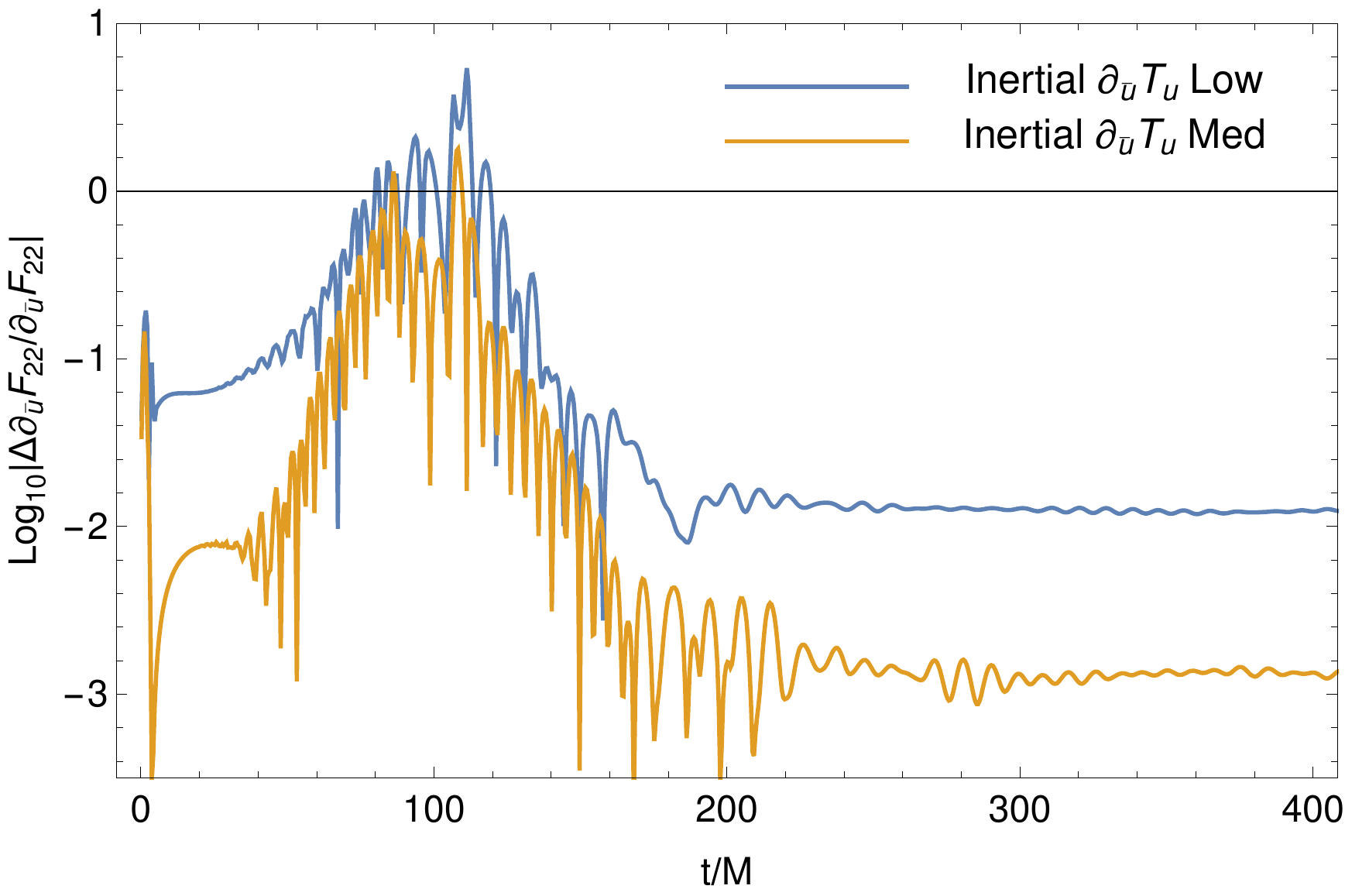}
    \label{fig:T0:FluxduConv}
  \end{subfigure}
  \caption{\small{ (\ref{fig:T0:KV}) The \(\tilde u\) component of
      \(\xi_{Tu}=1\) is uniform everywhere on the sphere and has no
      angular component.  (\ref{fig:T0:Flux}) The $(\ell=2,m=2)$
      spherical harmonic component of the flux, \(F_{Tu} = |N|^2\).
      (\ref{fig:T0:FluxConv}) Convergence of the flux is partially
      compromised by junk radiation, while the inertial time
      derivative of the flux (\ref{fig:T0:FluxduConv}) shows the
      appropriate $4^{th}$ order convergence following the junk phase.  }}
  \label{fig:T0}
\end{figure}

The time translation is described by the BMS generator
$\xi^\alpha_{[Tu]}$ with components $\alpha_{Tu}= 1$, $f^A=0$,
corresponding to an $\ell=0$ spherical harmonic. Figure~\ref{fig:T0}
shows the form of the generator (\ref{fig:T0:KV}), its associated
flux (\ref{fig:T0:Flux}), the flux convergence (\ref{fig:T0:FluxConv})
and the stronger convergence of the inertial time derivative of the
flux (\ref{fig:T0:FluxduConv}). 

The three momentum fluxes are derived from the three spatial
translations, described by the BMS generators $\xi^\alpha_{[Tx]}$,
$\xi^\alpha_{[Ty]}$ and $\xi^\alpha_{[Tz]}$, with $f^A=0$ and $\alpha$
constructed from \(l=1\) spherical harmonics,
\begin{equation}
 \alpha_{[Tx]} =\sin \tilde \theta \cos \tilde \phi \;, \quad
 \alpha_{[Ty]} =\sin \tilde \theta \sin \tilde \phi \;, \quad
 \alpha_{[Tz]} =\cos \tilde \theta \;,
\end{equation}
corresponding to the axes of the asymptotic inertial frame.  The
corresponding momentum fluxes $F_{Tx}$, $F_{Ty}$ and $F_{Tz}$ can also
be obtained directly from the energy flux by weighting it with the
corresponding $\ell=1$ harmonics,
\begin{equation}
 F_{Tx} = \sin\tilde\theta\cos\tilde\phi |N|^2, \quad F_{Ty} =
 \sin\tilde\theta\sin\tilde\phi |N|^2 , \quad F_{Tz} =\cos\tilde\theta
 |N|^2,
\end{equation}
in which case the clean $4^{th}$ order convergence obtained for 
the energy flux $|N|^2$ would also result following
the initial period of junk radiation.

\begin{figure}
  \centering
  \begin{subfigure}[b]{0.30\textwidth}
    \caption{} \includegraphics[width=\textwidth]{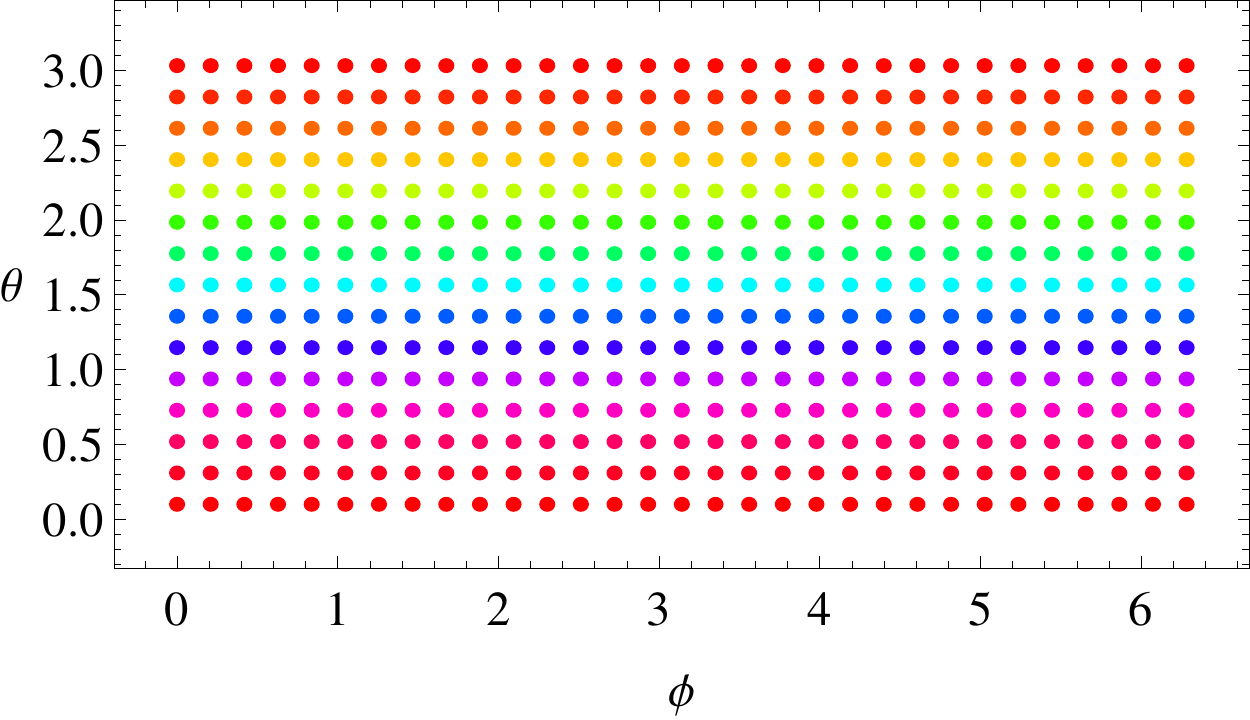}
    \label{fig:T1:KV}
  \end{subfigure}
  \begin{subfigure}[b]{0.30\textwidth}
    \caption{}
    \includegraphics[width=\textwidth]{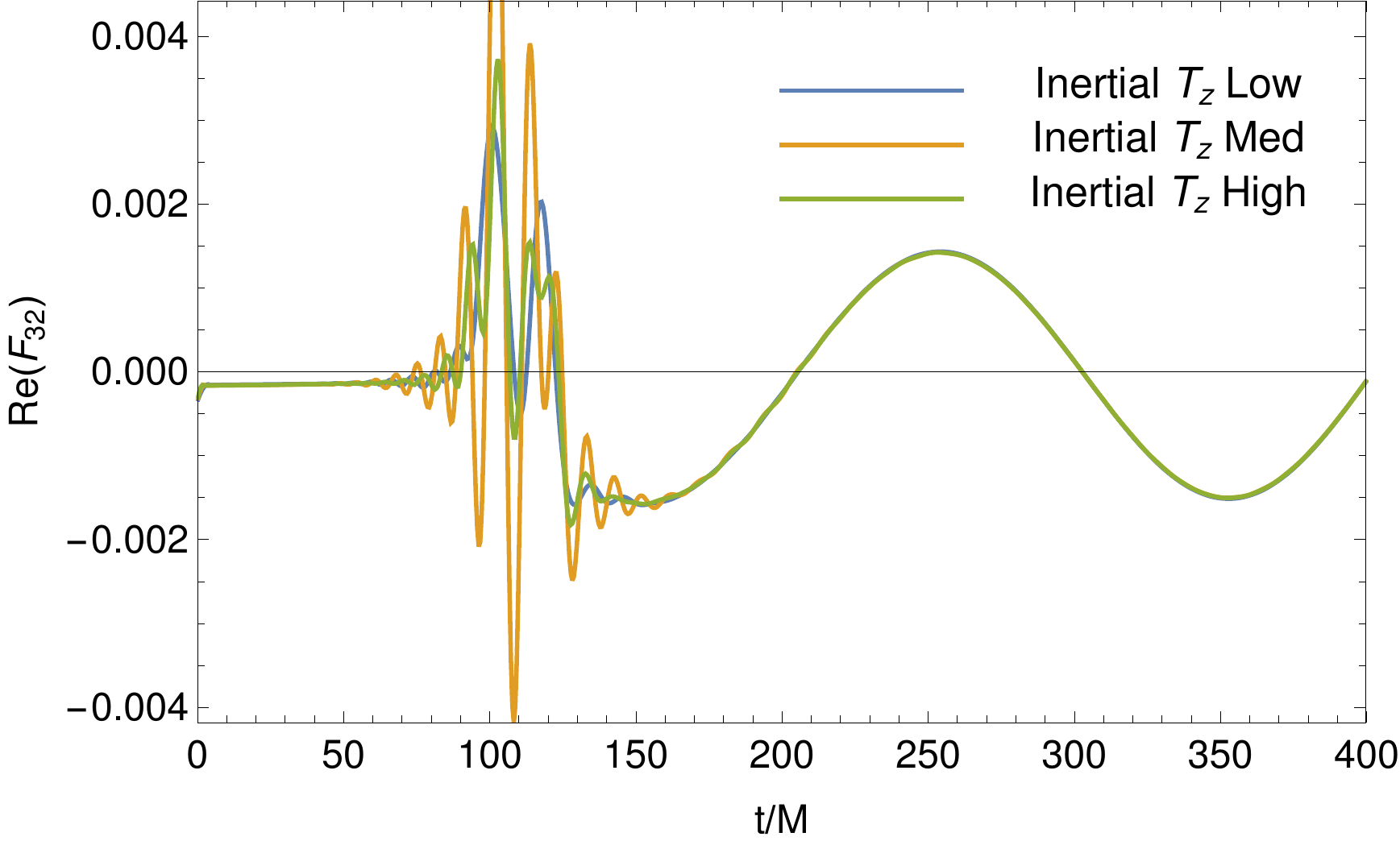}
    \label{fig:T1:Flux}
  \end{subfigure}

  \begin{subfigure}[b]{0.30\textwidth}
    \caption{}
    \includegraphics[width=\textwidth]{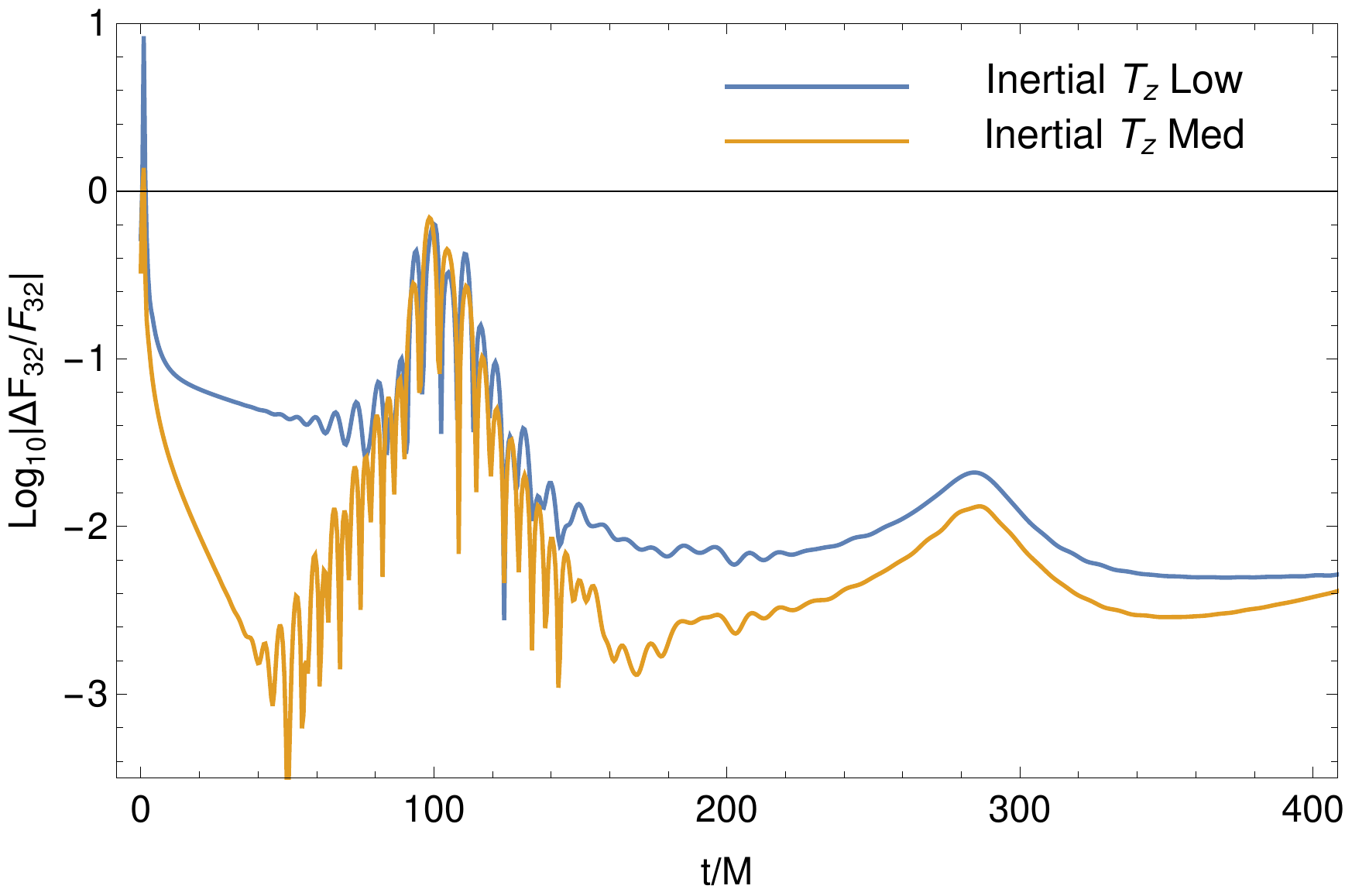}
    \label{fig:T1:FluxConv}
  \end{subfigure}
  \begin{subfigure}[b]{0.30\textwidth}
    \caption{}
    \includegraphics[width=\textwidth]{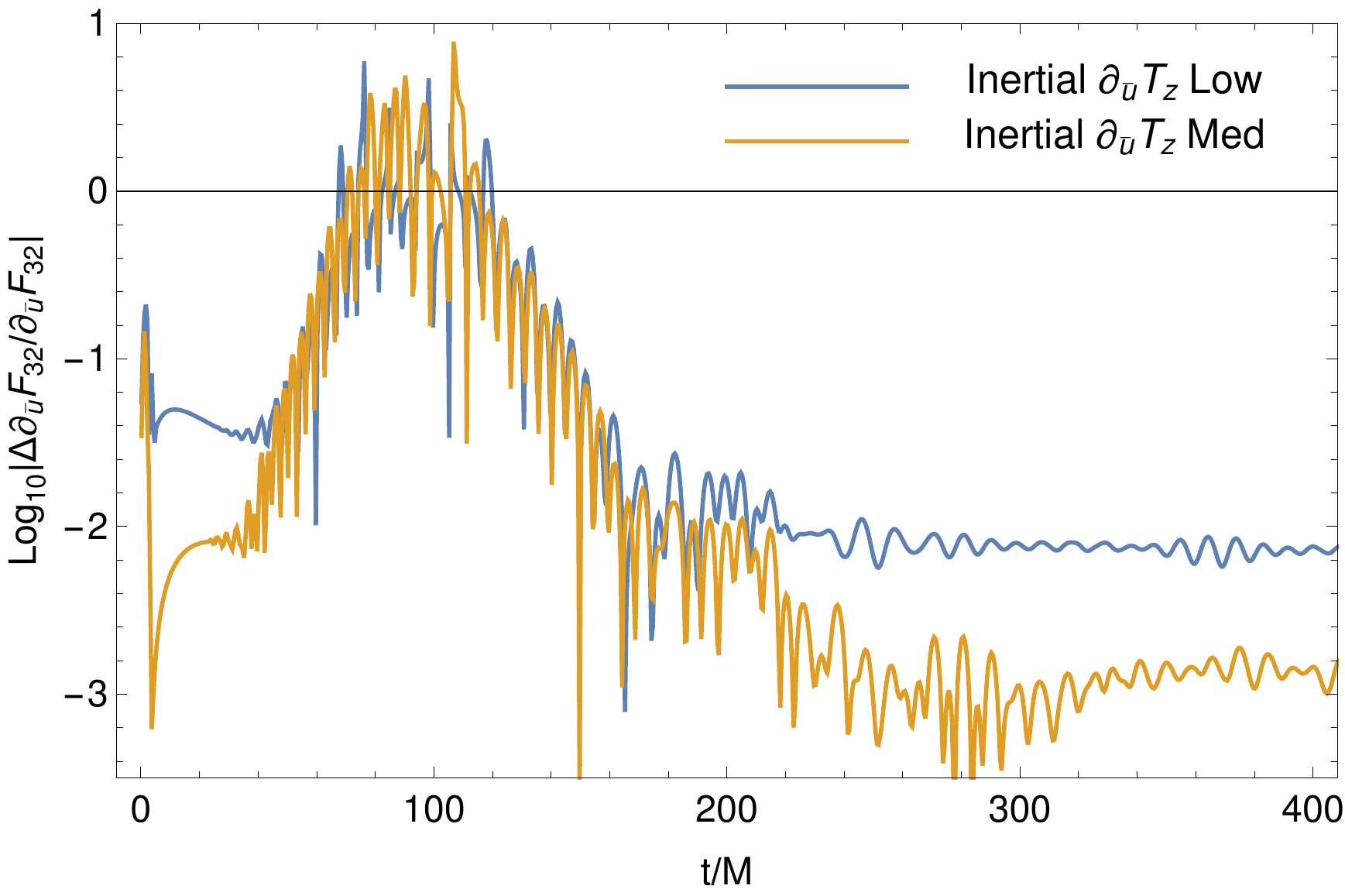}
    \label{fig:T1:FluxduConv}
  \end{subfigure}
  \caption{\small{ (\ref{fig:T1:KV}) The \(\tilde u\) component of
      \(\xi_{Tz}\) shifts from pole to pole, illustrating a
      translation in the $z$-direction.  (\ref{fig:T1:Flux}) The
      $(\ell=3,m=2)$ spherical harmonic component of the flux.
      (\ref{fig:T1:FluxConv}) Convergence of the flux is partially
      compromised by junk radiation, while the inertial time
      derivative of the flux (\ref{fig:T1:FluxduConv}) shows the
      appropriate $4^{th}$ order convergence following the junk phase. }}
  \label{fig:T1}
\end{figure}

Alternatively, these momentum fluxes can be obtained by a retarded
time integral, e.g.  $F_{Tz} =\int \dot F_{Tz}du$. However, although
the strain and energy flux both have dominant components in the
$(\ell=2,m=2)$ mode, the nonlinear effect of multiplication by an
$\ell=1$ harmonic shifts the momentum fluxes into other modes. For
this reason, we plot the $(\ell=3,m=2)$ mode.

The $z$-component of momentum flux \(F_{Tz}\) obtained this way is
shown in Fig.~\ref{fig:T1:Flux}. Convergence of the flux is shown in
Fig.~\ref{fig:T1:FluxConv} and its inertial time derivative in
Fig.~\ref{fig:T1:FluxduConv}.

\begin{figure}
  \centering
  \begin{subfigure}[b]{0.30\textwidth}
    \caption{} \includegraphics[width=\textwidth]{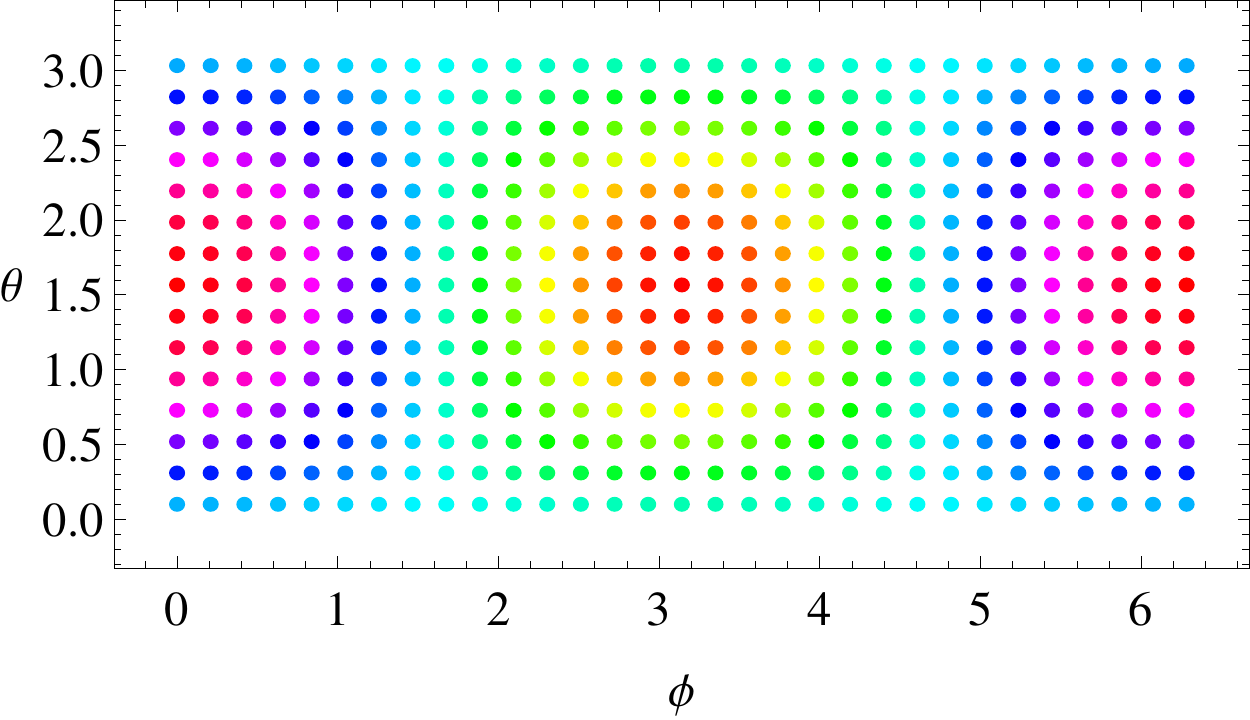}
    \label{fig:T2:KV}
  \end{subfigure}
  \begin{subfigure}[b]{0.30\textwidth}
    \caption{}
    \includegraphics[width=\textwidth]{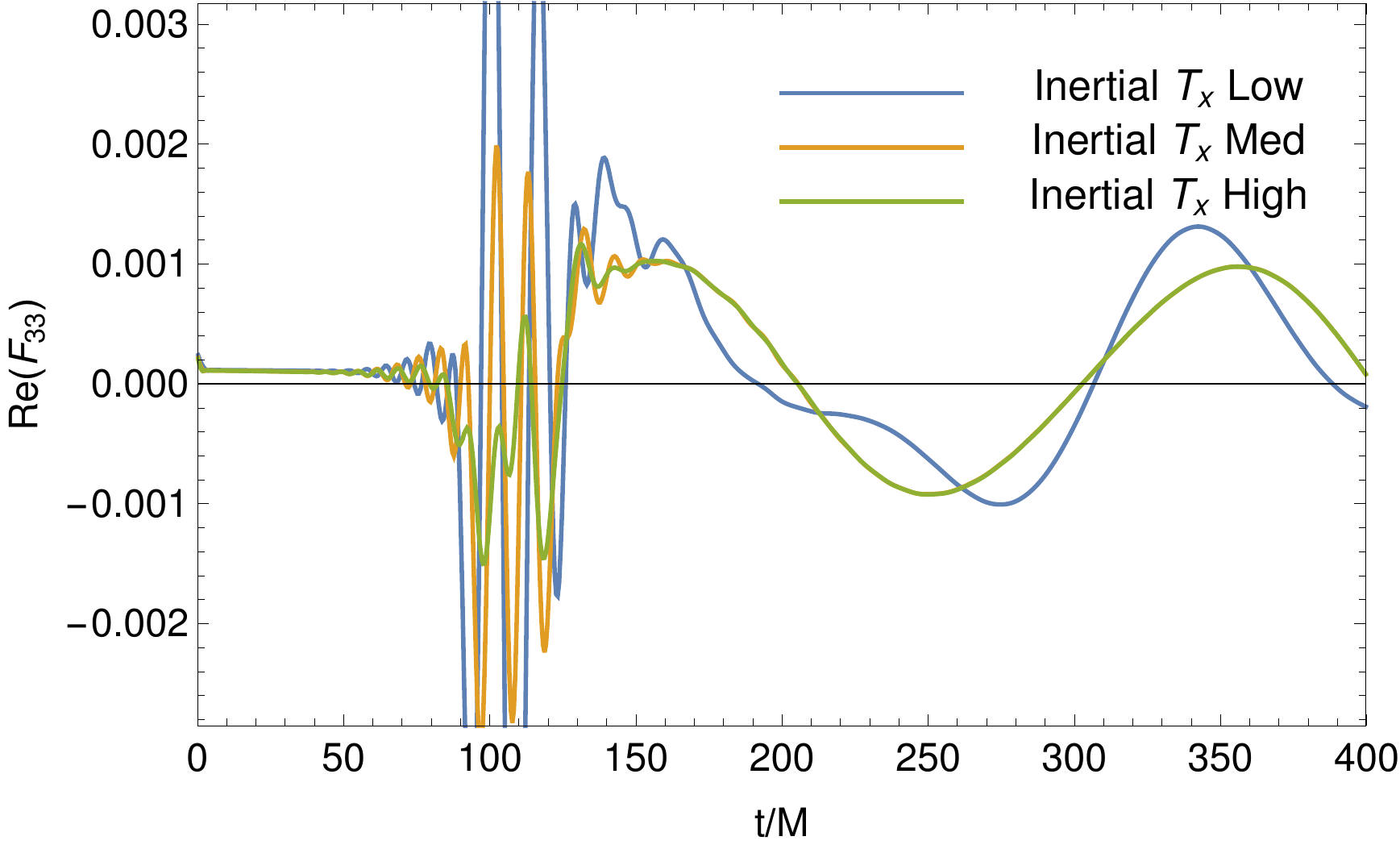}
    \label{fig:T2:Flux}
  \end{subfigure}

  \begin{subfigure}[b]{0.30\textwidth}
    \caption{}
    \includegraphics[width=\textwidth]{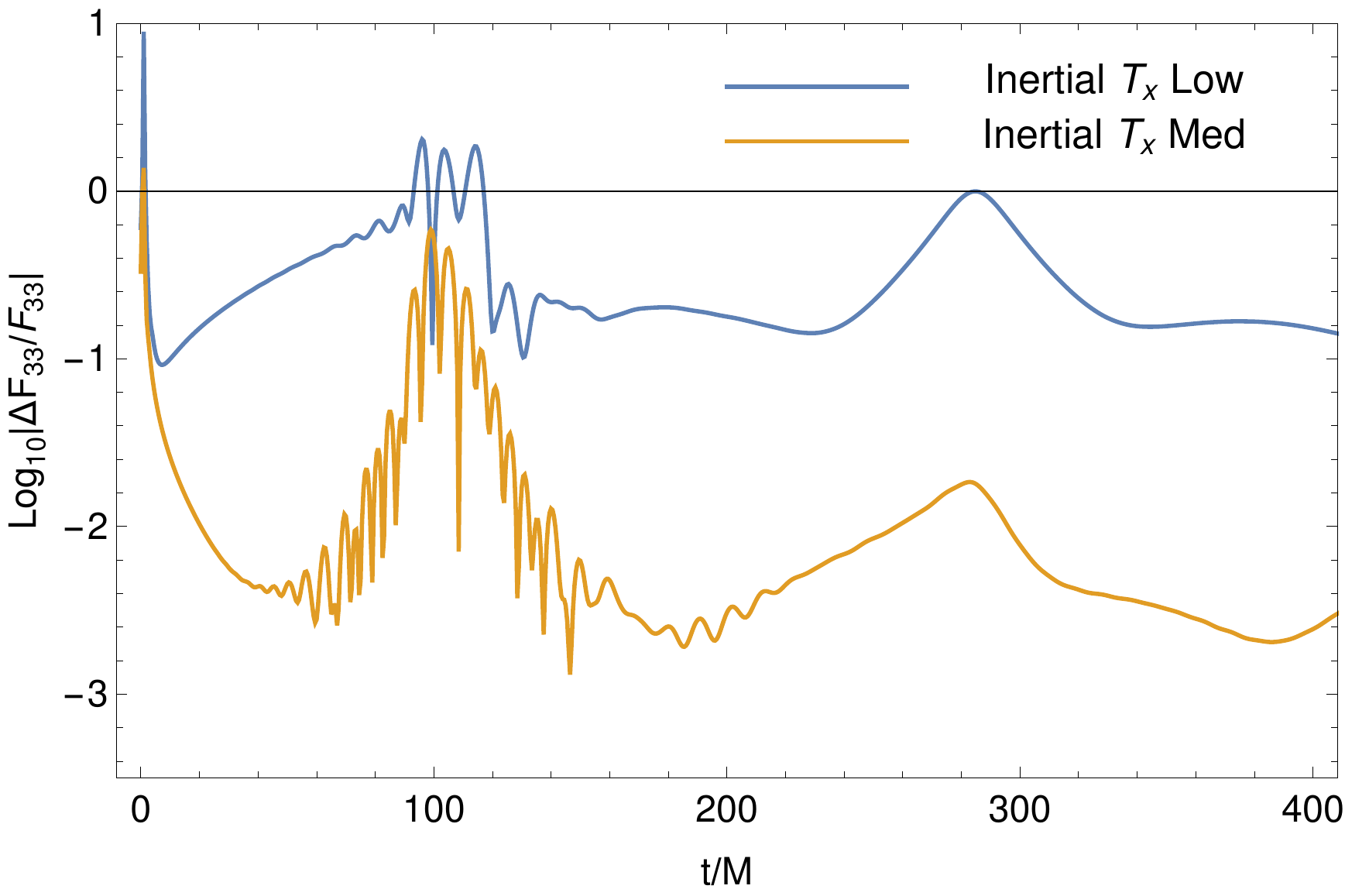}
    \label{fig:T2:FluxConv}
  \end{subfigure}
  \begin{subfigure}[b]{0.30\textwidth}
    \caption{}
    \includegraphics[width=\textwidth]{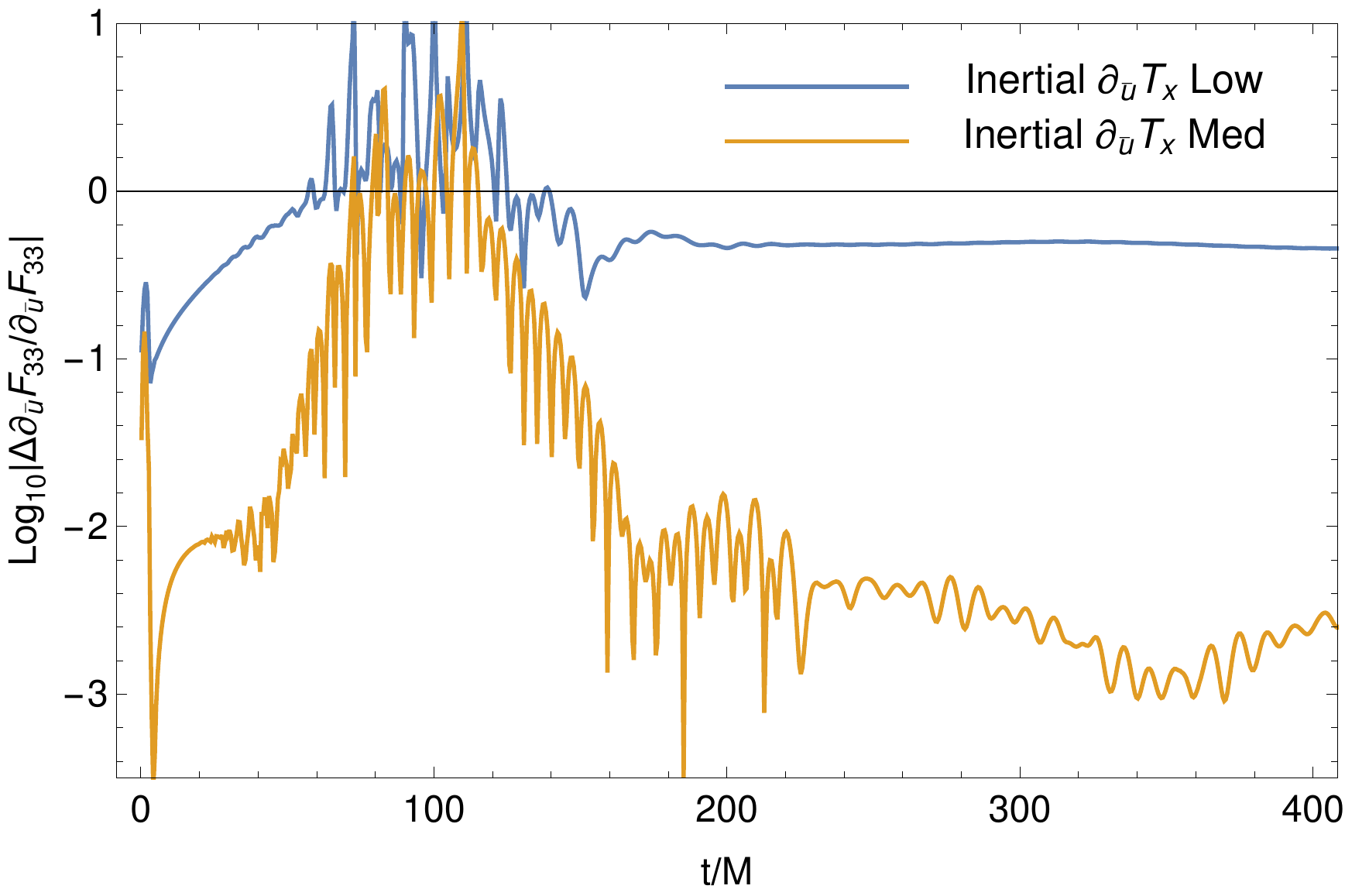}
    \label{fig:T2:FluxduConv}
  \end{subfigure}
  \caption{\small{ (\ref{fig:T2:KV}) The \(\tilde u\) component of
      \(\xi_{Tx}\) shifts with \(\phi\), showing a translation in the
      $x$ ($\phi=0$) direction.  (\ref{fig:T2:Flux}) The
      $(\ell=3,m=3)$ spherical harmonic component of the flux.
      (\ref{fig:T2:FluxConv}) Convergence of the flux is partially
      compromised by junk radiation, while the inertial time
      derivative of the flux (\ref{fig:T2:FluxduConv}) shows the
      appropriate $4^{th}$ order convergence following the junk phase.  }}
  \label{fig:T2}
\end{figure}

The $(\ell=3,m=3)$ mode of the $x$-component of the momentum flux $F_{Tx}$is shown in
Fig.~\ref{fig:T2:Flux}.  Convergence of the flux is shown in
Fig.~\ref{fig:T2:FluxConv} and its inertial time derivative in
Fig.~\ref{fig:T2:FluxduConv}.

\begin{figure}
  \centering
  \begin{subfigure}[b]{0.30\textwidth}
    \caption{} \includegraphics[width=\textwidth]{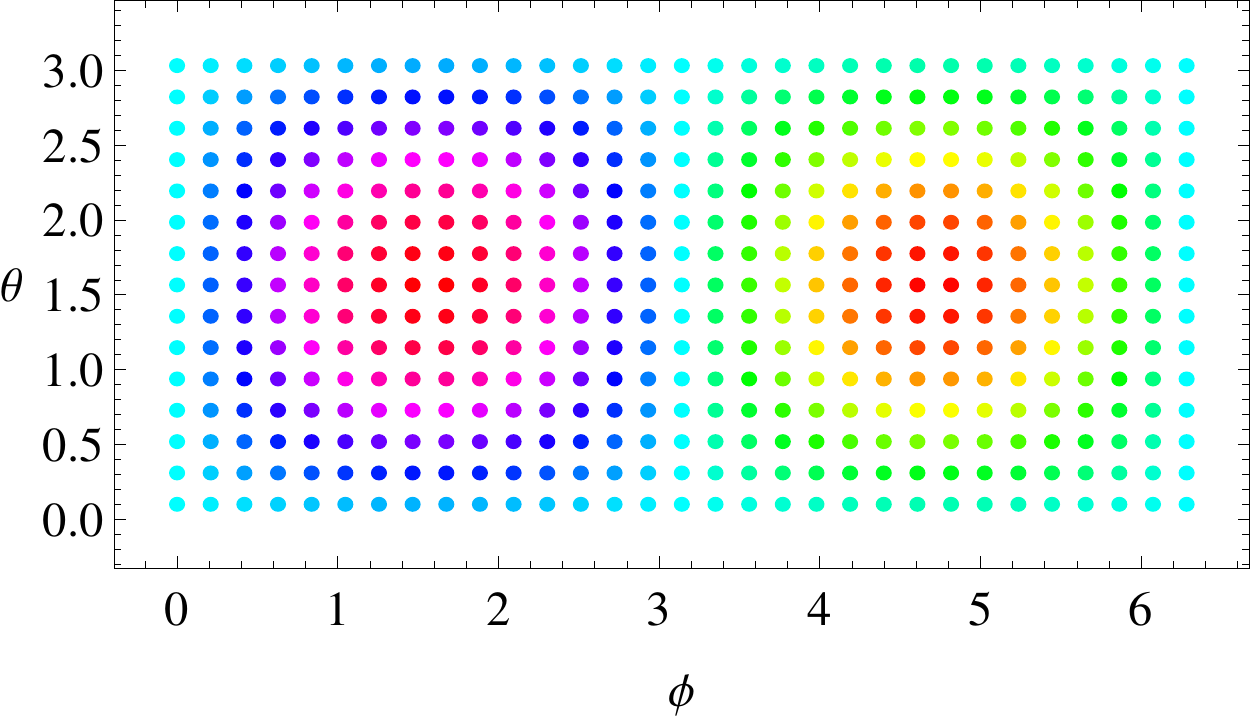}
    \label{fig:T3:KV}
  \end{subfigure}
  \begin{subfigure}[b]{0.30\textwidth}
    \caption{}
    \includegraphics[width=\textwidth]{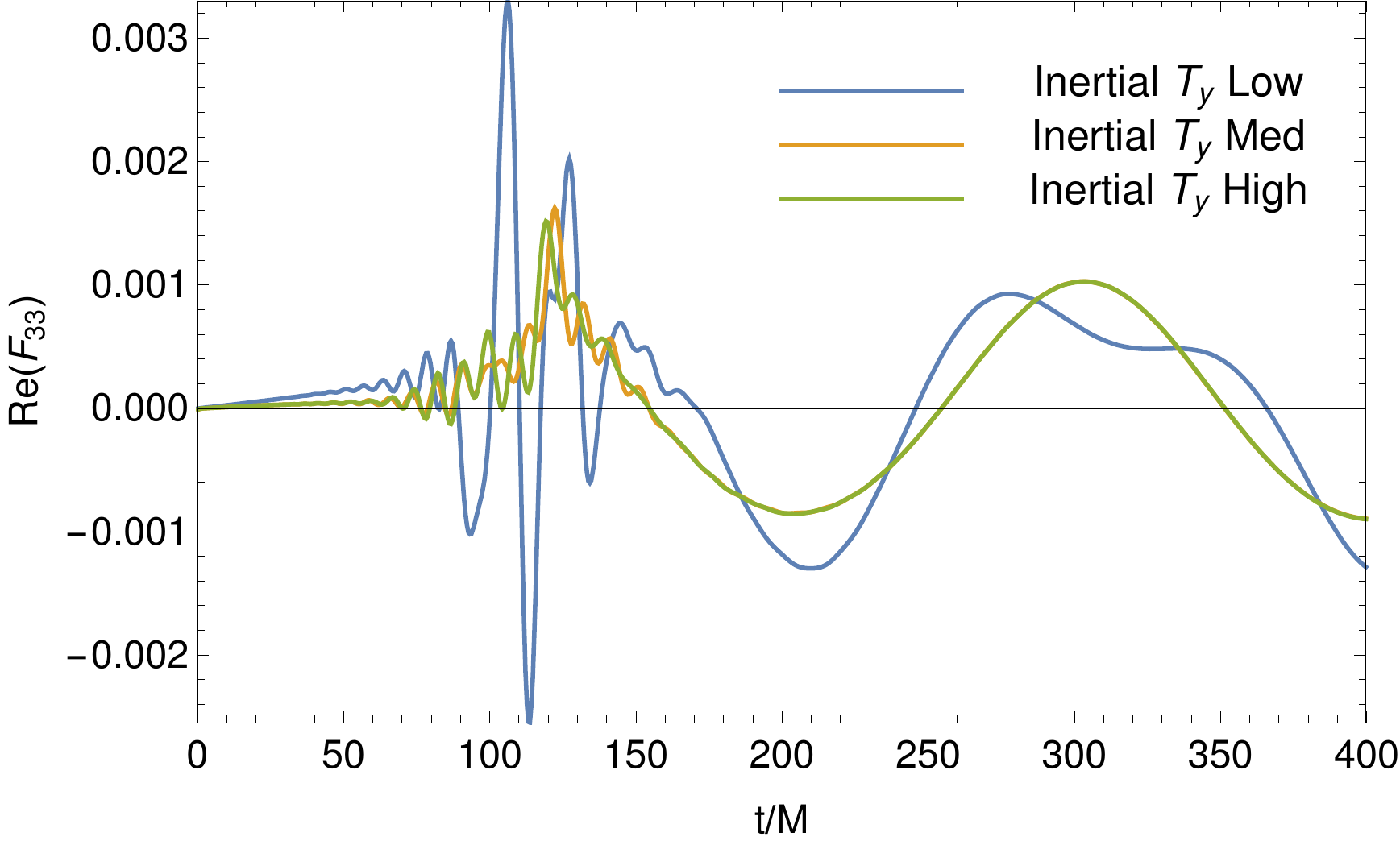}
    \label{fig:T3:Flux}
  \end{subfigure}

  \begin{subfigure}[b]{0.30\textwidth}
    \caption{}
    \includegraphics[width=\textwidth]{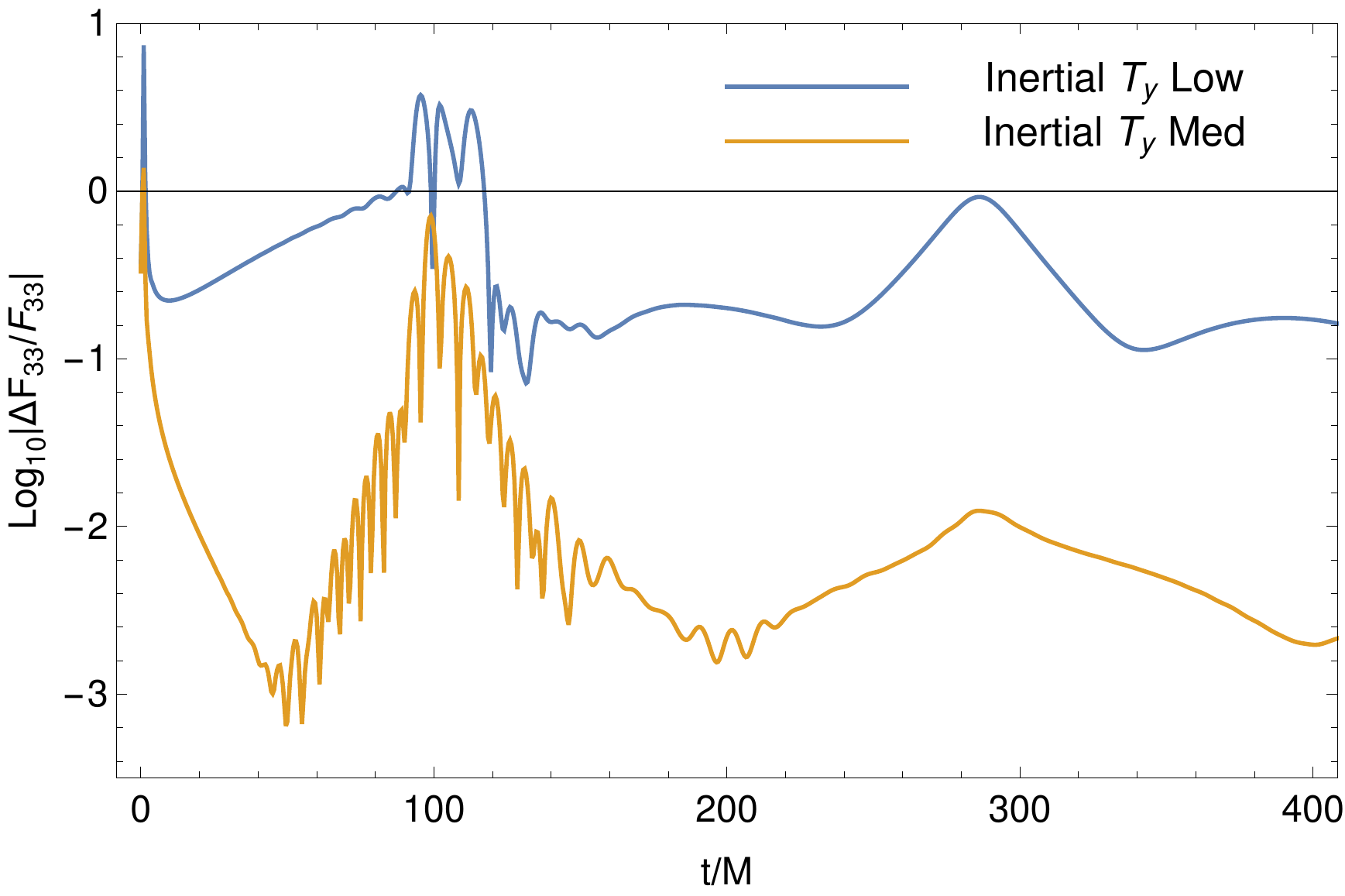}
    \label{fig:T3:FluxConv}
  \end{subfigure}
  \begin{subfigure}[b]{0.30\textwidth}
    \caption{}
    \includegraphics[width=\textwidth]{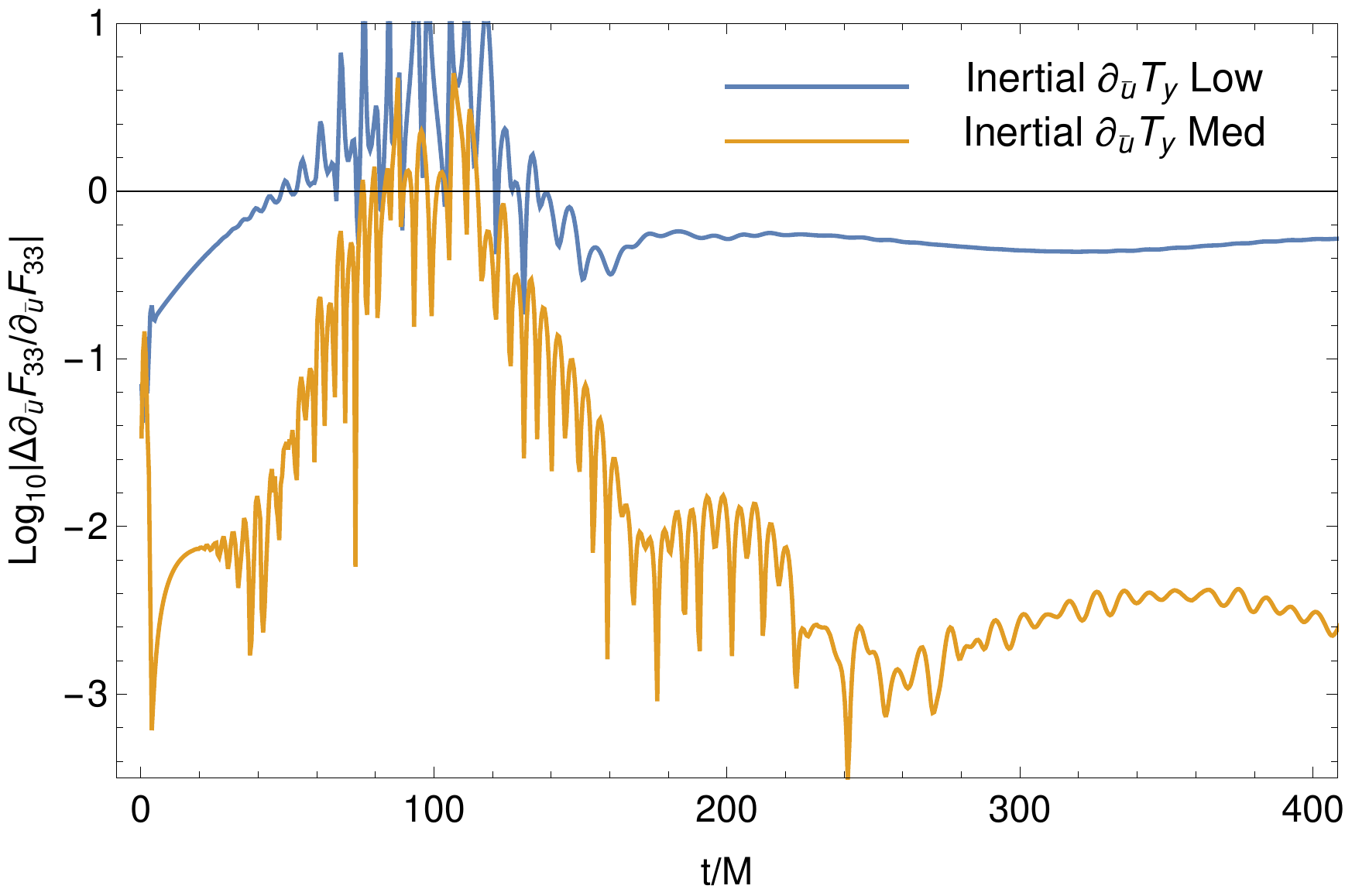}
    \label{fig:T3:FluxduConv}
  \end{subfigure}
  \caption{\small{ (\ref{fig:T3:KV}) The \(\tilde u\) component of
      \(\xi_{Ty}\) shifts with \(\phi\), showing a translation in the
      $y$ ($\phi=\pi/2$) direction.  (\ref{fig:T3:Flux}) The
      $(\ell=3,m=3)$ spherical harmonic component of the flux.
      (\ref{fig:T3:FluxConv}) Convergence of the flux is partially
      compromised by junk radiation, while the inertial time
      derivative of the flux (\ref{fig:T3:FluxduConv}) shows the
      appropriate $4^{th}$ order convergence following the junk phase.  }}
  \label{fig:T3}
\end{figure}

Similarly, the $(\ell=3,m=3)$ mode of the $y$-component of the momentum flux \(F_{Ty}\) is shown in
Fig.~\ref{fig:T3:Flux}.  Convergence of the flux is shown in
Fig.~\ref{fig:T3:FluxConv} and its inertial time derivative in
Fig.~\ref{fig:T3:FluxduConv}.

\subsubsection{Rotations}

\begin{figure}
  \centering
  \begin{subfigure}[b]{0.30\textwidth}
    \caption{} \includegraphics[width=\textwidth]{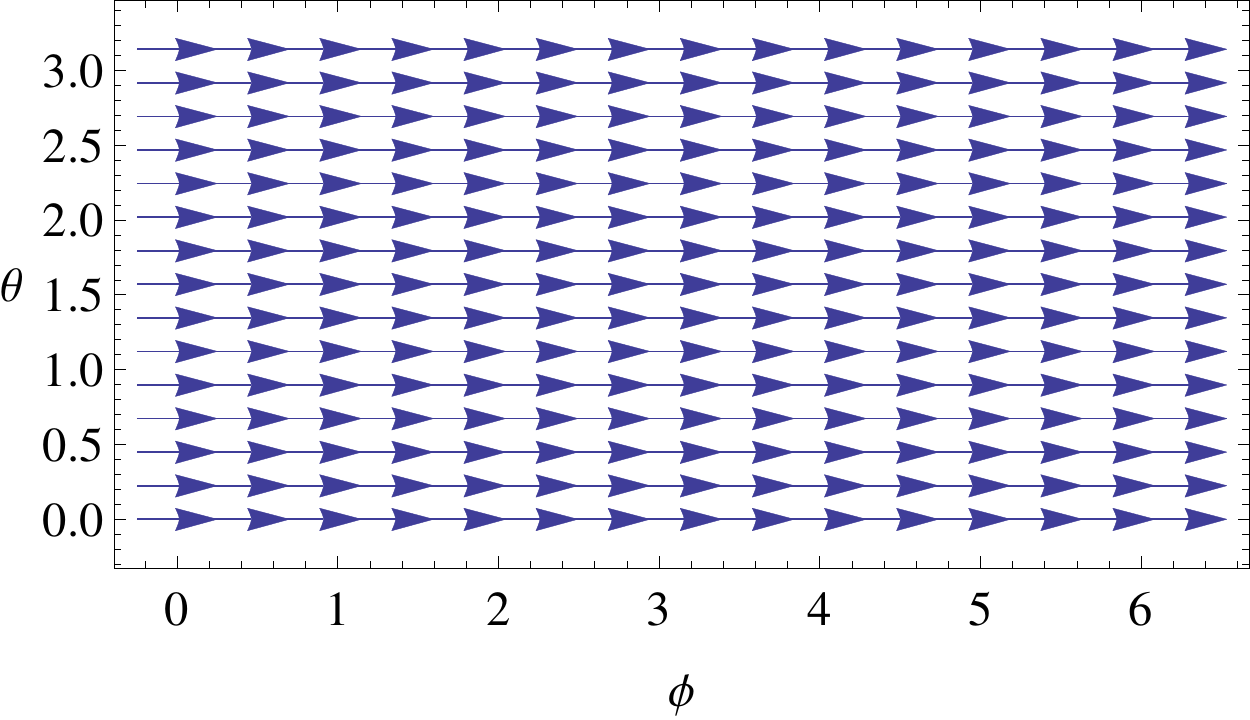}
    \label{fig:R1:KV}
  \end{subfigure}
  \begin{subfigure}[b]{0.30\textwidth}
    \caption{} \includegraphics[width=\textwidth]{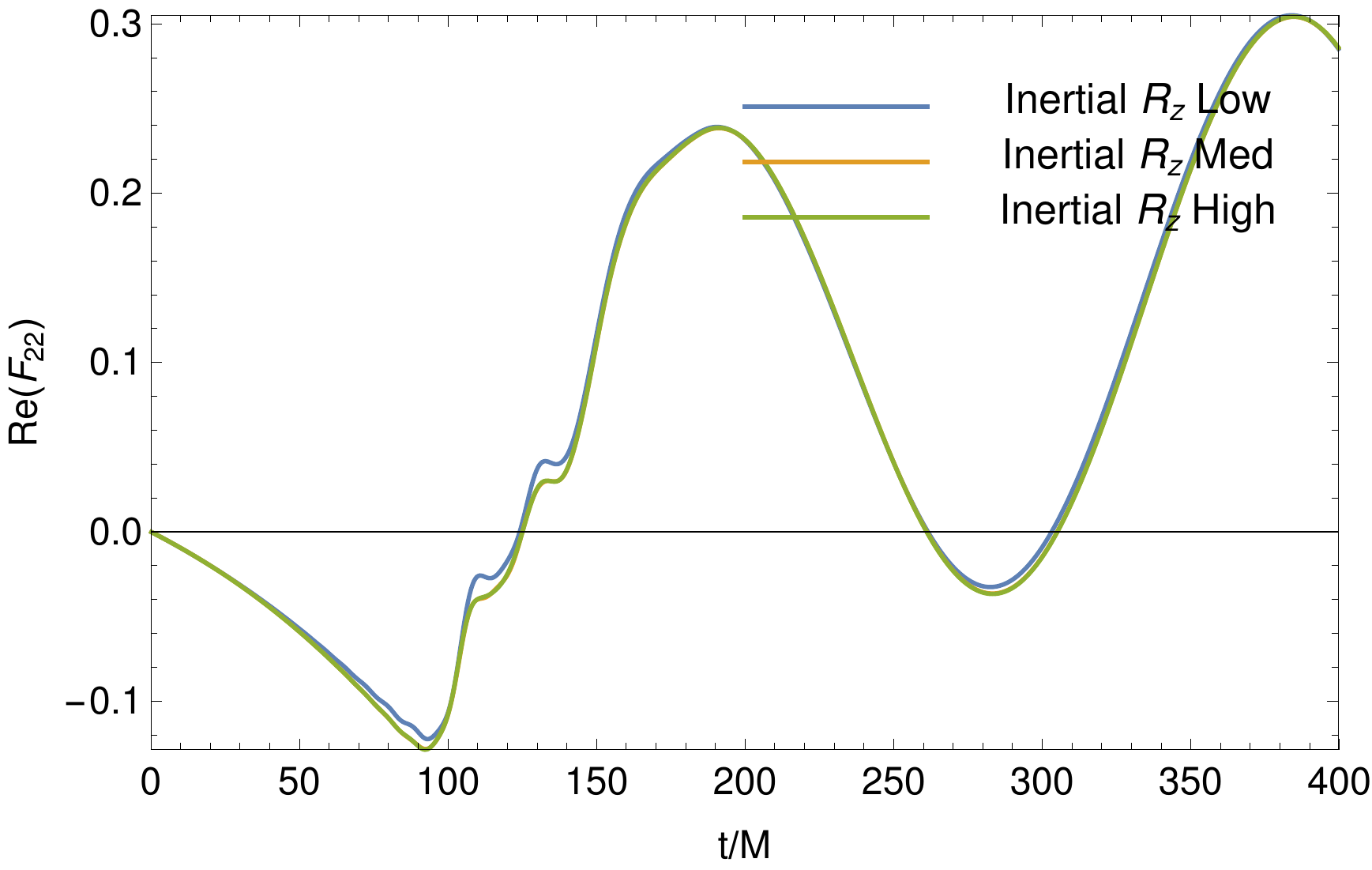}
    \label{fig:R1:Flux}
  \end{subfigure}

  \begin{subfigure}[b]{0.30\textwidth}
    \caption{} \includegraphics[width=\textwidth]{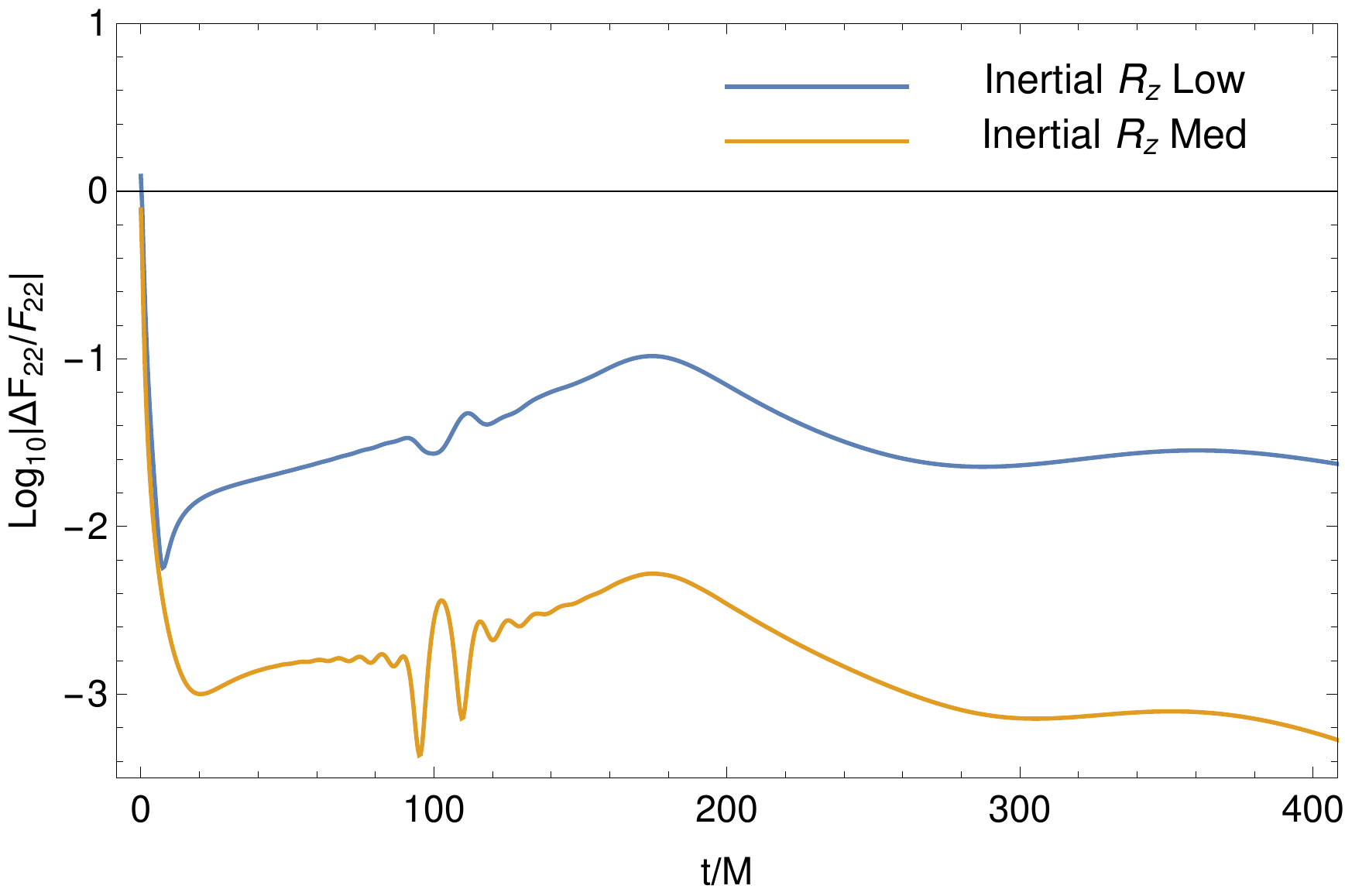}
    \label{fig:R1:FluxConv}
  \end{subfigure}
  \begin{subfigure}[b]{0.30\textwidth}
    \caption{} \includegraphics[width=\textwidth]{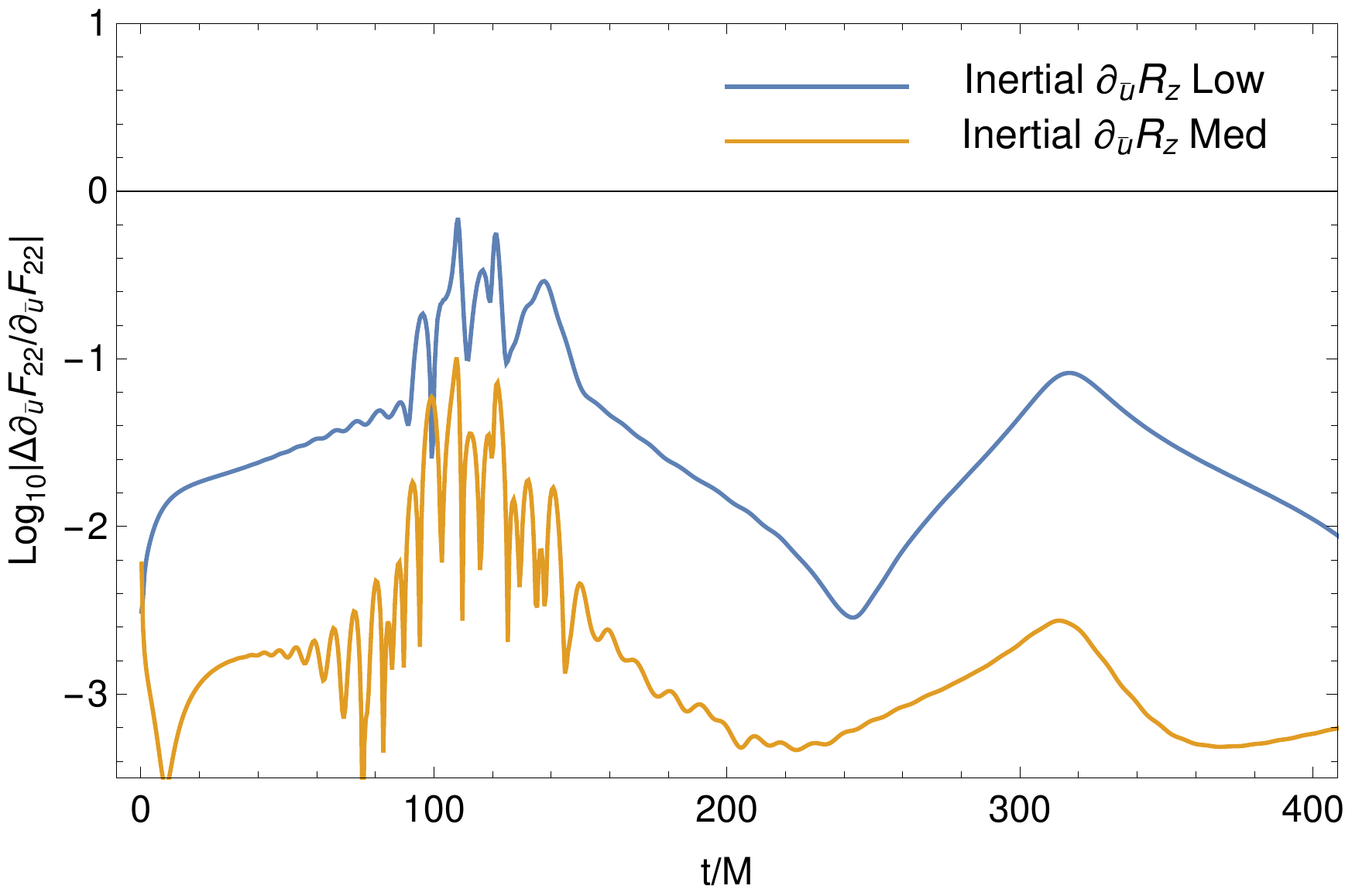}
    \label{fig:R1:FluxduConv}
  \end{subfigure}
  \caption{\small{ (\ref{fig:R1:KV}) Vectors illustrate the nature of
      \(\xi_{Rz}\), a rotation about $z$.  (\ref{fig:R1:Flux}) The
      $(\ell=2,m=2)$ spherical harmonic component of the flux.
      (\ref{fig:R1:FluxConv}) Convergence of the flux is partially
      compromised by junk radiation, while the inertial time
      derivative of the flux (\ref{fig:R1:FluxduConv}) shows the
      appropriate $4^{th}$ order convergence following the junk phase.  }}
  \label{fig:R1}
\end{figure}

The three spatial rotations with respect to the inertial axes are
described by the BMS generators $\xi^\alpha_{[Rx]}$,
$\xi^\alpha_{[Ry]}$ and $\xi^\alpha_{[Rz]}$ with \(\alpha=0\) and
$f^{\tilde A}=\epsilon^{\tilde A \tilde B} \Phi_{:\tilde B}$, where
$\Phi$ is constructed from \(\ell=1\) spherical harmonics. A rotation
$R_z$ about the $z$-axis corresponds to the spherical harmonic $\Phi =
\cos \tilde \theta$, so that $f^{\tilde A }= (0,1)$.  Most of the
motion of the orbiting black holes is about this axis, so we expect to
see a greater flux of the corresponding $z$-component of angular
momentum \(F_{Rz}\). 
The $(\ell=2,m=2)$ mode of the $z$ component of the angular momentum
flux \(F_{Rz}\) is shown in
Fig.~\ref{fig:R1:Flux}.  Convergence of the flux is shown in
Fig.~\ref{fig:R1:FluxConv} and its inertial time derivative in
Fig.~\ref{fig:R1:FluxduConv}.

A rotation $R_x$ about the $x$-axis corresponds to the spherical
harmonic $\Phi= \sin \tilde \theta \cos \tilde \phi$ so that
$f^{\tilde A }= (-\sin \tilde \phi, -\cot \tilde \theta \cos \tilde
\phi)$. Similarly, a rotation $R_y$ about the $y$-axis corresponds to the
spherical harmonic $\Phi= \sin \tilde \theta \sin \tilde \phi$ so that
$f^{\tilde A }= (\cos \tilde \phi,- \cot \tilde \theta \sin \tilde
\phi)$.

\begin{figure}
  \centering
  \begin{subfigure}[b]{0.30\textwidth}
    \caption{} \includegraphics[width=\textwidth]{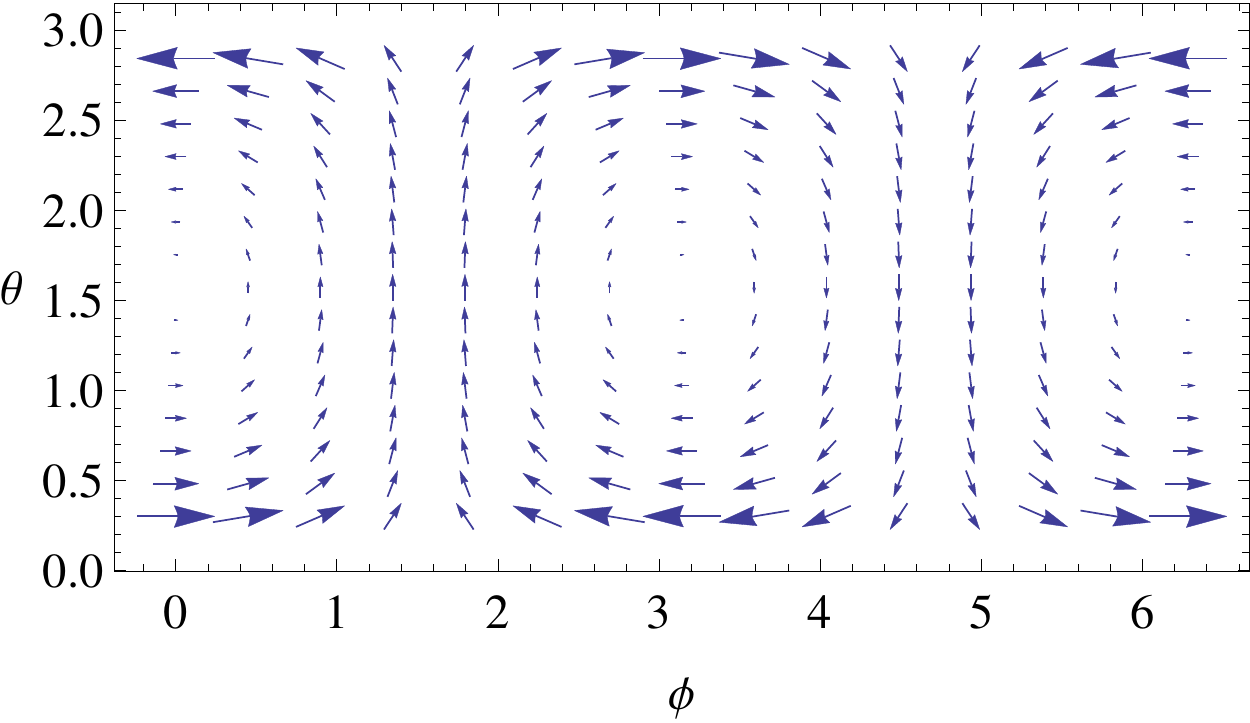}
    \label{fig:R3:KV}
  \end{subfigure}
  \begin{subfigure}[b]{0.30\textwidth}
    \caption{}
    \includegraphics[width=\textwidth]{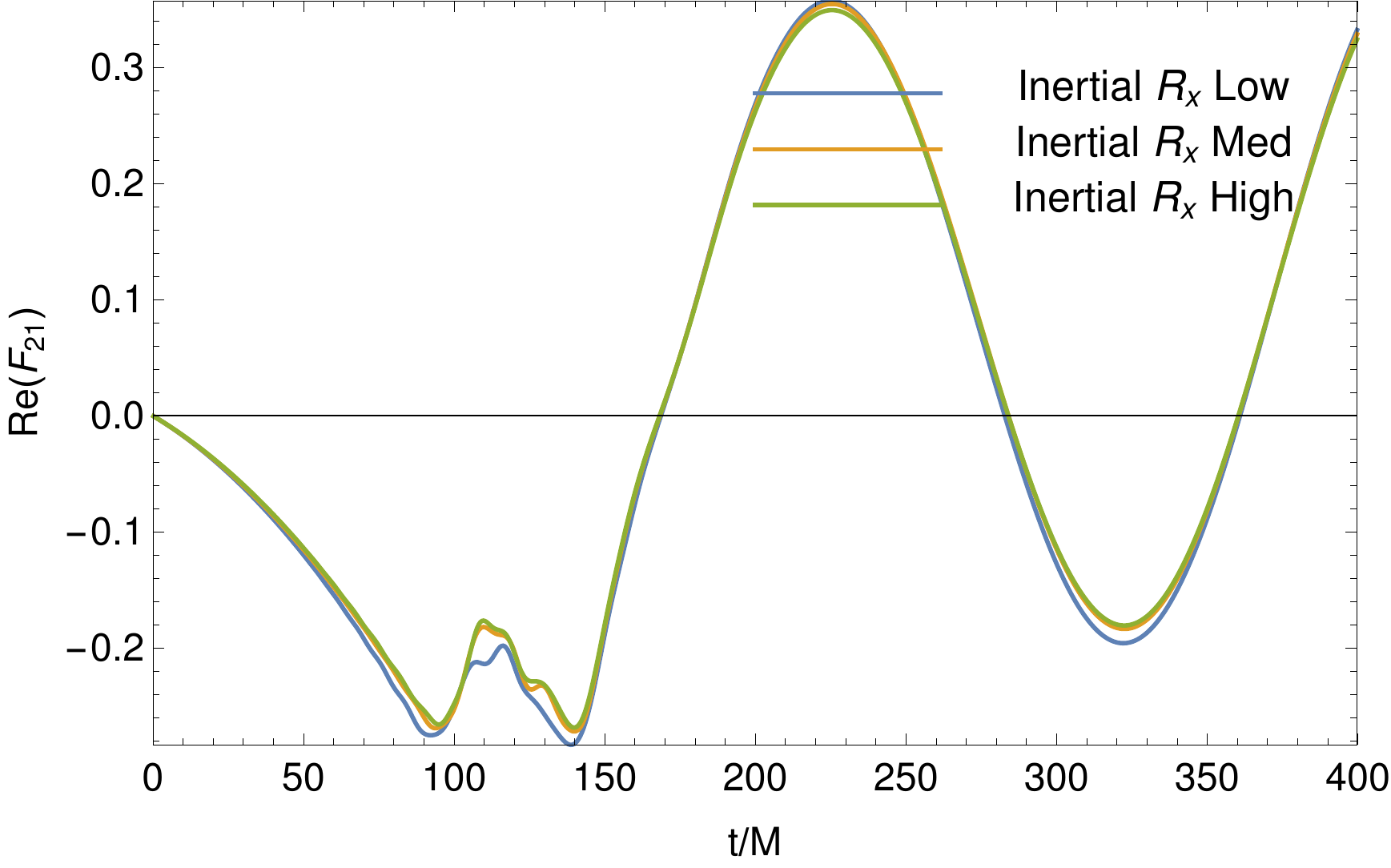}
    \label{fig:R3:Flux}
  \end{subfigure}

  \begin{subfigure}[b]{0.30\textwidth}
    \caption{}
    \includegraphics[width=\textwidth]{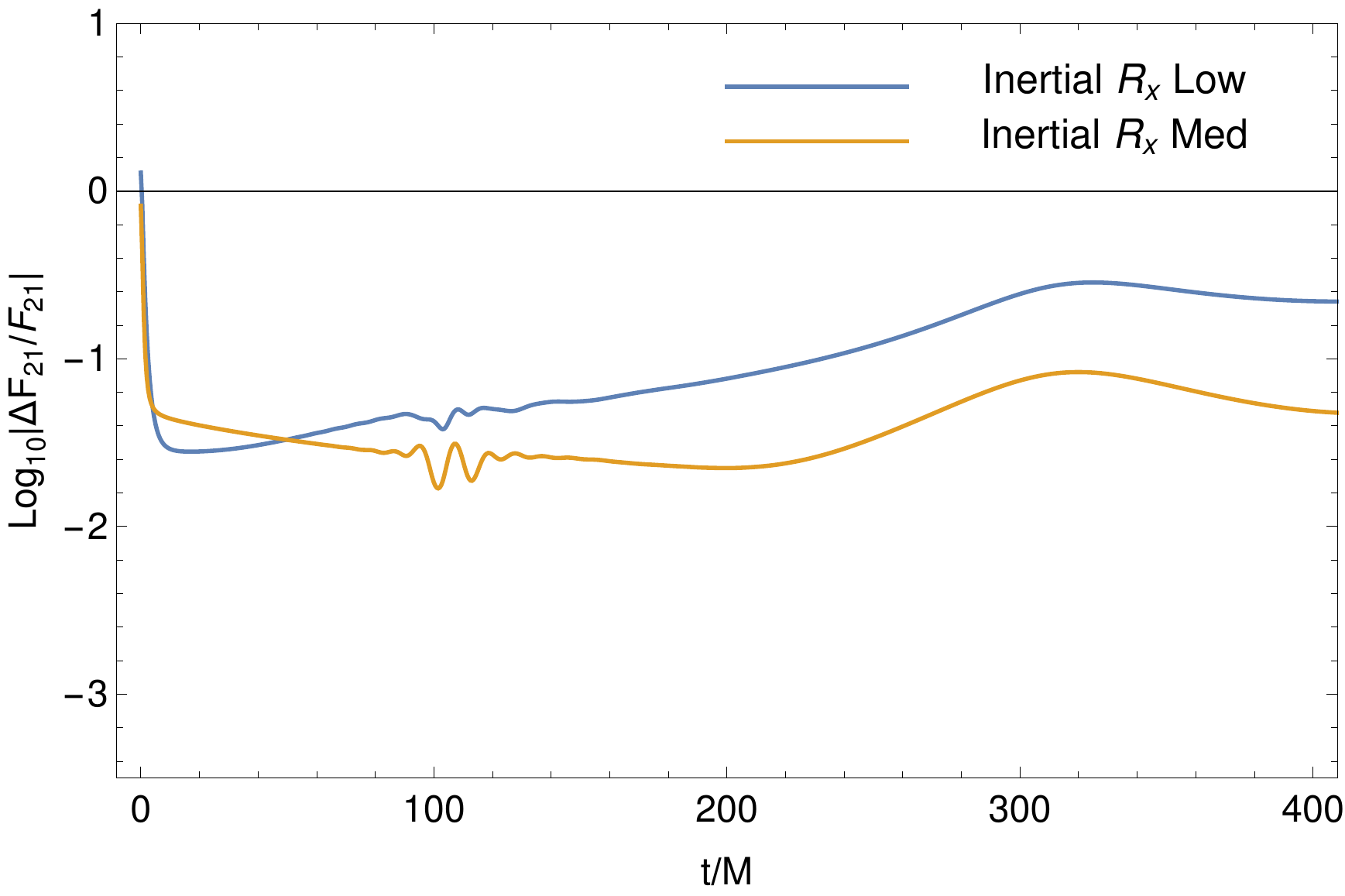}
    \label{fig:R3:FluxConv}
  \end{subfigure}
  \begin{subfigure}[b]{0.30\textwidth}
    \caption{}
    \includegraphics[width=\textwidth]{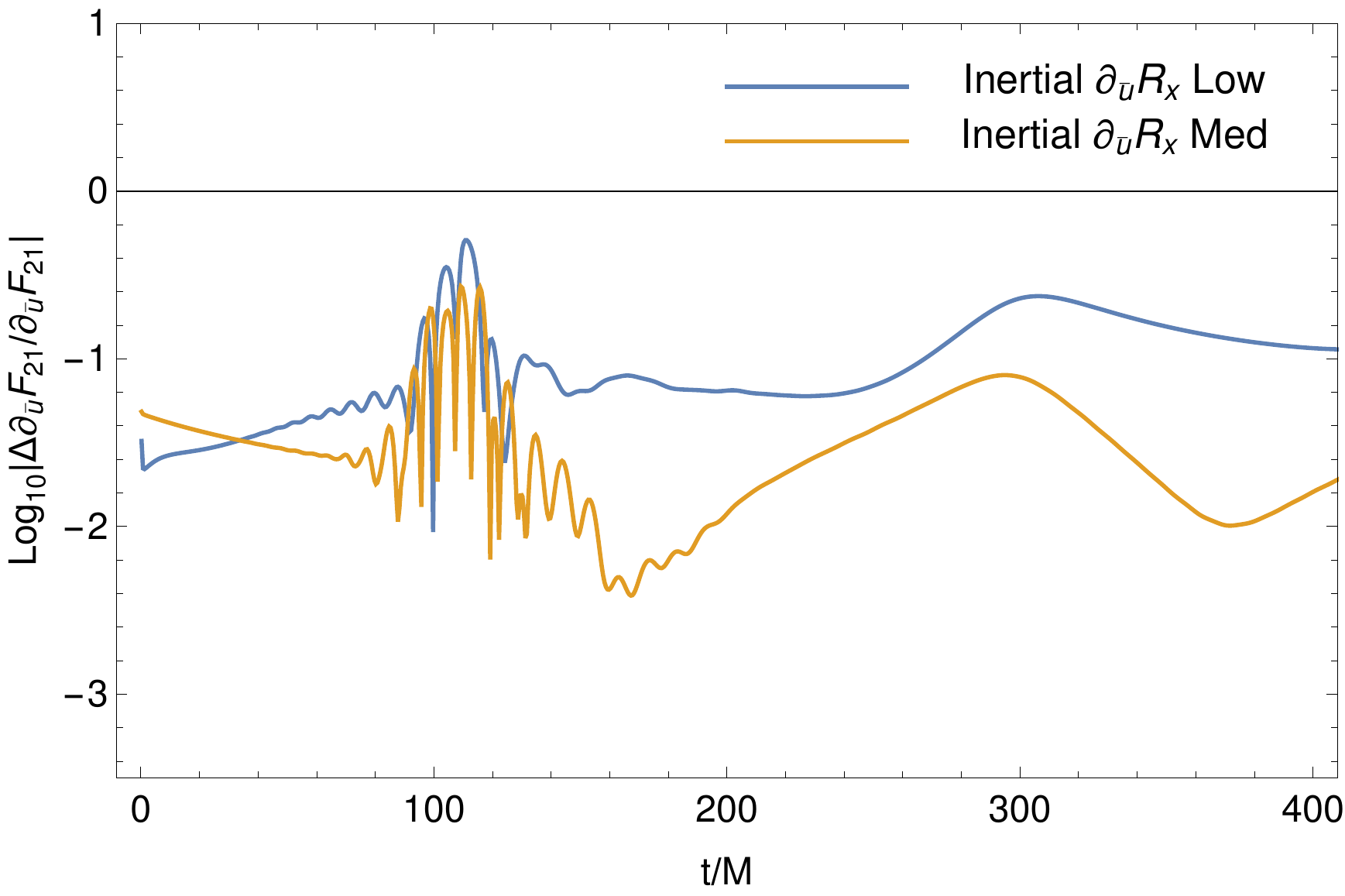}
    \label{fig:R3:FluxduConv}
  \end{subfigure}
  \caption{\small{ (\ref{fig:R3:KV}) Vectors illustrate the nature of \(\xi_{Rx}\), a rotation
      about the $x$-axis ($\phi = 0$).  (\ref{fig:R3:Flux}) The
      $(\ell=2,m=1)$ spherical harmonic component of the flux.
      (\ref{fig:R3:FluxConv}) Convergence of the flux is partially
      compromised by junk radiation, while the inertial time
      derivative of the flux (\ref{fig:R3:FluxduConv}) shows the
      appropriate $4^{th}$ order convergence following the junk phase.  }}
  \label{fig:R3}
\end{figure}

\begin{figure}
  \centering
  \begin{subfigure}[b]{0.30\textwidth}
    \caption{} \includegraphics[width=\textwidth]{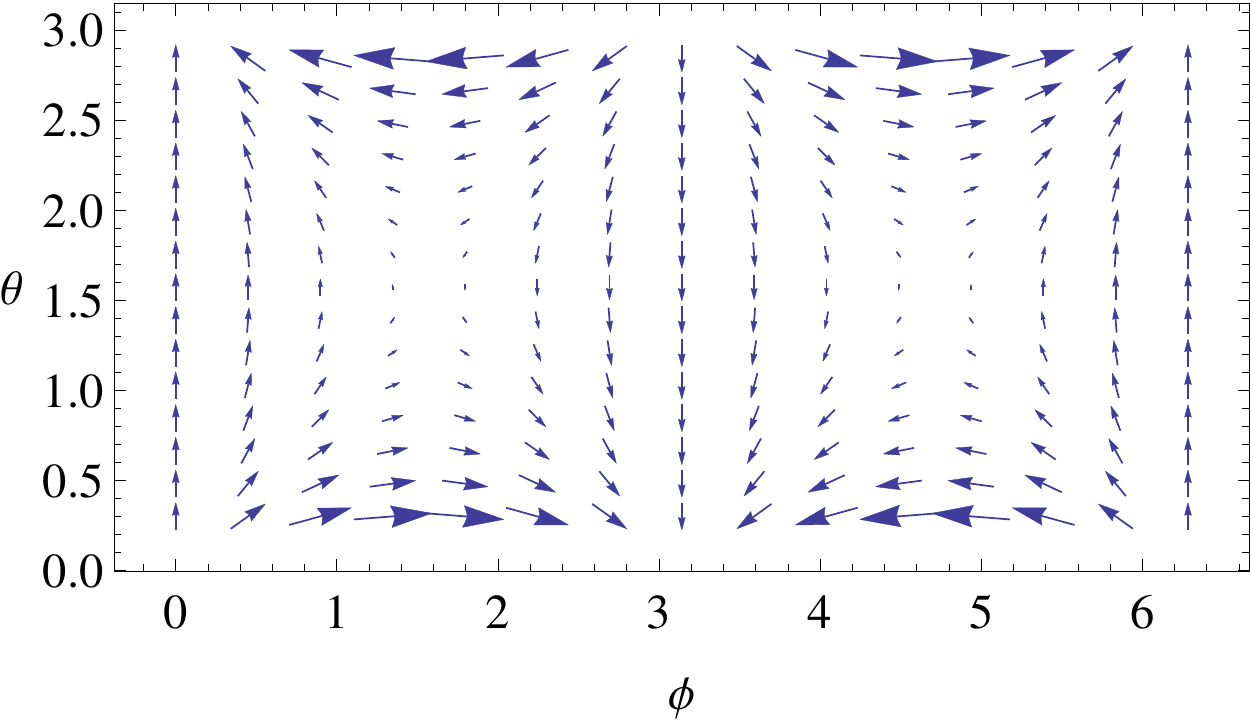}
    \label{fig:R2:KV}
  \end{subfigure}
  \begin{subfigure}[b]{0.30\textwidth}
    \caption{}
    \includegraphics[width=\textwidth]{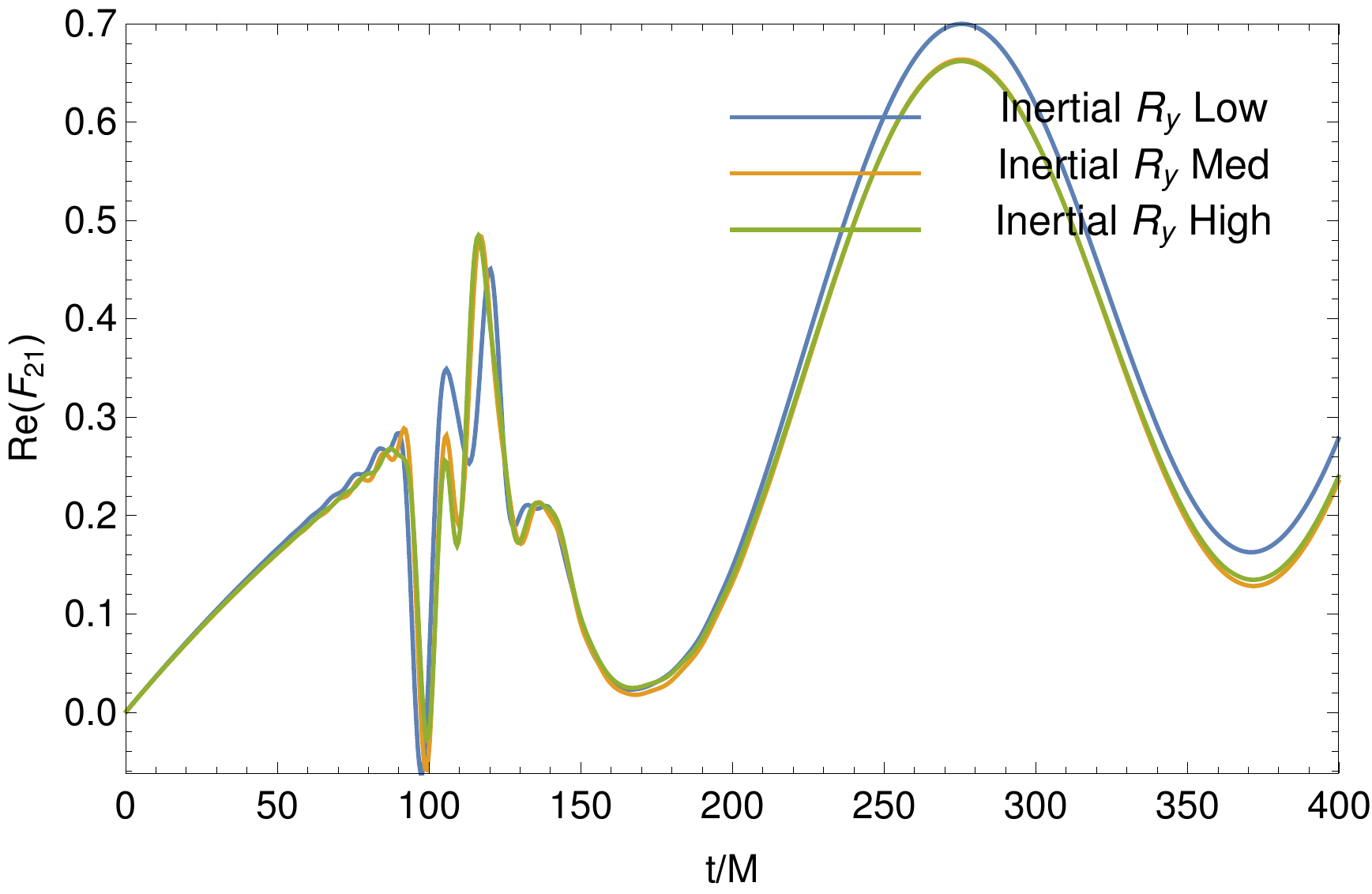}
    \label{fig:R2:Flux}
  \end{subfigure}

  \begin{subfigure}[b]{0.30\textwidth}
    \caption{}
    \includegraphics[width=\textwidth]{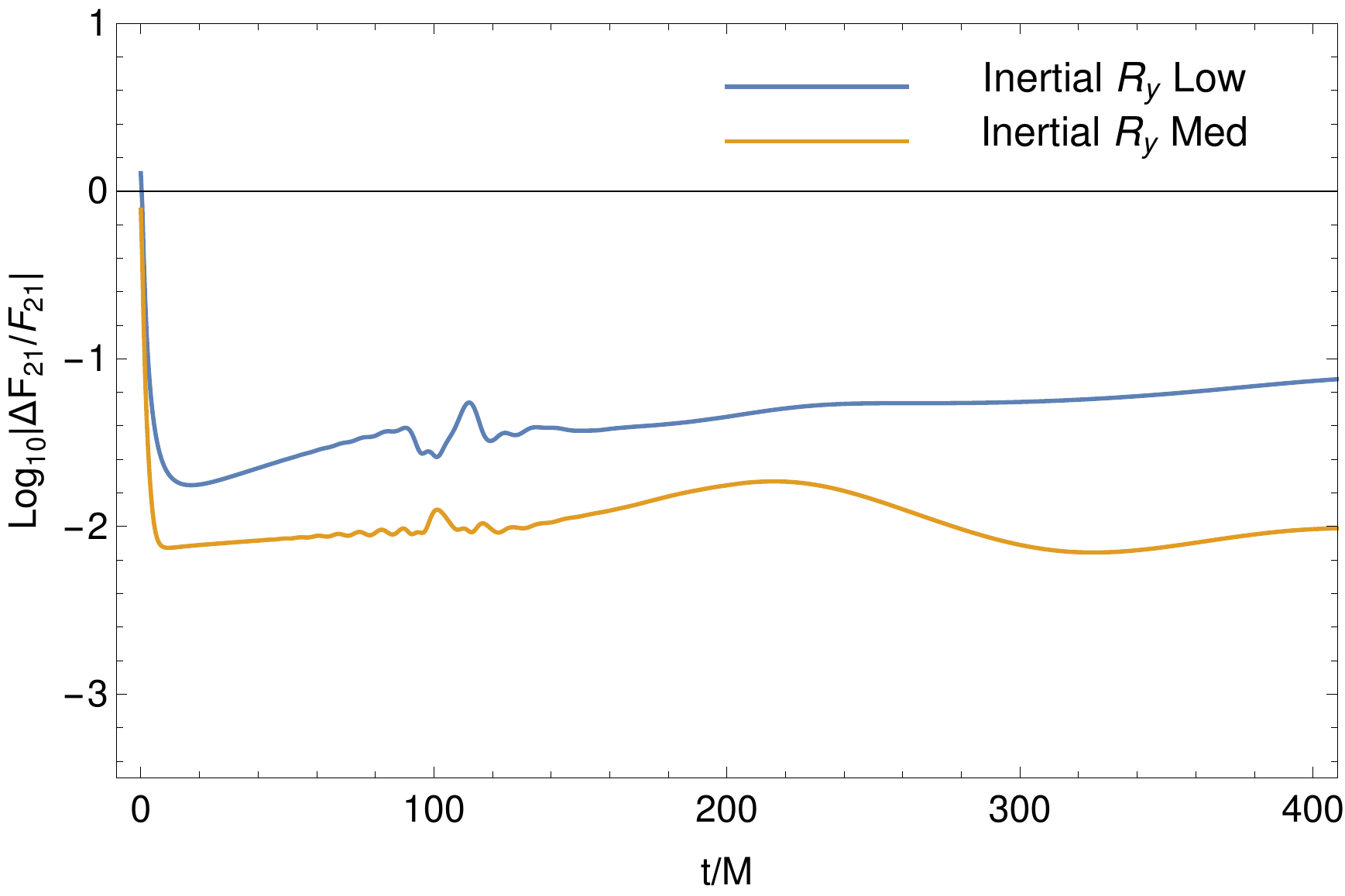}
    \label{fig:R2:FluxConv}
  \end{subfigure}
  \begin{subfigure}[b]{0.30\textwidth}
    \caption{}
    \includegraphics[width=\textwidth]{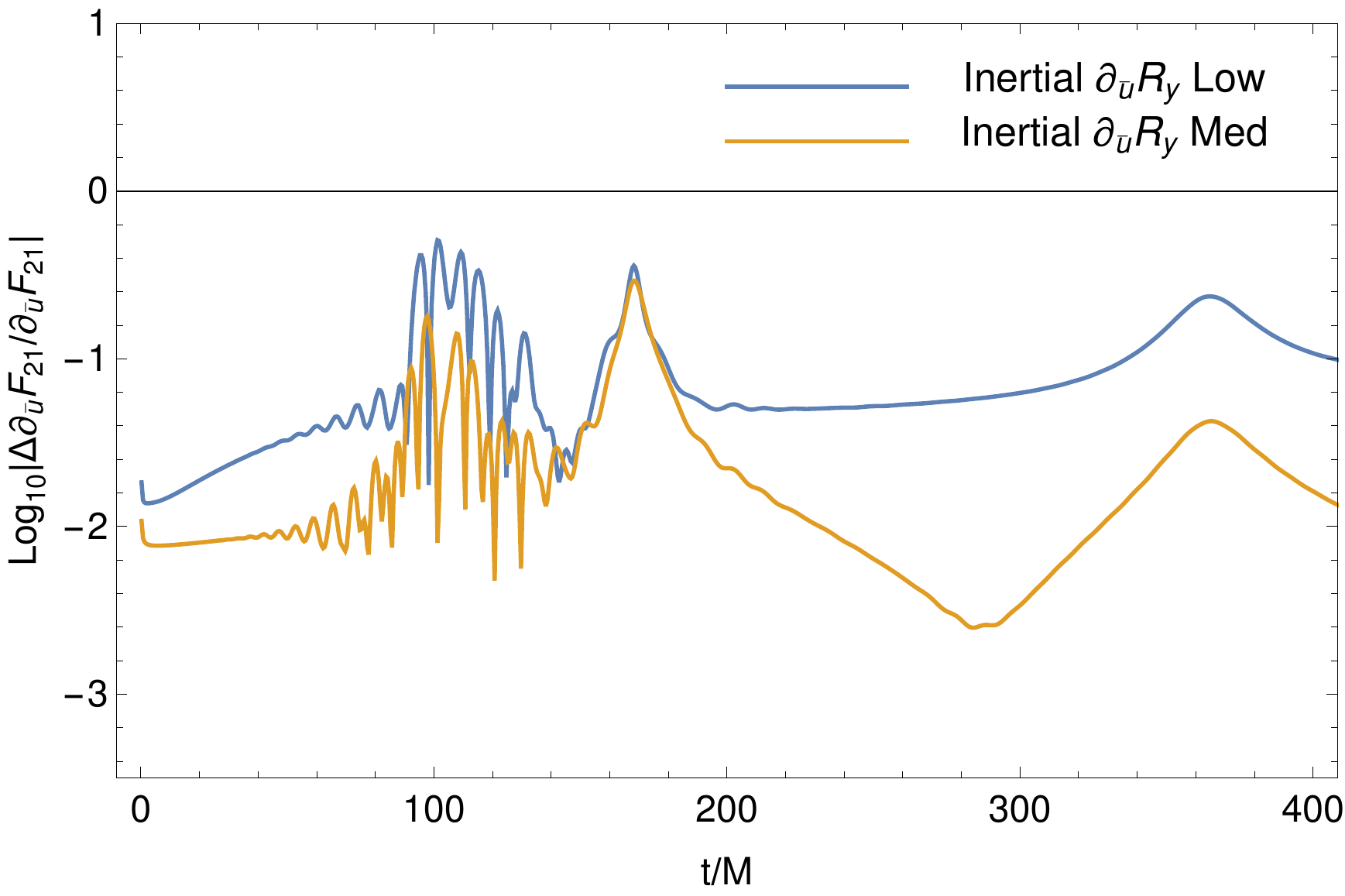}
    \label{fig:R2:FluxduConv}
  \end{subfigure}
  \caption{\small{ (\ref{fig:R2:KV}) Vectors illustrate the nature of \(\xi_{Ry}\),
       a rotation about $y$ ($\phi = \pi/2$).  (\ref{fig:R2:Flux}) The
      $(\ell=2,m=1)$ spherical harmonic component of the flux.
      (\ref{fig:R2:FluxConv}) Convergence of the flux is partially
      compromised by junk radiation, while the inertial time
      derivative of the flux (\ref{fig:R2:FluxduConv}) shows the
      appropriate $4^{th}$ order convergence following the junk phase.  }}
  \label{fig:R2}
\end{figure}

The $(\ell=2,m=1)$ modes of the $x$ and $y$-components of the angular momentum flux,
\(F_{Rx}\) and \(F_{Ry}\), are shown in Fig.~\ref{fig:R3:Flux} and Fig.~\ref{fig:R2:Flux},
respectively. 
Convergence of these fluxes is shown in
Fig.~\ref{fig:R3:FluxConv} and  Fig.~\ref{fig:R2:FluxConv};
and convergence of their inertial time derivative in
Fig.~\ref{fig:R3:FluxduConv} and Fig.~\ref{fig:R2:FluxduConv}.

\subsubsection{Boosts}

In addition to the rotations, the other transformations of the Lorentz
group are the three boosts with BMS generators $\xi^\alpha_{[Bx]}$,
$\xi^\alpha_{[By]}$ and $\xi^\alpha_{[Bz]}$ for which \(f^{\tilde
  A}_{:\tilde A} \neq 0\).  For these boost generators, $\alpha=0$,
$f^{\tilde A}= \Gamma^{:\tilde A}$, where $\Gamma$ consists of
$\ell=1$ spherical harmonics, so that $\Gamma^{:\tilde A}{}_{: \tilde
  A}=-2\Gamma$ and $\xi^{\tilde u} = -u \Gamma$.  As a result, the
boosts acquire a \(\xi^{\tilde u}\) component with linear dependence
on \(\tilde u\), as well as $\ell=1$ angular dependence.
The corresponding physical quantities describe the dipole moment
of the system corresponding to the center-of-mass integrals
in Lorentz covariant theories.

\begin{figure}
  \centering
  \begin{subfigure}[b]{0.30\textwidth}
    \caption{} \includegraphics[width=\textwidth]{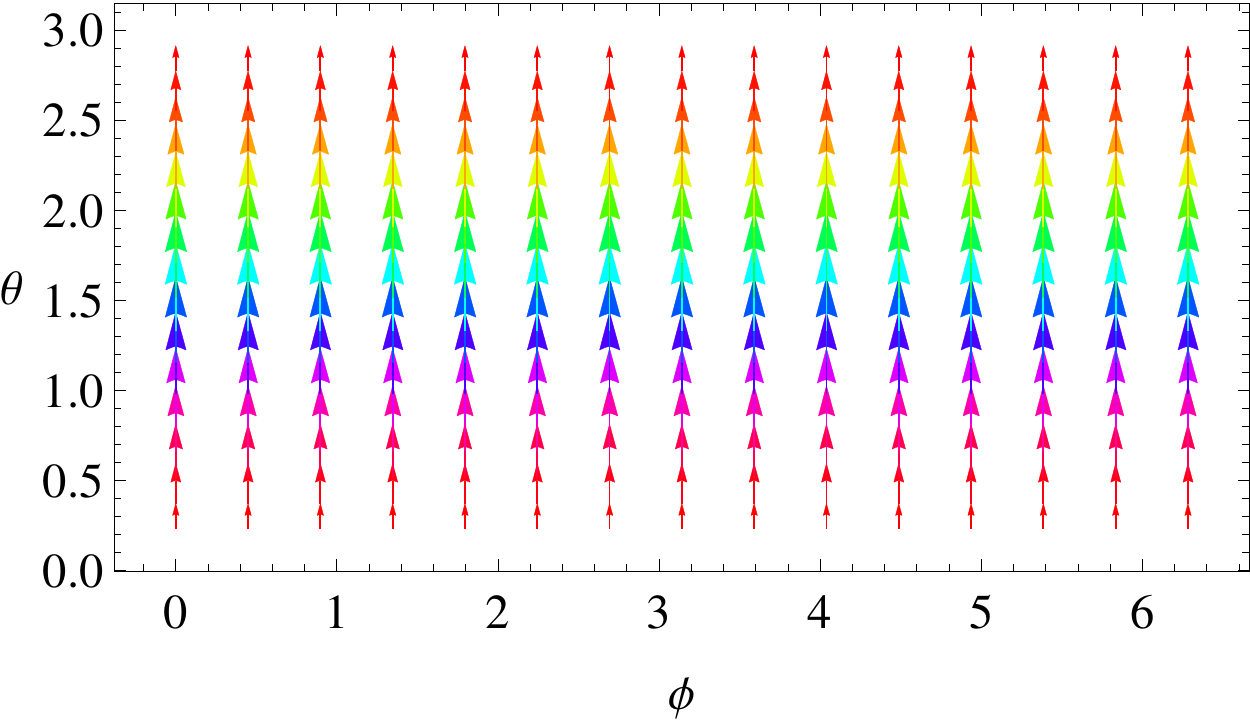}
    \label{fig:B1:KV}
  \end{subfigure}
  \begin{subfigure}[b]{0.30\textwidth}
    \caption{}
    \includegraphics[width=\textwidth]{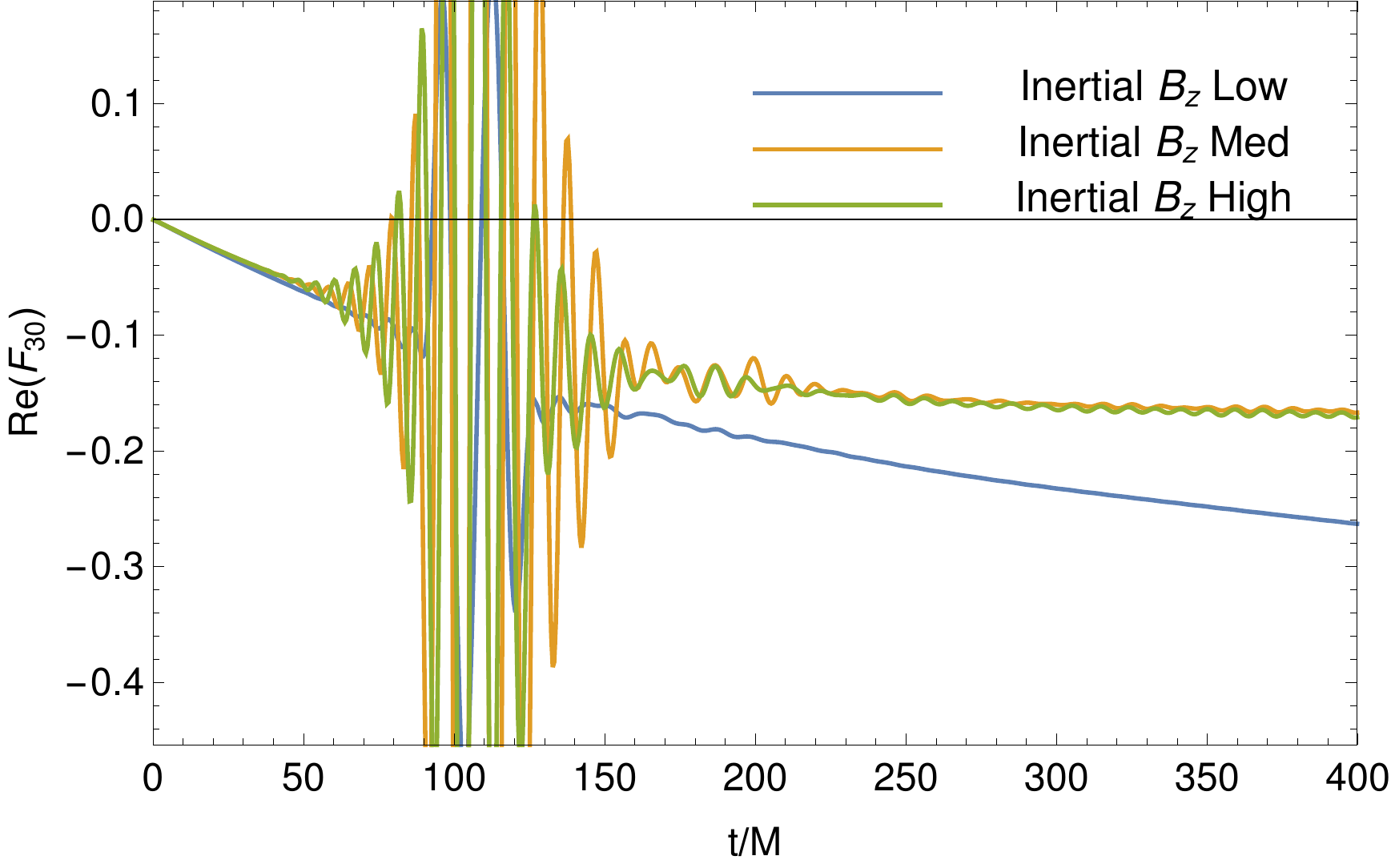}
    \label{fig:B1:Flux}
  \end{subfigure}

  \begin{subfigure}[b]{0.30\textwidth}
    \caption{}
    \includegraphics[width=\textwidth]{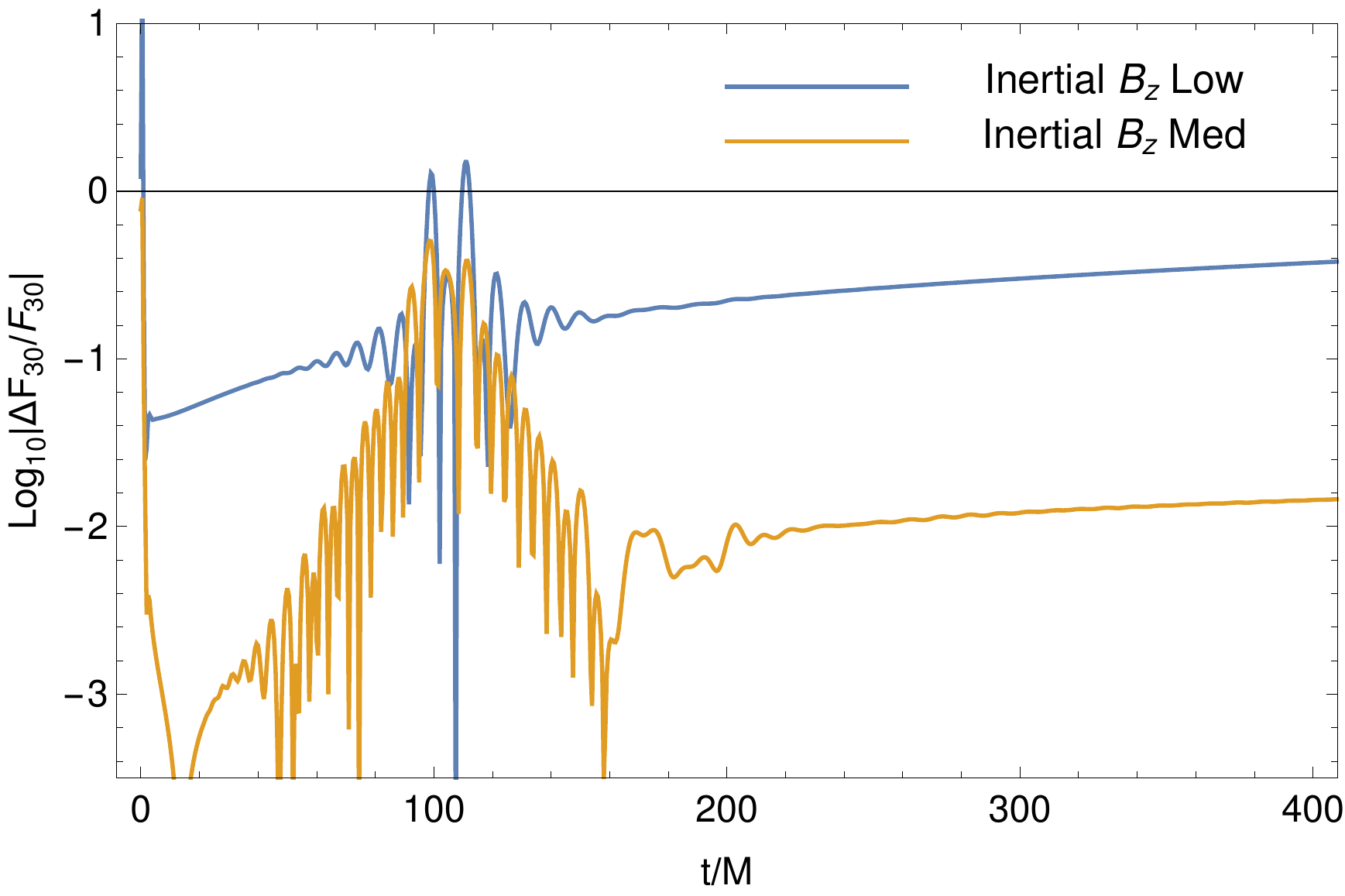}
    \label{fig:B1:FluxConv}
  \end{subfigure}
  \begin{subfigure}[b]{0.30\textwidth}
    \caption{}
    \includegraphics[width=\textwidth]{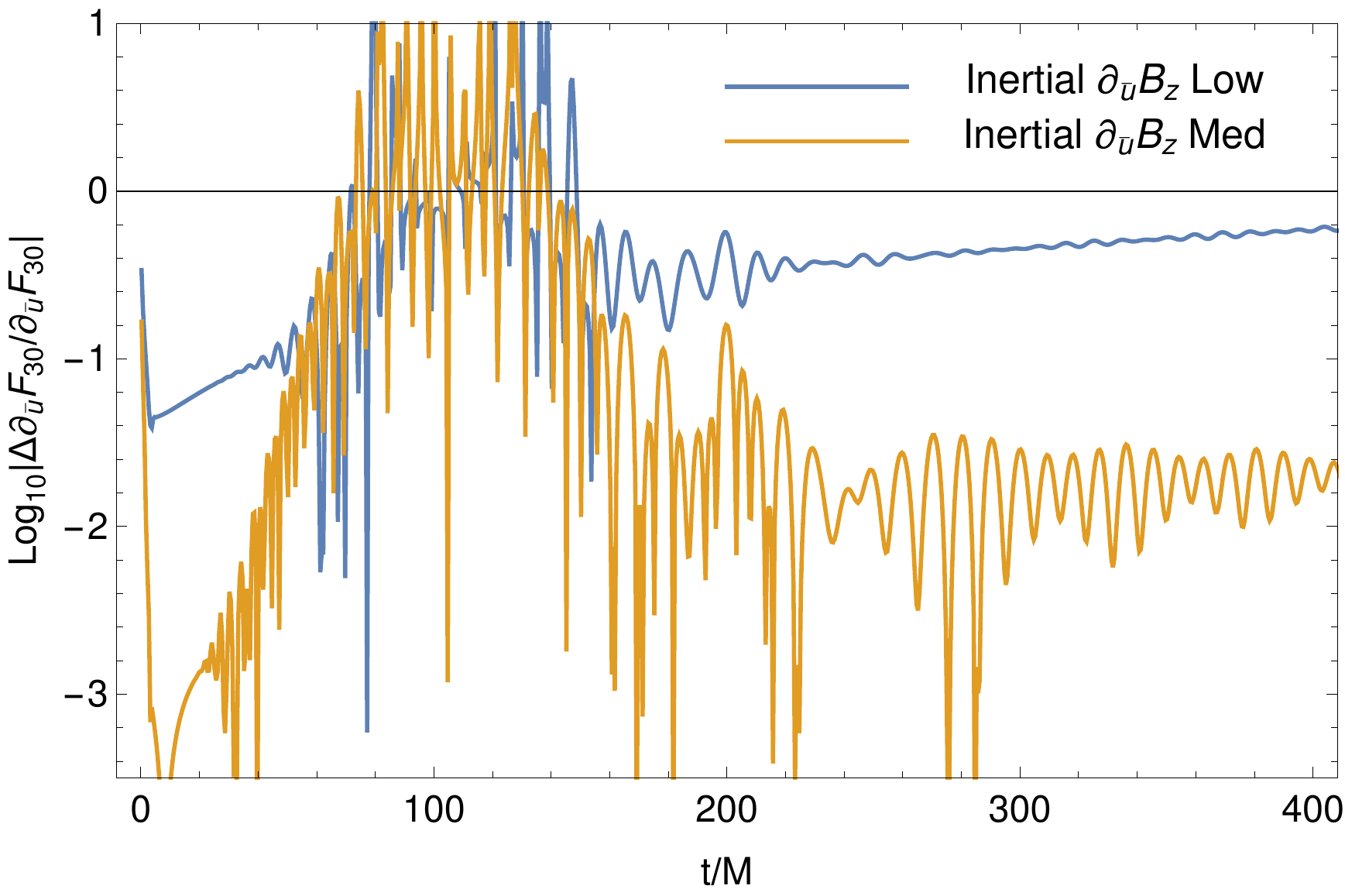}
    \label{fig:B1:FluxduConv}
  \end{subfigure}
  \caption{\small{ (\ref{fig:B1:KV}) Vectors illustrate the nature of \(\xi_{Bz}^{\tilde A}\),
  while colors illustrate the gradient of \(\xi_{Bz}^{\tilde u}\), illustrating a Lorentz boost
  $Bz$ in the $z$-direction.  (\ref{fig:B1:Flux}) The
      $(\ell=3,m=0)$ spherical harmonic component of the flux.
      (\ref{fig:B1:FluxConv}) Convergence of the flux is partially
      compromised by junk radiation, while the inertial time
      derivative of the flux (\ref{fig:B1:FluxduConv}) shows the
      appropriate $4^{th}$ order convergence following the junk phase. }}
  \label{fig:B1}
\end{figure}

For a boost $B_z$ in the inertial \(z\)-direction, $\Gamma=\cos \tilde
\theta$ so that $f^{\tilde A}=(-\sin \tilde \theta,0)$ and
$\xi_{[Bz]}^{\tilde u} =- \tilde u \cos \tilde \theta$.
The $(\ell=3,m=0)$ mode of the $z$-component of the boost flux \(F_{Bz}\) is shown in
Fig.~\ref{fig:B1:Flux}.  Convergence of the flux is shown in
Fig.~\ref{fig:B1:FluxConv} and its inertial time derivative in
Fig.~\ref{fig:B1:FluxduConv}.

\begin{figure}
  \centering
  \begin{subfigure}[b]{0.30\textwidth}
    \caption{} \includegraphics[width=\textwidth]{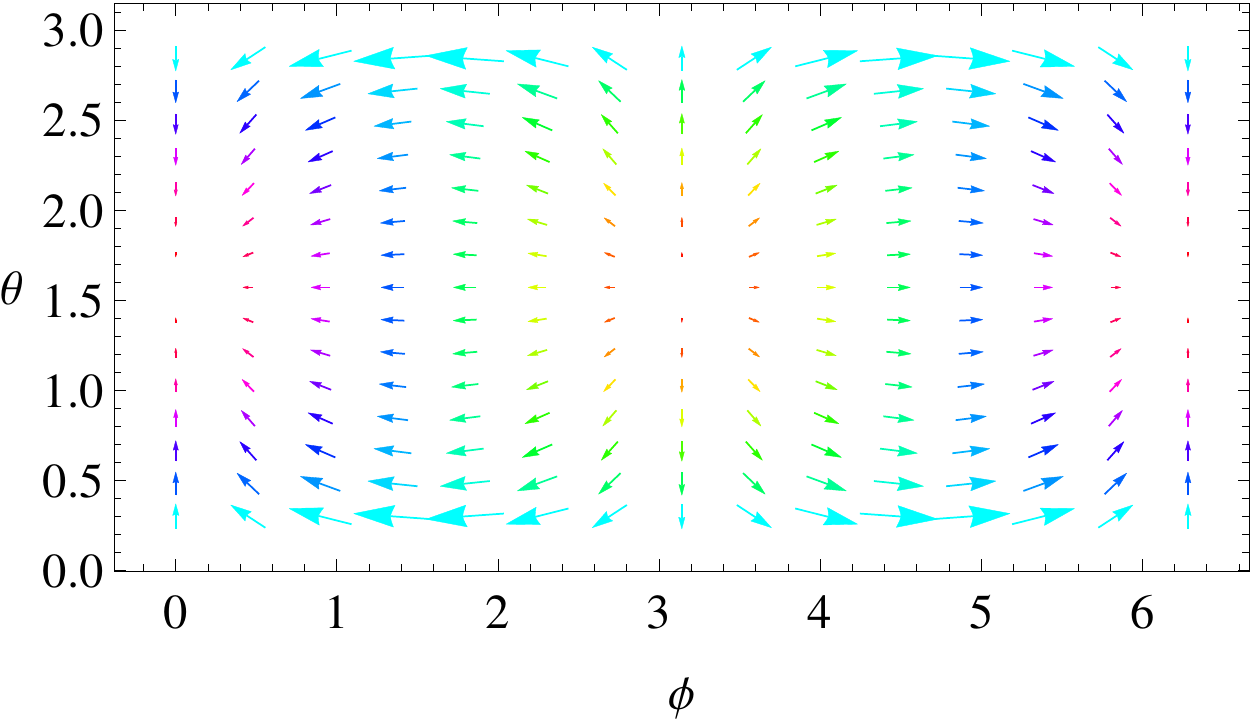}
    \label{fig:B2:KV}
  \end{subfigure}
  \begin{subfigure}[b]{0.30\textwidth}
    \caption{}
    \includegraphics[width=\textwidth]{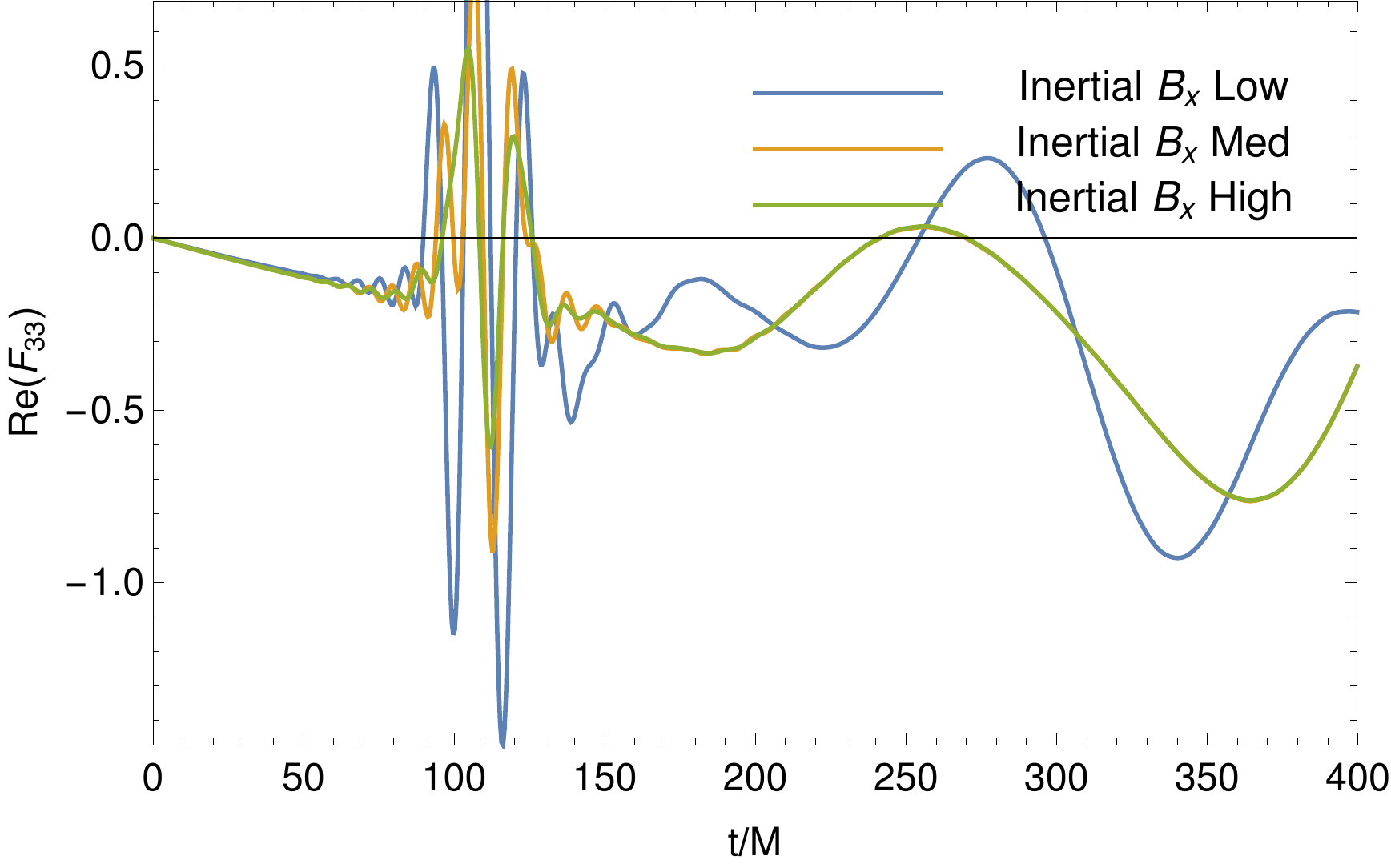}
    \label{fig:B2:Flux}
  \end{subfigure}

  \begin{subfigure}[b]{0.30\textwidth}
    \caption{}
    \includegraphics[width=\textwidth]{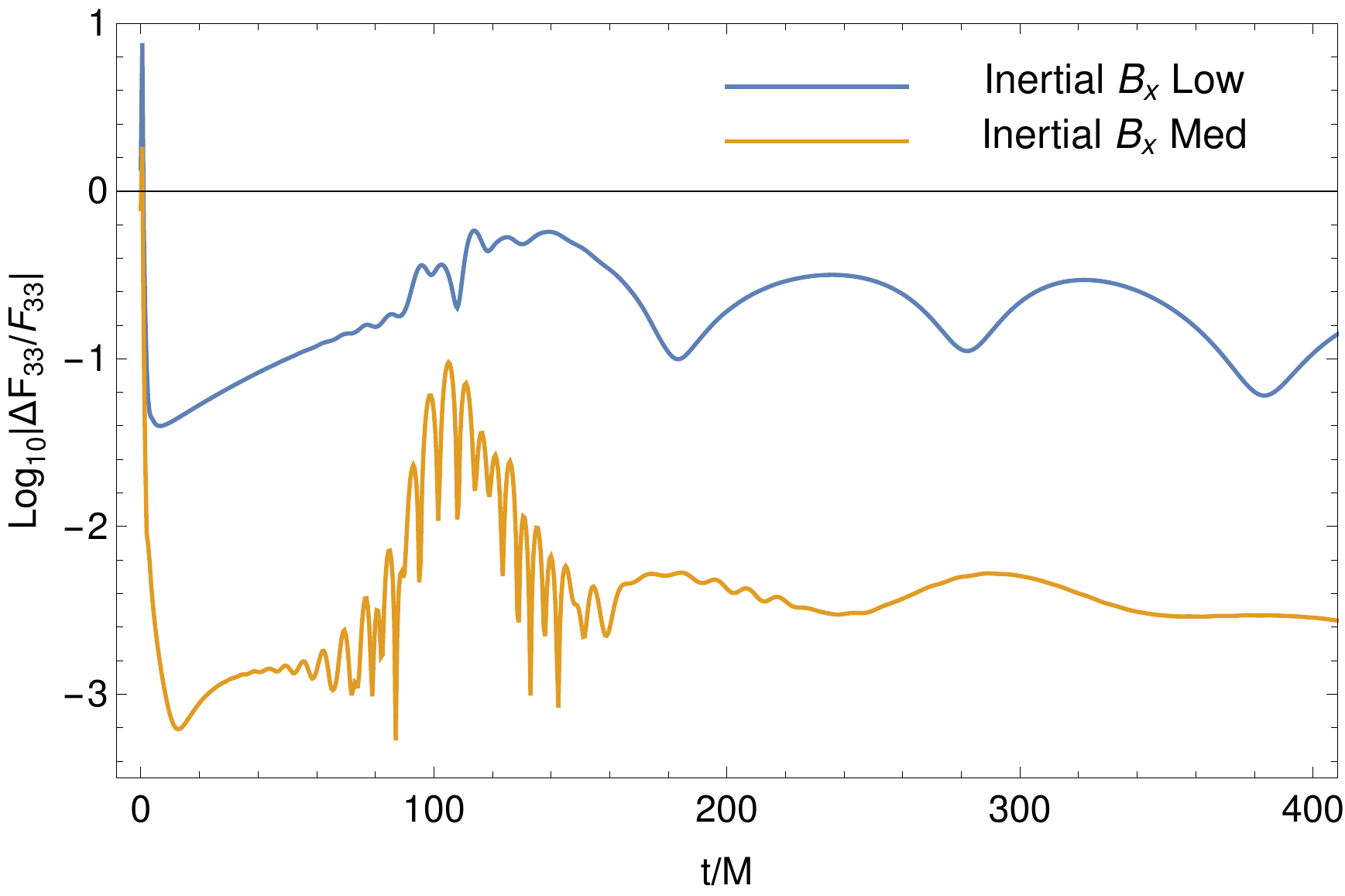}
    \label{fig:B2:FluxConv}
  \end{subfigure}
  \begin{subfigure}[b]{0.30\textwidth}
    \caption{}
    \includegraphics[width=\textwidth]{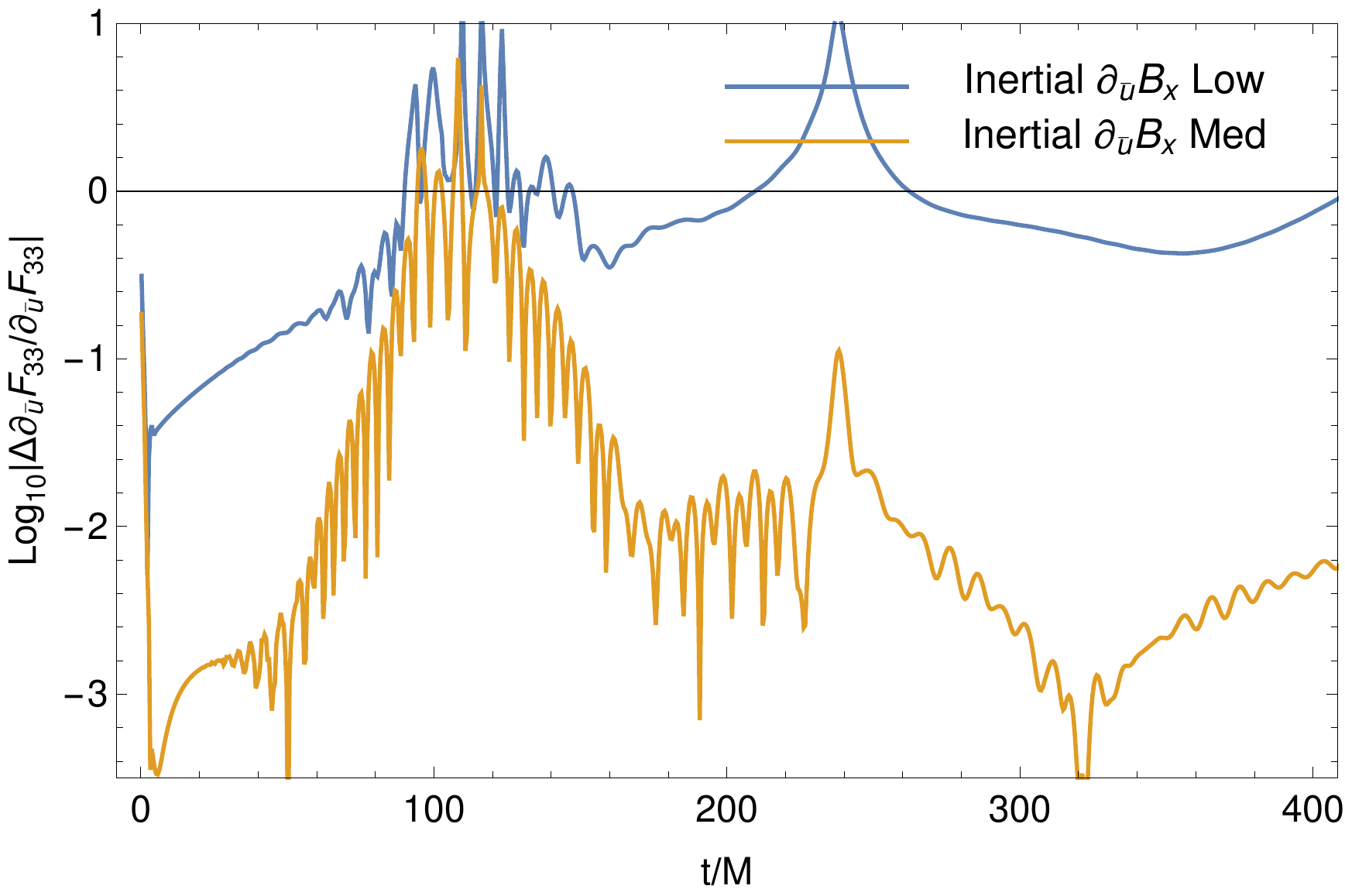}
    \label{fig:B2:FluxduConv}
  \end{subfigure}
  \caption{\small{ (\ref{fig:B2:KV}) Vectors illustrate the nature of \(\xi_{Bx}^{\tilde A}\),
  while colors illustrate the gradient of \(\xi_{Bx}^{\tilde u}\), illustrating a boost
  in the $x$ ($\phi=0$) direction.  (\ref{fig:B2:Flux}) The
      $(\ell=3,m=3)$ spherical harmonic component of the flux.
      (\ref{fig:B2:FluxConv}) Convergence of the flux is partially
      compromised by junk radiation, while the inertial time
      derivative of the flux (\ref{fig:B2:FluxduConv}) shows the
      appropriate $4^{th}$ order convergence following the junk phase. }}
  \label{fig:B2}
\end{figure}

\begin{figure}
  \centering
  \begin{subfigure}[b]{0.30\textwidth}
    \caption{} \includegraphics[width=\textwidth]{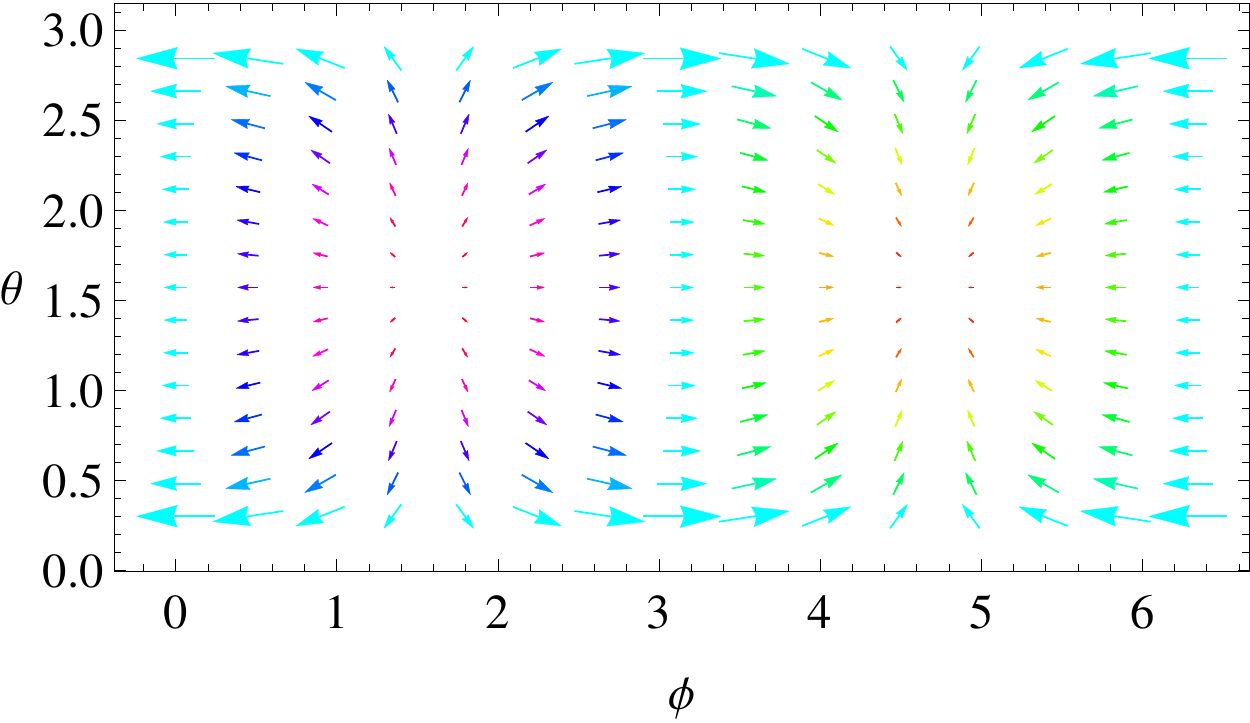}
    \label{fig:B3:KV}
  \end{subfigure}
  \begin{subfigure}[b]{0.30\textwidth}
    \caption{}
    \includegraphics[width=\textwidth]{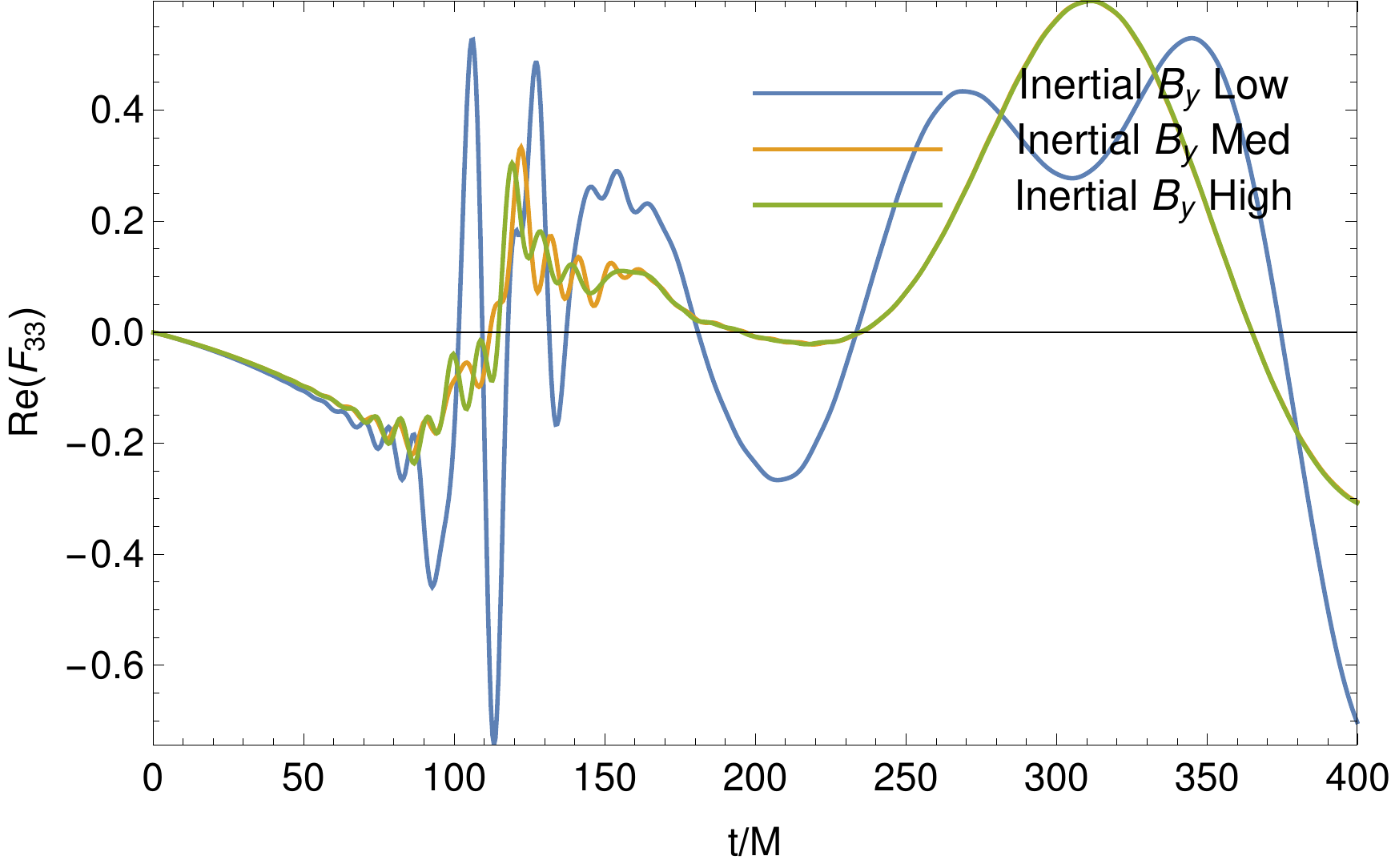}
    \label{fig:B3:Flux}
  \end{subfigure}

  \begin{subfigure}[b]{0.30\textwidth}
    \caption{}
    \includegraphics[width=\textwidth]{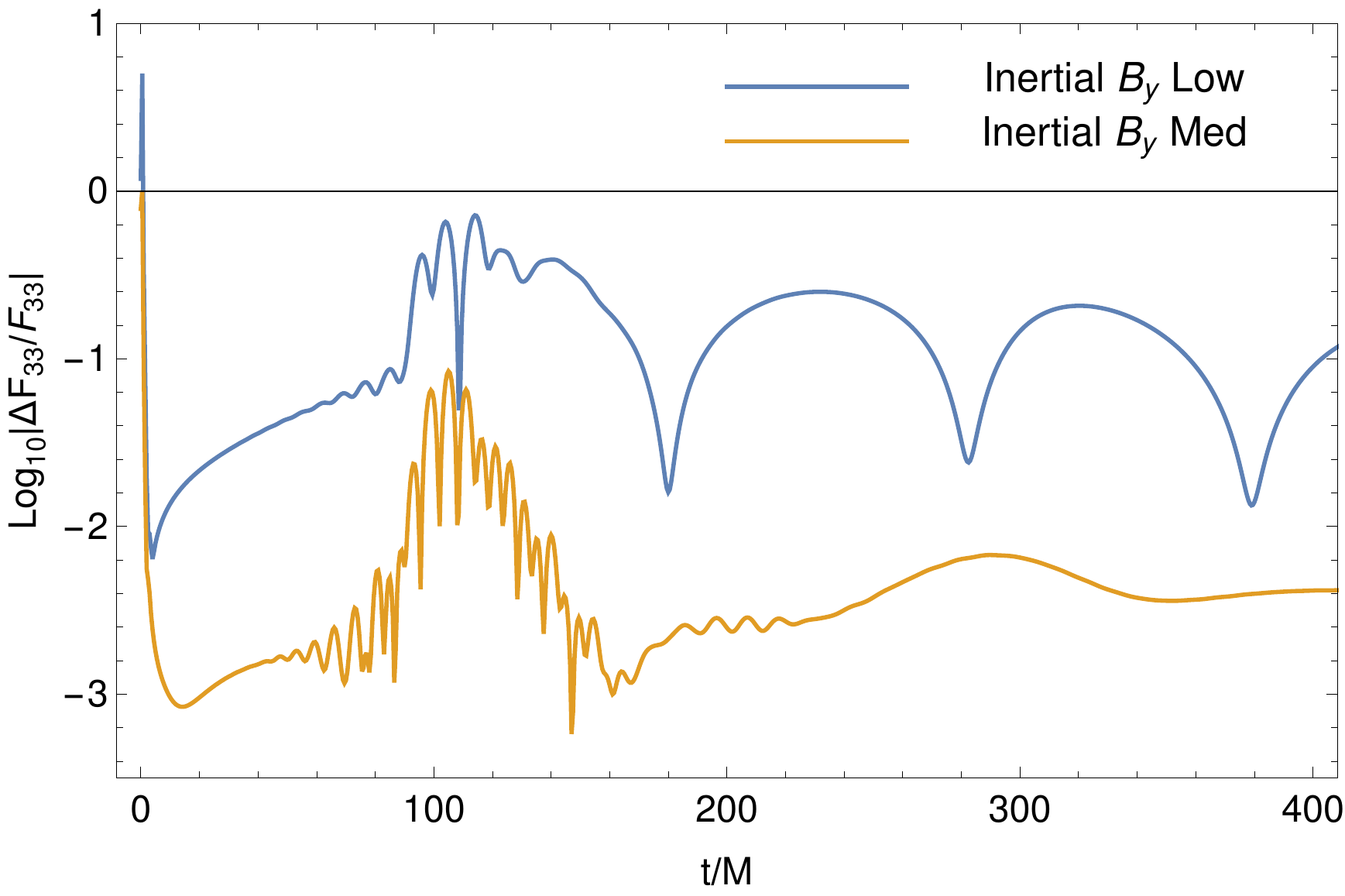}
    \label{fig:B3:FluxConv}
  \end{subfigure}
  \begin{subfigure}[b]{0.30\textwidth}
    \caption{}
    \includegraphics[width=\textwidth]{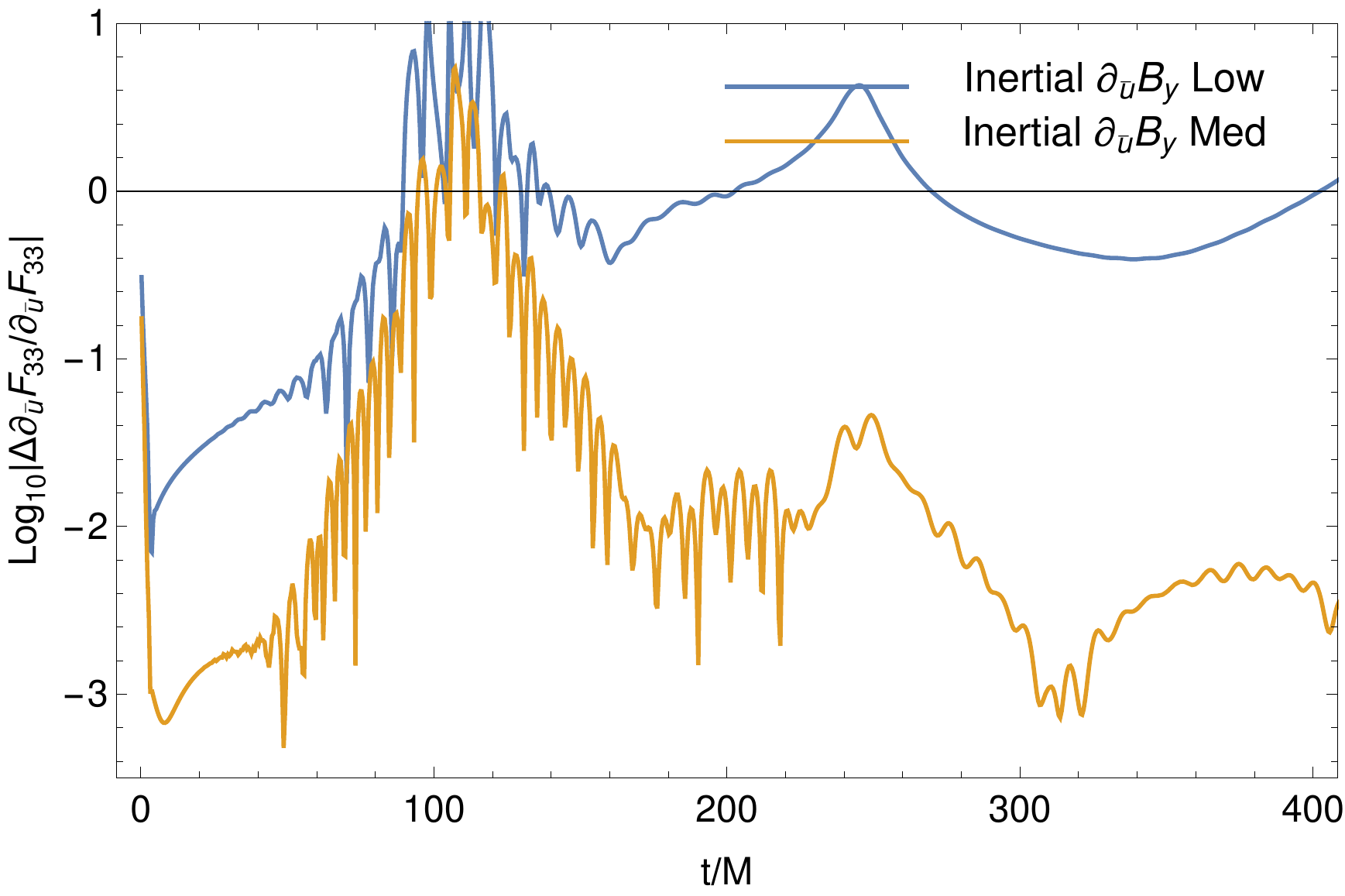}
    \label{fig:B3:FluxduConv}
  \end{subfigure}
  \caption{\small{ (\ref{fig:B3:KV}) Vectors illustrate the nature of \(\xi_{By}^{\tilde A}\),
  while colors illustrate the gradient of \(\xi_{By}^{\tilde u}\), illustrating a boost in
  the $y$ ($\phi=\pi/2$) direction. (\ref{fig:B3:Flux}) The
      $(\ell=3,m=3)$ spherical harmonic component of the flux.
      (\ref{fig:B3:FluxConv}) Convergence of the flux is partially
      compromised by junk radiation, while the inertial time
      derivative of the flux (\ref{fig:B3:FluxduConv}) shows the
      appropriate $4^{th}$ order convergence following the junk phase.  }}
  \label{fig:B3}
\end{figure}

For a boost $B_x$ in the inertial \(x\)-direction,
$\Gamma=\sin \tilde \theta \cos \tilde \phi$ so that
$f^{\tilde A}=(\cos \tilde \theta\cos \tilde \phi,- \csc \tilde \theta \sin \tilde \phi)$.
Similarly, for a boost $B_y$ in the inertial \(y\)-direction,
$\Gamma=\sin \tilde \theta \sin \tilde \phi$ so that
$f^{\tilde A}=(\cos \tilde \theta\sin \tilde \phi,\csc \tilde \theta \cos \tilde \phi)$. 
The $(\ell=3,m=3)$ mode of the $x$ and $y$-components of the boost flux,
 \(F_{Bx}\) and \(F_{By}\), is shown in
Fig.~\ref{fig:B2:Flux} and Fig.~\ref{fig:B3:Flux}, respectively.  Convergence of these fluxes
is shown in Fig.~\ref{fig:B2:FluxConv} and Fig.~\ref{fig:B3:FluxConv}; and
convergence of their inertial time derivative in
Fig.~\ref{fig:B2:FluxduConv} and Fig.~\ref{fig:B3:FluxduConv}.

\subsubsection{Supertranslations}

The BMS supertranslation generators $\xi_{[ST]}^{\tilde \alpha}$ are
described by $f^{\tilde A}=0$, with $\alpha (\tilde x^B)$ constructed
from spherical harmonics with \(l>1\).  This leads to an infinite set
of transformations, which extend well beyond the limit of code
resolution. Here we concentrate on the supertranslation corresponding
to a \(|Y_{22}|\) spherical harmonic, $\alpha_{ST}= \sin^2 \tilde \theta
\cos 2 \tilde \phi$. 

\begin{figure}
  \centering
  \begin{subfigure}[b]{0.30\textwidth}
    \caption{} \includegraphics[width=\textwidth]{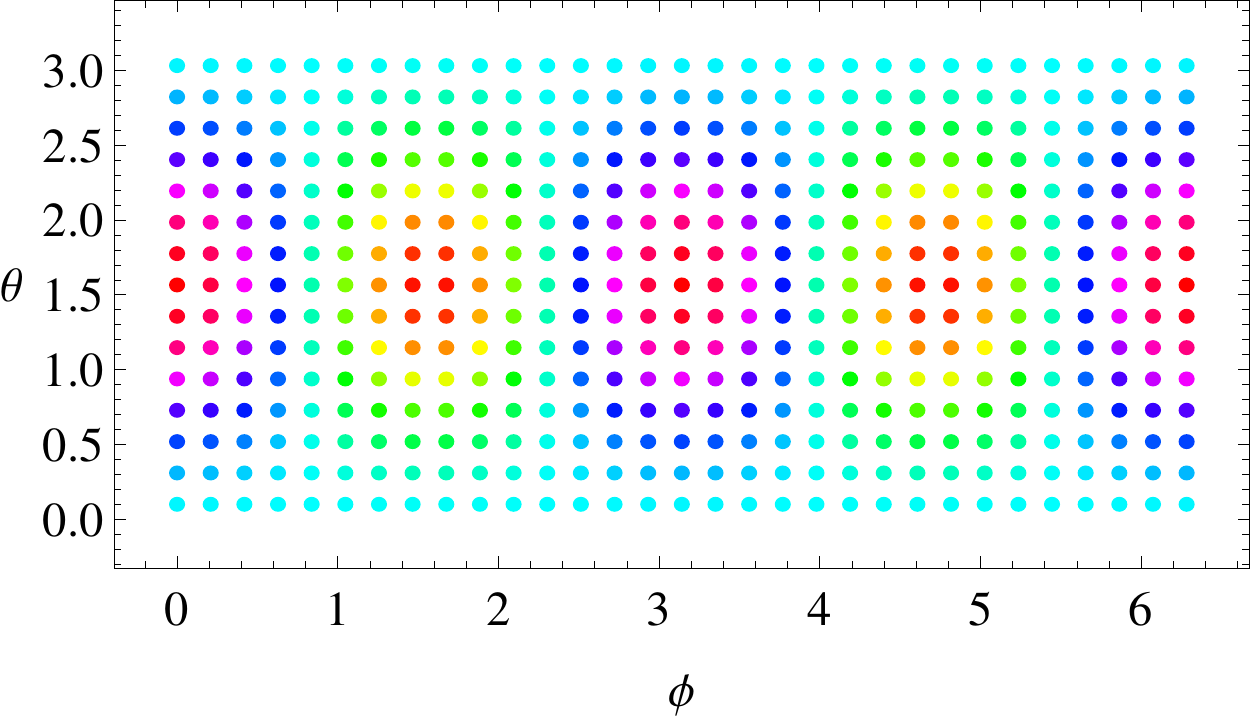}
    \label{fig:STS:KV}
  \end{subfigure}
  \begin{subfigure}[b]{0.30\textwidth}
    \caption{}
    \includegraphics[width=\textwidth]{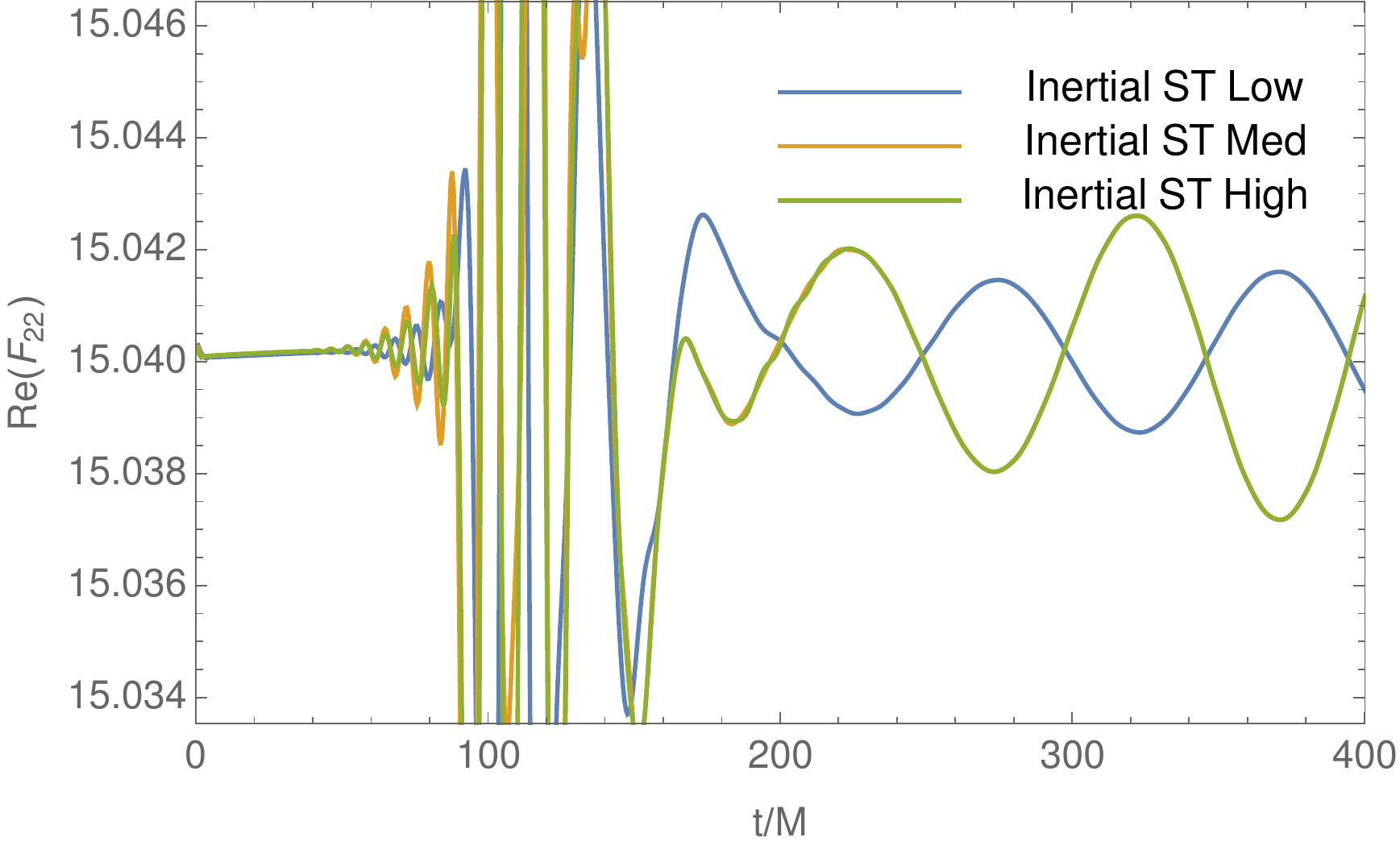}
    \label{fig:STS:Flux}
  \end{subfigure}

  \begin{subfigure}[b]{0.30\textwidth}
    \caption{}
    \includegraphics[width=\textwidth]{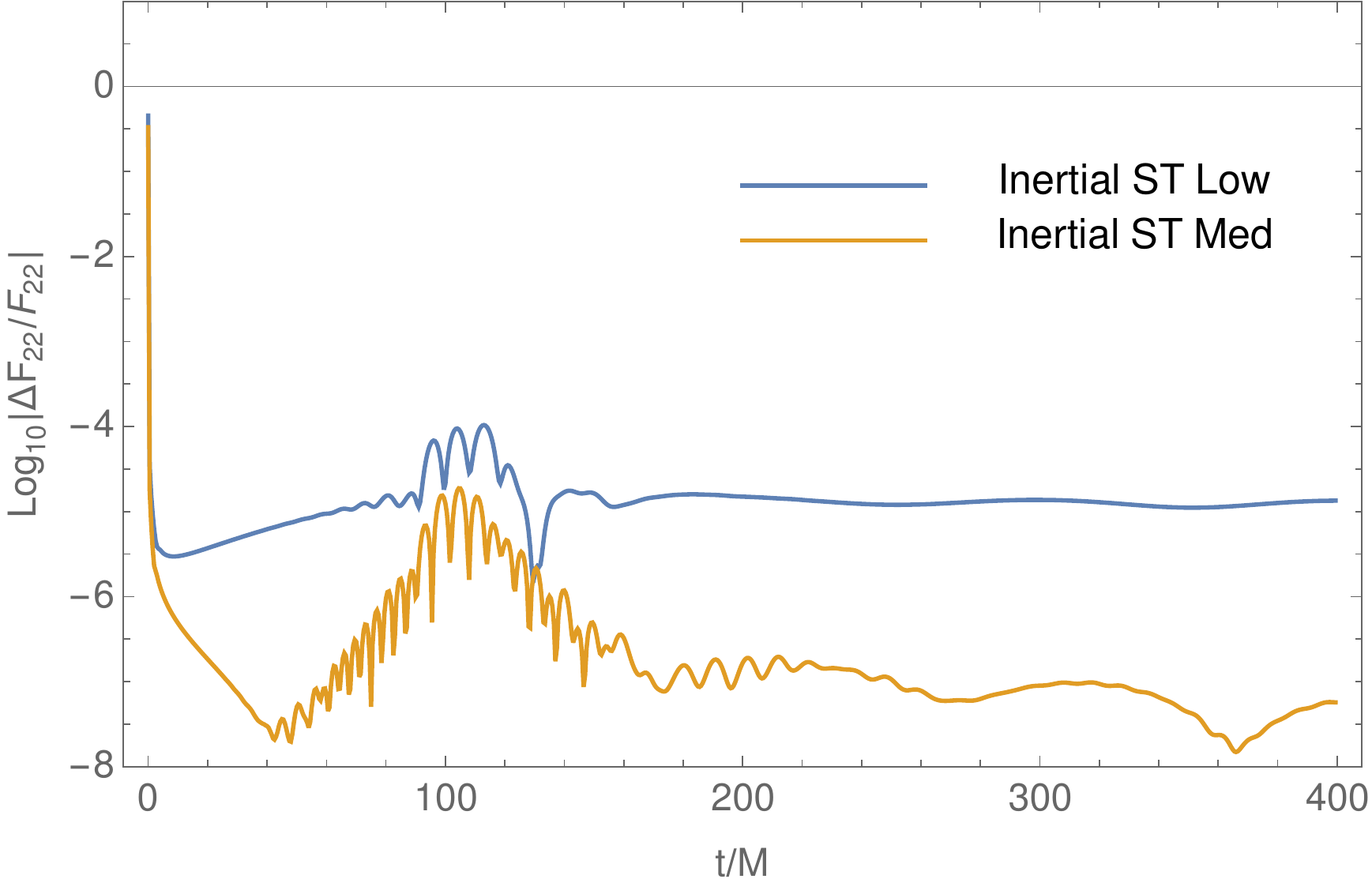}
    \label{fig:STS:FluxConv}
  \end{subfigure}
  \begin{subfigure}[b]{0.30\textwidth}
    \caption{}
    \includegraphics[width=\textwidth]{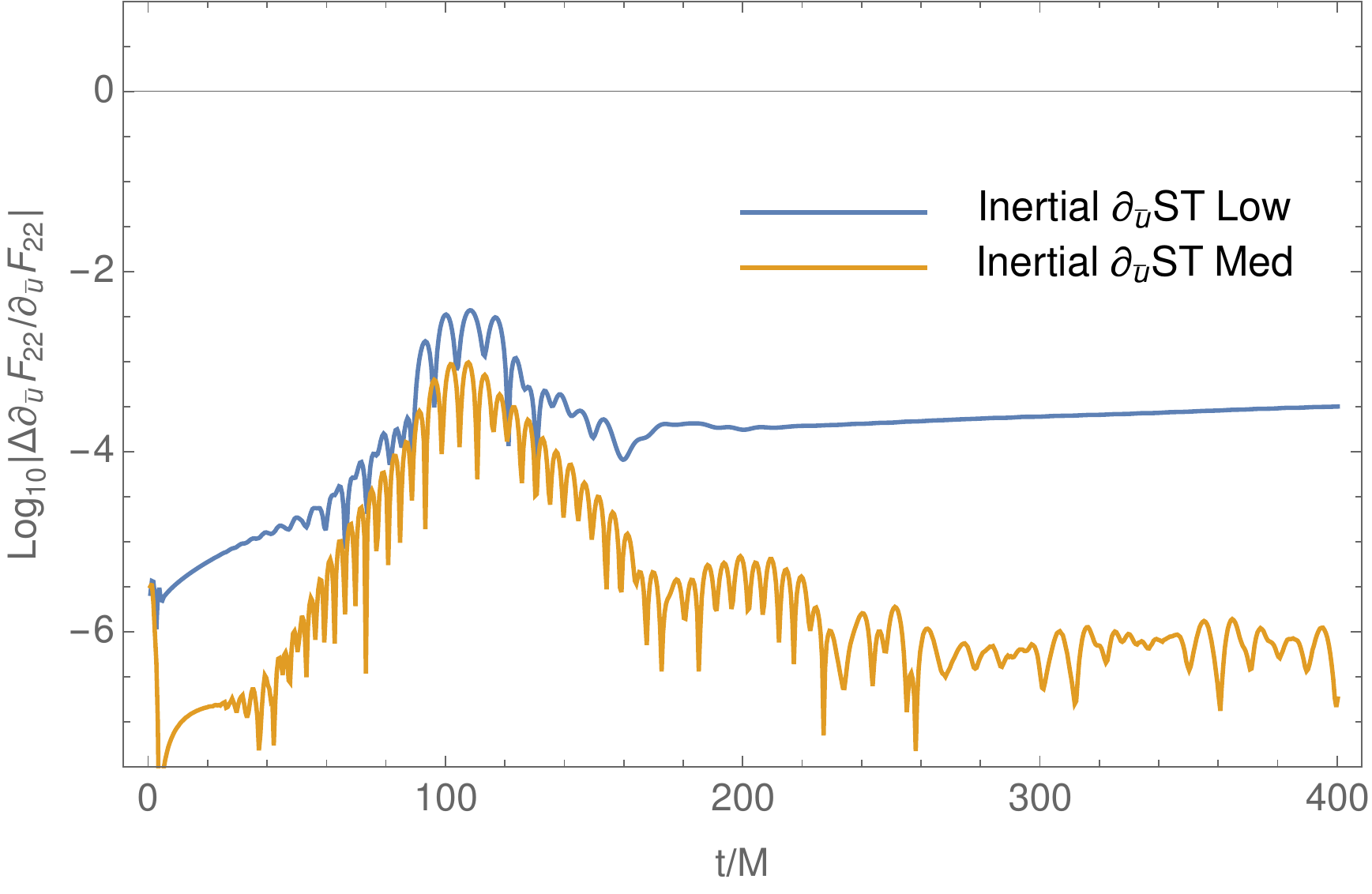}
    \label{fig:STS:FluxduConv}
  \end{subfigure}
  \caption{\small{ (\ref{fig:STS:KV}) A color gradient illustrates the nature of \(\xi_{ST}^{\tilde u}\),
  a \(|Y_{22}|\) supertranslation. (\ref{fig:STS:Flux}) The
      $(\ell=2,m=2)$ spherical harmonic component of the flux.
      (\ref{fig:STS:FluxConv}) Convergence of the flux is partially
      compromised by junk radiation, while the inertial time
      derivative of the flux (\ref{fig:STS:FluxduConv}) shows the
      appropriate $4^{th}$ order convergence following the junk phase. }}
  \label{fig:STS}
\end{figure}

The $(\ell=2,m=2)$ mode of the \(|Y_{22}|\)-derived supertranslation flux \(F_{ST}\) is shown in
Fig.~\ref{fig:STS:Flux}.  Convergence of the flux is shown in
Fig.~\ref{fig:STS:FluxConv} and its inertial time derivative in
Fig.~\ref{fig:STS:FluxduConv}.

\section{Conclusion}
%Hello @OverheardOnAph. Always check the power spectrum. Always.

In the context of Cauchy-characteristic evolution, we have developed the mathematical formalism
for computing the gravitational radiation fluxes to $\scri^+$ of energy-momentum,
angular-momentum-dipole-moment and supermomentum, associated with
the asymptotic symmetries of the BMS group. We have implemented this algorithm
as part of the Spectral Einstein Code (\texttt{SpEC}). The resulting code supplies a
uniform computation of the radiation strain, news function, Newman-Penrose radiative
$\psi_4^0$ curvature component and BMS fluxes in terms of inertial coordinates at $\scri^+$.
It  is a stable, convergent, and highly efficient code for determining all the physical attributes of the
gravitational radiation field.

Convergence tests were carried out based upon the simulation
of a generic precessing binary black hole. These tests showed that the numerical accuracy was
limited by the $4^{th}$ order time integrator, as opposed to the exponential convergence rate
expected of the spatial spectral code. The main source of error arose from the artificial junk radiation
introduced by the binary black hole initial data.

The accuracy for radiation strain, news function $N$ and $\psi_4^0$ were comparable to computations
using a prior version of the \texttt{SpEC} characteristic code. The same was also found for the computation
of energy-momentum flux, which is determined by the $\ell=0$ and $\ell=1$ components of $|N|^2$.

The computation of the angular momentum and supermomentum fluxes is more complicated than
the energy-momentum flux. In addition, there are ambiguities in their underlying construction, which
we base here upon the linkage integrals. However, these ambiguities are not as serious in
the case of the retarded time derivatives of the linkage flux $\dot F_\xi$, which only depend upon a
product of $\psi_4^0$ with the choice of BMS generator, according to (\ref{eq:fluxdot}). An
idealized strategy for studying, say, angular momentum would be to base its initial
value either on the Wald-Zoupas Hamiltonian approach~\cite{waldzoupas}
in the infinite past retarded time $u\rightarrow -\infty$
or on its unambiguous definition at spatial infinity,
where the initial flux should vanish. The dynamical properties of angular momentum can then be
studied by retarded time integrals of  $\dot F_\xi$. However, in practice this strategy would
require binary black hole data devoid of junk radiation, which is not true at least for the
generic precessing system simulated here. Following the
the initial phase of junk radiation, tests of $\dot F_\xi$ showed clear $4^{th}$ order convergence. 
However, the high derivatives involved in the calculation of $\dot F_\xi$ magnify the
effect of the junk radiation. 
 
 These considerations add to other important reasons to develop methods
 for obtaining binary black hole data which suppress junk radiation. In particular,
 this would allow application of our code to study the interesting question
 of how a supertranslation shift between the preferred Poincar{\' e} groups
 at $u=\pm\infty$ might affect angular momentum loss. 

\ack{We thank Nicholas Taylor for his generic precessing
  binary black hole run that we used to test and baseline code
  performance. We thank Mark Scheel, Yanbei Chen, and Christian
  Reisswig for their advice, support, and technical expertise.  This
  research used the Spectral Einstein Code
  (\texttt{SpEC})\cite{Mroue:2013PRL}. The Caltech cluster
  \texttt{zwicky.cacr.caltech.edu} is an essential resource for
  \texttt{\texttt{SpEC}} related research, supported by the Sherman
  Fairchild Foundation and by NSF award PHY-0960291. This research
  also used the Extreme Science and Engineering Discovery Environment
  (XSEDE) under grant TG-PHY990002. The UCSD cluster
  \texttt{ccom-boom.ucsd.edu} was used during code development. This
  project was supported by the Sherman Fairchild Foundation, and by
  NSF Grants PHY-1068881, AST-1333520, and CAREER Grant PHY-0956189 at
  Caltech. JW's research was supported by NSF grant PHY-1505965 to the
  University of Pittsburgh.

\section*{References}

\bibliographystyle{utphys} \bibliography{References/References.bib}

\providecommand{\href}[2]{#2}\begingroup\raggedright\begin{thebibliography}{10}

\bibitem{gw2016}
B.~P.~A. et~al. (LIGO Scientific~Collaboration and V.~Collaboration),
  ``Observation of gravitational waves from a binary black hole merger,'' {\em
  Phy. Rev. Lett.} {\bfseries 116} (2016) 061102.

\bibitem{Waldman2011}
S.~J. Waldman, ``The advanced ligo gravitational wave detector,'' Tech. Rep.
  LIGO-P0900115-v2, {LIGO} Project, 2011.

\bibitem{Accadia:2009zz}
T.~Accadia, F.~Acernese, F.~Antonucci, P.~Astone, G.~Ballardin, {\em et~al.},
  \href{http://dx.doi.org/10.1142/9789814374552_0313}{``{Plans for the upgrade
  of the gravitational wave detector VIRGO: Advanced VIRGO},''} in {\em
  Proceedings of the Twelfth Marcel Grossmann Meeting on General Relativity},
  T.~Damour, R.~T. Jantzen, and R.~Ruffini, eds., pp.~1738--1742.
\newblock 2009.

\bibitem{Grote:2010zz}
{\bfseries LIGO Scientific Collaboration} Collaboration, H.~Grote, ``{The GEO
  600 status},''
\href{http://dx.doi.org/10.1088/0264-9381/27/8/084003}{{\em Class.\ Quantum
  Grav.} {\bfseries 27} (2010) 084003}.
%%CITATION = CQGRD,27,084003;%%.

\bibitem{Somiya:2012}
K.~Somiya and the {KAGRA}~Collaboration, ``Detector configuration of
  {KAGRA}--the japanese cryogenic gravitational-wave detector,''
  \href{http://dx.doi.org/10.1088/0264-9381/29/12/124007}{{\em Class.\ Quantum
  Grav.} {\bfseries 29} no.~12, (2012) 124007}.

\bibitem{Tichy:2007hk}
W.~Tichy and P.~Marronetti, ``{Binary black hole mergers: Large kicks for
  generic spin orientations},'' {\em Phys.\ Rev.\ D} {\bfseries 76} (2007)
  061502(R).

\bibitem{Lousto:2011kp}
C.~O. Lousto and Y.~Zlochower, ``{Hangup Kicks: Still Larger Recoils by Partial
  Spin/Orbit Alignment of Black-Hole Binaries},''
  \href{http://dx.doi.org/10.1103/PhysRevLett.107.231102}{{\em Phys.\ Rev.\
  Lett.} {\bfseries 107} (2011) 231102},
\href{http://arxiv.org/abs/1108.2009}{{\ttfamily arXiv:1108.2009 [gr-qc]}}.
%%CITATION = ARXIV:1108.2009;%%.

\bibitem{Gonzalez2007}
J.~A. Gonz\'{a}lez, U.~Sperhake, B.~Br{\"u}gmann, M.~Hannam, and S.~Husa,
  ``Maximum kick from nonspinning black-hole binary inspiral,'' {\em Phys.\
  Rev.\ Lett.} {\bfseries 98} (2007) 091101,
  \href{http://arxiv.org/abs/gr-qc/0610154}{{\ttfamily gr-qc/0610154}}.

\bibitem{Favata2004}
M.~Favata, S.~A. Hughes, and D.~E. Holz, ``How black holes get their kicks:
  Gravitational radiation recoil revisited,'' {\em Astrophys. J.} {\bfseries
  607} (2004) L5--L8.

\bibitem{Baker2008}
J.~G. Baker, W.~D. Boggs, J.~Centrella, B.~J. Kelly, S.~T. McWilliams, M.~C.
  Miller, and J.~R. van Meter, ``Modeling kicks from the merger of generic
  black-hole binaries,'' {\em Astrophys.\ J.} {\bfseries 682} (2008) L29,
  \href{http://arxiv.org/abs/arXiv:0802.0416}{{\ttfamily arXiv:0802.0416}}.

\bibitem{Healy2008}
J.~Healy, F.~Herrmann, I.~Hinder, D.~M. Shoemaker, P.~Laguna, and R.~A.
  Matzner, ``Superkicks in hyperbolic encounters of binary black holes,'' {\em
  Phys. Rev. Lett.} {\bfseries 102} (2009) 041101,
  \href{http://arxiv.org/abs/0807.3292}{{\ttfamily arXiv:0807.3292 [gr-qc]}}.

\bibitem{Bondi1962}
H.~Bondi, M.~G.~J. van~der Burg, and A.~W.~K. Metzner, ``Gravitational waves in
  general relativity {VII}. {W}aves from axi-symmetric isolated systems,'' {\em
  Proc. R. Soc. Lond. A} {\bfseries 269} (1962) 21--52.

\bibitem{Sachs1962}
R.~K. Sachs, ``Gravitational waves in general relativity. {VIII}. waves in
  asymptotically flat space-time,'' {\em Proc. R. Soc. Lond. A} {\bfseries 270}
  no.~1340, (October, 1962) 103--126.
  \url{http://www.jstor.org/stable/2416200}.

\bibitem{Penrose1963}
R.~Penrose, ``Asymptotic properties of fields and space-times,'' {\em Phys.\
  Rev.\ Lett.} {\bfseries 10} no.~2, (1963) 66--68.

\bibitem{Winicour2009}
J.~Winicour, ``Characteristic evolution and matching,'' {\em Living Rev.~Rel.}
  {\bfseries 15} no.~2, (2012) . \url{http://www.livingreviews.org/lrr-2012-2}.

\bibitem{TamburinoWinicour1966}
L.~A. Tamburino and J.~H. Winicour, ``Gravitational fields in finite and
  conformal {B}ondi frames,'' {\em Phys. Rev.} {\bfseries 150} (1966)
  1039--1053. \url{http://link.aps.org/doi/10.1103/PhysRev.150.1039}.

\bibitem{Isaacson:1983}
R.~A. Isaacson, J.~S. Welling, and J.~Winicour, ``{Null cone computation of
  gravitational radiation},'' {\em {J. Math. Phys.}} {\bfseries {24}} (1983)
  1824. \url{http://iopscience.iop.org/0264-9381/30/7/075017}.

\bibitem{Bishop:1997ik}
N.~T. Bishop, R.~Gomez, L.~Lehner, M.~Maharaj, and J.~Winicour, ``{High-powered
  gravitational news},'' \href{http://dx.doi.org/10.1103/PhysRevD.56.6298}{{\em
  Phys. Rev.} {\bfseries D56} (1997) 6298--6309},
\href{http://arxiv.org/abs/gr-qc/9708065}{{\ttfamily arXiv:gr-qc/9708065}}.
%%CITATION = GR-QC/9708065;%%.

\bibitem{BabiucEtAl2008}
M.~C. Babiuc, N.~T. Bishop, B.~Szil{\'a}gyi, and J.~Winicour, ``Strategies for
  the characteristic extraction of gravitational waveforms,'' {\em Phys.\ Rev.\
  D} {\bfseries 79} (2009) 084011,
  \href{http://arxiv.org/abs/0808.0861}{{\ttfamily arXiv:0808.0861}}.

\bibitem{Babiuc:2010ze}
M.~C. Babiuc, B.~Szil\'agyi, J.~Winicour, and Y.~Zlochower, ``A characteristic
  extraction tool for gravitational waveforms,''
  \href{http://dx.doi.org/10.1103/PhysRevD.84.044057}{{\em Phys.\ Rev.\ D}
  {\bfseries 84} (Aug, 2011) 044057},
  \href{http://arxiv.org/abs/1011.4223}{{\ttfamily arXiv:1011.4223 [gr-qc]}}.
  \url{http://link.aps.org/doi/10.1103/PhysRevD.84.044057}.

\bibitem{Handmer:2014}
C.~J. Handmer and B.~Szil\'{a}gyi, ``Spectral characteristic evolution: A new
  algorithm for gravitational wave propagation,'' {\em Classical and Quantum
  Gravity} {\bfseries 32} (2015) 025008,
  \href{http://arxiv.org/abs/1406.7029}{{\ttfamily arXiv:1406.7029}}.

\bibitem{Handmer:2015}
C.~J. Handmer, B.~Szil\'{a}gyi, and J.~Winicour, ``Gauge invariant spectral
  characteristic extraction,''
  \href{http://arxiv.org/abs/1502.06987}{{\ttfamily arXiv:1502.06987}}.

\bibitem{Geroch1977}
R.~Geroch, {\em Asymptotic Structure of Spacetime}.
\newblock Plenum, New York, 1995.

\bibitem{Szabados2004}
L.~B. Szabados, ``Quasi-local energy-momentum and angular momentum in {G}eneral
  {R}elativity: A review article,'' {\em Living Rev.~Rel.} {\bfseries 12}
  no.~4, (2009) . \url{http://www.livingreviews.org/lrr-2009-4}.

\bibitem{Ashtekar-Hansen}
A.~Ashtekar and R.~O. Hansen, ``A unified treatment of null and spatial
  infinity in general relativity. i. universal structure, asymptotic
  symmetries, and conserved quantities at spatial infinity,'' {\em J. Math.
  Phys.} {\bfseries 19} (1978) 1542--1566.

\bibitem{sachsbms}
R.~K. Sachs, ``Asymptotic symmetries in gravitational theory,'' {\em Phys.
  Rev.} {\bfseries 128} (1962) 2851--2864.

\bibitem{NewmanPenrose1966}
E.~T. Newman and R.~Penrose, ``Note on the {B}ondi--{M}etzner--{S}achs group,''
  \href{http://dx.doi.org/10.1063/1.1931221}{{\em J.\ Math.\ Phys.} {\bfseries
  7} (1966) 863--870}. \url{http://link.aip.org/link/?JMP/7/863/1}.

\bibitem{jwangular}
J.~Winicour, ``Angular momentum in general relativity,'' in {\em General
  Relativity and Gravitation}, A.~Held, ed., vol.~2, pp.~71--96.
\newblock Plenum Press, New York, 1980.

\bibitem{Geroch1981}
R.~Geroch and J.~Winicour, ``Linkages in general relativity,'' {\em J.\ Math.\
  Phys.} {\bfseries 22} (1981) 803.

\bibitem{jwemem}
J.~Winicour, ``Global aspects of radiation memory,'' {\em Classical and quantum
  gravity} {\bfseries 31} no.~20, (2014) 205003.

\bibitem{win1968}
J.~Winicour, ``Some total invariants of asymptotically flat space-times,'' {\em
  J. Math. Phys.} {\bfseries 9} no.~6, (1968) 861--867.

\bibitem{Komar:1958wp}
A.~Komar, ``{Covariant conservation laws in general relativity},''
\href{http://dx.doi.org/10.1103/PhysRev.113.934}{{\em Phys.~Rev.} {\bfseries
  113} (1959) 934--936}.
%%CITATION = PHRVA,113,934;%%.

\bibitem{ashtekar-streubel}
A.~Ashtekar and M.~Streubel, ``Symplectic geometry of radiative modes and
  conserved quantities at null infinity,'' {\em Proc. R. Soc. Lond. A}
  {\bfseries 376} no.~1767, (1981) 585--607.

\bibitem{waldzoupas}
R.~Wald and A.~Zoupas, ``General definition of "conserved quantities" in
  general relativity and other theories of gravity,'' {\em Phys. Rev. D}
  {\bfseries 61} (2000) 084027.

\bibitem{dray1984angular}
T.~Dray and M.~Streubel, ``Angular momentum at null infinity,'' {\em Classical
  and Quantum Gravity} {\bfseries 1} no.~1, (1984) 15.

\bibitem{ashtekar-win}
A.~Ashtekar and J.~Winicour, ``Linkages and hamiltonians at null infinity,''
  {\em Journal of Mathematical Physics} {\bfseries 23} no.~12, (1982)
  2410--2417.

\bibitem{helfer}
A.~D. Helfer, ``Angular momentum of isolated systems,'' {\em General Relativity
  and Gravitation} {\bfseries 39} no.~12, (2007) 2125--2147.

\bibitem{penrose1982quasi}
R.~Penrose, ``Quasi-local mass and angular momentum in general relativity,''
  {\em Proceedings of the Royal Society of London A} {\bfseries 381} no.~1780,
  (1982) 53--63.

\bibitem{nesterov1997quasigroups}
A.~I. Nesterov, ``Quasigroups, asymptotic symmetries, and conservation laws in
  general relativity,'' {\em Phys. Rev. D} {\bfseries 56} no.~12, (1997) R7498.

\bibitem{deshbish}
N.~Bishop and S.~Deshingkar, ``New approach to calculating the news,'' {\em
  Phys. Rev. D.} {\bfseries 68} (2003) 024031.

\bibitem{Bishop:2013}
N.~T. Bishop and C.~Reisswig, ``{The gravitational wave strain in the
  characteristic formalism of numerical relativity},'' {\em Gen. Rel. Grav.}
  {\bfseries 46} (2014) 1843.

\bibitem{boyle16}
M.~Boyle, ``Transformations of asymptotic gravitational-wave data,'' {\em Phys.
  Rev. D} {\bfseries 93} (2016) 084031.

\bibitem{helfer10}
A.~D. Helfer, ``Estimating energy-momentum and angular momentum near null
  infinity,'' {\em Phys. Rev. D} {\bfseries 81} (2010) 084001.

\bibitem{Winicour1983}
J.~Winicour, ``Newtonian gravity on the null cone,'' {\em J. Math. Phys.}
  {\bfseries 1193} (1983) .

\bibitem{Taylor:2013zia}
N.~W. Taylor, M.~Boyle, C.~Reisswig, M.~A. Scheel, T.~Chu, L.~E. Kidder, and
  B.~Szil{\'a}gyi, ``Comparing gravitational waveform extrapolation to
  {C}auchy-characteristic extraction in binary black hole simulations,''
  \href{http://dx.doi.org/10.1103/PhysRevD.88.124010}{{\em Phys. Rev. D}
  {\bfseries 88} (Dec, 2013) 124010},
  \href{http://arxiv.org/abs/1309.3605}{{\ttfamily arXiv:1309.3605 [gr-qc]}}.
  \url{http://link.aps.org/doi/10.1103/PhysRevD.88.124010}.

\bibitem{Mroue:2013PRL}
A.~H. Mroue, M.~A. Scheel, B.~Szilagyi, H.~P. Pfeiffer, M.~Boyle, D.~A.
  Hemberger, L.~E. Kidder, G.~Lovelace, S.~Ossokine, N.~W. Taylor,
  A.~Zenginoglu, L.~T. Buchman, T.~Chu, E.~Foley, M.~Giesler, R.~Owen, and
  S.~A. Teukolsky, ``A catalog of 174 binary black hole simulations for
  gravitational wave astronomy,'' {\em Phys.\ Rev.\ Lett.} {\bfseries 111}
  (2013) 241104, \href{http://arxiv.org/abs/1304.6077}{{\ttfamily
  arXiv:1304.6077 [gr-qc]}}.

\end{thebibliography}\endgroup

\end{document}